\documentclass[onecolumn,amsmath,amssymb,amsfonts,showkeys,preprintnumbers,a4paper,nofootinbib,superscriptaddress]{revtex4}

\usepackage{graphicx,wasysym}
\usepackage{rotating}
\usepackage{color}
\usepackage{bm}
\usepackage{hyperref}
\usepackage{url}
\usepackage{multirow}

\newcommand{\be}{\begin{equation}}
\newcommand{\ee}{\end{equation}}

\newcommand{\MeV}{{\rm MeV}}
\newcommand{\GeV}{{\rm GeV}}
\newcommand{\GV}{{\rm GV}}

\newcommand{\erg}{{\rm ergs}}

\newcommand{\kpc}{{\rm kpc}}

\newcommand{\cm}{{\rm cm}}
\newcommand{\km}{{\rm km}}

\newcommand{\s}{{\rm s}}
\newcommand{\yr}{{\rm yr}}
\newcommand{\sr}{{\rm sr}}

\begin{document}

\title{Diffuse Galactic Gamma Rays at intermediate and high latitudes.\\ 
I.~Constraints on the ISM properties}% Force line breaks with \\

\author{Ilias Cholis}
\email{ilias.cholis@sissa.it}
\affiliation{SISSA, Via Bonomea, 265, 34136 Trieste, Italy}
\affiliation{INFN, Sezione di Trieste, Via Bonomea 265, 34136 Trieste, Italy}
\author{Maryam Tavakoli}
\email{tavakoli@sissa.it}
\affiliation{SISSA, Via Bonomea, 265, 34136 Trieste, Italy}
\affiliation{INFN, Sezione di Trieste, Via Bonomea 265, 34136 Trieste, Italy}
\author{Carmelo Evoli}
\email{carmelo.evoli@me.com}
\affiliation{SISSA, Via Bonomea, 265, 34136 Trieste, Italy}
\affiliation{National Astronomical Observatories, Chinese Academy of Sciences, 20A Datun Road, Beijing 100012, P.R. China} 
\author{Luca Maccione}
\email{luca.maccione@lmu.de}
\affiliation{DESY, Theory Group, Notkestra{\ss}e 85, D-22607 Hamburg, Germany}
\affiliation{Max Planck Institut, F\"ohringer Ring 6, D-80805, M\"unchen, Germany}
\author{Piero Ullio}
\email{ullio@sissa.it}
\affiliation{SISSA, Via Bonomea, 265, 34136 Trieste, Italy}
\affiliation{INFN, Sezione di Trieste, Via Bonomea 265, 34136 Trieste, Italy}

\date{\today}

\preprint{DESY 11-106}

\begin{abstract}
We study the high latitude ($|b|>10^\circ$) diffuse $\gamma$-ray emission in the Galaxy in light of the recently published 
data from the \textit{Fermi} collaboration at energies between 100 MeV and 100 GeV.
The unprecedented accuracy in these measurements allows to probe and constrain the properties of sources and propagation 
of cosmic rays (CRs) in the Galaxy, as well as confirming conventional assumptions 
made on the interstellar medium (ISM). 
Using the publicly available DRAGON code, 
that has been shown to reproduce local measurements of CRs,
we study assumptions 
made in the literature on atomic (HI) and molecular hydrogen (H2) gas distributions in the ISM, and non spatially uniform models of 
diffusion in the Galaxy.
By performing a combined analysis of CR and $\gamma$-ray spectra, we derive constraints on the properties of the 
ISM gas distribution and the vertical scale height of galactic CR diffusion, which may have implications also on indirect Dark Matter detection.
We also discuss some of the possible interpretations 
of the break at high rigidity in CR protons and helium spectra, recently observed by \textit{PAMELA} and 
their impact on $\gamma$-rays. 
\end{abstract}


\maketitle
%------------------------------------------------------------------------------
\section{Introduction}

The study of the physics of Galactic cosmic rays (CRs) is one of the most active research areas at present.
Sensible advances in the field have come in connection to the wealth of high-accuracy data
recently collected by several new instruments, with further progresses expected in the upcoming
future. In particular, since its launch three years ago, the \textit{Fermi Gamma-Ray Telescope} \cite{1999APh....11..277G}
has been producing the most detailed and precise maps of the $\gamma$-ray sky ever, given
its wide energy coverage and excellent energy resolution, its large effective area and field of view,
as well as the best angular resolution for a $\gamma$-ray detector in space (for details on
the performances of the instrument see \cite{Rando:2009yq}). Since the interaction of Galactic CRs with
the interstellar medium give rise to a copious $\gamma$-ray yield, the diffuse emission in the Milky Way is by
far the brightest source detected by \textit{Fermi}. It is then expected that \textit{Fermi} data will drive a significant
improvement in understanding the origin and propagation of CRs.

A key feature that would be particularly important to establish is whether there is room for (or eventually even need for)
an exotic source of $\gamma$-rays and/or CR leptons in the Milky Way, on top of the astrophysical sources
most plausibly providing the bulk of Galactic CRs, namely supernova remnants (SNRs) and, possibly, pulsars.
Such exotic component is predicted, e.g., in connection to the dark matter halo of the Galaxy for several dark matter
candidates, the prime example being
Weakly Interacting Massive Particles (WIMPs), i.e.~early Universe relics which have a small but finite probability
of annihilating in pairs and produce multi-GeV (or -TeV) yields (for a recent review on WIMP dark matter, see, e.g.
\cite{2010pdmo.book.....B}; the issue of probing and constraining dark matter models with \textit{Fermi}-LAT data was
recently discussed, e.g.,~\cite{Bergstrom:2001jj, Ullio:2002pj, Bertone:2006kr, Baltz:2008wd, Cholis:2008wq,
Barger:2009yt, Buchmuller:2009xv, Ibarra:2009dr}). Suggestions of an improvement in the fit of \textit{Fermi}
$\gamma$-ray data at intermediate latitudes taking into account the spectral hardening due to a signal
from annihilating dark matter have already been proposed in the literature, see, e.g.,~\cite{Barger:2010mc}.
On the other hand it is evident that for such kind of analysis it is crucial a very accurate modeling of background
components, whether of galactic, extragalactic, or instrumental origin.

The question we wish to address in this work is to what extent the cross correlation of the $\gamma$-ray data with
other local CR probes, namely the local measurements of the flux of CR primaries, the ratio of secondary
to primary CR nuclei and the flux of electrons and positrons, helps disentangle degeneracies between the type and
distribution of sources and the propagation modeling 
similarly to what was done in the past with the \textit{EGRET} data by \cite{Strong:1998fr, Strong:2004de}. The emphasis on the issue of locality comes from the
observation that, in the vast majority of works in the literature, propagation of CRs in the Galaxy is treated
with an effective approach in terms of a diffusive/convective equation encoding, on average, through a set of simplifying
assumptions and few parameters to be tuned to the data, the physical process of charged particles making a
random walk in the regular and turbulent Galactic magnetic fields. Local measurements give fairly good tests
of average properties of the local medium, with the physical averaging scale depending
on the species considered, and with the caveat that some of the parameters in the propagation model show
patterns of degeneracy, as studied at length in the literature. After selecting models fulfilling these local constraints,
covering a wide range of different physical propagation regimes, as well as gas distributions and, to
some extent, galactic (but excluding dark matter) source distributions, we will derive predictions for the
$\gamma$-ray flux at intermediate and high latitudes, compare against \textit{Fermi} data and discuss whether a
discrimination among the different models is possible. The choice of excluding from the analysis low latitude data is connected
again to the issue of focussing on local properties, since although $\gamma$-ray data naturally reflect a global
observable, summing along the line of sight contributions to the emissivity from all regions of the Galaxy, high latitude
fluxes are dominated by the local terms.

The \textit{Fermi} Collaboration has first published spectral data at intermediate latitudes in \cite{Abdo:2009mr} disproving the presence of a GeV excess in the diffuse $\gamma$-ray spectra suggested by \textit{EGRET} \cite{Hunger:1997we}.  
We will consider in the following the updated published spectral data from the \textit{Fermi} Collaboration
at latitudes $10^{\circ} < \mid b \mid < 20^{\circ}$, $20^{\circ} < \mid b \mid < 60^{\circ}$,
$\mid b \mid > 60^{\circ}$ and $0^{\circ} < l < 360^{\circ}$, and energies up to $\sim 100$
GeV~\cite{Abdo:2010nz}. Such spectral data have been evaluated using a more clean sample of
$\gamma$-rays than the publicly available ``Data Clean 4'' (P6V11)~\cite{fermidata};
furthermore the CR contamination to that remaining data-set has been modeled in \cite{Abdo:2010nz}
and its contribution (isotropic) to the spectra has been subtracted, thus making these spectra ideal
for studying the large scale properties of the Galaxy.
For $E_{\gamma} > 100$ GeV, statistical errors and contamination of CR electrons and nuclei
result in great uncertainty on the exact spectrum of the $\gamma$-rays \cite{Abdo:2010nz}
(see also \cite{fermicaveats}).

Since we do not include the data from $\mid b \mid < 10^{\circ}$ in our analysis, we
mainly probe the properties of CR propagation and to some extent the gasses but we are less
sensitive to the properties of the distributions of galactic sources.
The assumptions made on how CRs diffuse away
from the disk, can have a significant effect on the $\gamma$-ray distribution from inverse Compton
scattering (ICS) by electrons/positrons that are either diffusive shock accelerated ISM $e^{-}$
at SNRs,
secondaries from inelastic pp and pHe collisions (predominantly), or produced in Pulsars magnetospheres and accelerated within
Pulsar Wind Nebulae (PWNe).
Also since magnetic fields decrease as we move radially away from the galactic center, we
expect a radial profile as well in the diffusion \cite{Dobler:2011mk} of CRs which can also have an
effect on the observed $\gamma$-ray spectra.
Gamma-rays from $\pi^{0}$ decays produced in inelastic collisions of CR protons and heavier nuclei
with ISM nuclei, and bremsstrahlung off electrons are dominant contributions to the $\gamma$-ray
spectrum from 100 MeV up to at least 50 GeV. Thus the observed $\gamma$-ray spectra can also be used
to probe the ISM properties, such as confirming models for the HI and H2 gas distributions.

Studies of the sky observed by \textit{Fermi}, at latitudes $\mid b \mid > 5^{\circ}$
have shown an excess of diffuse $\gamma$-rays towards the center of the Galaxy and up to latitudes
of $\mid b \mid \approx 50^{\circ}$ \cite{Dobler:2009xz, Su:2010qj} known as the ``Fermi haze'' or
``Fermi bubbles'' that can be of either
astrophysical \cite{Su:2010qj, Crocker:2010dg, Guo:2011eg, Mertsch:2011es} or DM origin
\cite{Dobler:2011mk,Cholis:2009va}.
Even though this feature(s) extends up to high latitudes, since it is confined in longitude
within $\mid l \mid  \lesssim 20^{\circ}$ its effect on the spectra in the regions of our study is washed out.
Thus our analysis can not probe the properties of the Galaxy in that region.
Nor it can in the region of ($\mid b \mid < 10^{\circ}$), where
\cite{Goodenough:2009gk, Hooper:2010mq, Cholis:2009gv} have suggested the presence of a DM signal.

This paper is organized as follows. In section~\ref{sec:DRAGON_Assump}, we summarize
the assumptions that we make on the primary CR sources, diffusion, magnetic fields
and gas models, as well as briefly present the tool we use to solve numerically the CR propagation
equation, namely the DRAGON code~\cite{Evoli:2008dv}.
In section~\ref{sec:astro_analysis} we describe the analysis that we conduct in constraining the CR
propagation and ISM properties, by fitting to the local fluxes of CRs.
Section~\ref{sec:astro_results} is devoted to studying the effects that the various
assumptions on diffusion of CRs, ISM gasses, distribution of CR sources and
$\gamma$-ray production cross-section from pp-collisions have on the $\gamma$-ray
spectra.
We also place constraints imposed on CR propagation models and on models of ISM gasses
distributions using the diffuse $\gamma$-ray data\cite{Abdo:2010nz}, as well as discuss the
implications of our findings. Finally in section~\ref{sec:Conclusions} we present our conclusions.

%--------------------------------------------------------------------------------------------------
\section{Cosmic Ray propagation}
\label{sec:DRAGON_Assump}

As suggested in \cite{1990acr..book.....B} the propagation of CRs in the Galaxy at energies below $10^{17}$eV can be described by:
\begin{eqnarray} 
\frac{\partial \psi(\vec{r},p,t)}{\partial t} &=&q(\vec{r},p,t)+\vec{\nabla}.
(D_{xx}\vec{\nabla}\psi)+\frac{\partial}{\partial p}\Big[p^2D_{pp}\frac{\partial}{\partial p}
(\frac{\psi}{p^2})\Big]-\frac{\partial}{\partial p}(\dot{p}\psi) \nonumber \\
&-&\vec{\nabla}.(\vec{V}\psi)+
\frac{\partial}{\partial p}\Big[\frac{p}{3}(\vec{\nabla}.\vec{V})
\psi\Big]-\frac{\psi}{\tau_{frag}}-\frac{\psi}{\tau_{decay}}
\label{eq:CR_trans}
\end{eqnarray}
where $\psi(\vec{r},p,t)$ is the CR density per unit particle momentum, or in terms of phase 
space density $f(p)$, $\psi(p)dp=4 \pi p^2 f(p)dp$, $q(\vec{r},p,t)$ is the source term including primary, spallation and decay of heavier CR species. $D_{xx}(\vec{r})$ is the diffusion tensor in physical space and $D_{pp}(\vec{r})$ 
the diffusion coefficient in momentum space. $\dot{p}$ is the momentum 
loss rate due to interactions with ISM, the Galactic magnetic field 
or the interstellar radiation field (ISRF), $\vec{V}$ is the convection 
velocity, and 
$\tau_{frag}$ and $\tau_{decay}$ are the time scales for fragmentation loss 
and radioactive decay respectively.

For our simulations we use DRAGON \cite{Evoli:2008dv, DiBernardo:2009ku, DiBernardo:2010is, DRAGONweb} that numerically solves Eq.~\ref{eq:CR_trans} in the steady state approximation $\partial \psi / \partial t = 0$, assuming cylindrical symmetry, 
in a 2+1-D grid where each point is described by its galactocentric radial distance $r \in (0,20)$ kpc, the distance from the galactic plane $z \in (-L, L)$ 
with $L \leq 20$ kpc and momentum $p$.
%---------------------------------------------------------------------------------
\subsection{Primary Sources of CRs}
\label{sec:CRSources}

We consider SNRs as the primary CR sources up to energies of $\sim$100TeV \cite{Aharonian:2006ws}.
For each nucleus $i$ of charge $Z$, the source term describing the injection of CRs in the ISM is given as a function of rigidity, $R= p/Z$, by:
\be 
q_i(r,z,E)=f_s(r,z)q_{0,i}\left(\frac{R(E)}{R_0}\right)^{-\gamma^i} 
\label{eq:injNucl}
\ee
where $q_{0,i}$ is the normalization of the injected CR species, and $f_s(r,z)$ traces the spatial distribution of SNRs. In our reference model we consider the distribution by \cite{Ferriere:2001rg}, which is derived on the basis of pulsar and progenitor star surveys. Alternative models will be considered as well (see section~\ref{sec:SRN}). 

Being interested in high energy $\gamma$-ray fluxes, we pay particular attention to the proton spectrum, given that protons provide the dominant contribution to the diffuse $\gamma$-ray spectra in the whole energy range we consider (see~\ref{sec:ElossDiffGammas}). 
We allow for a CR proton injection spectrum described by a broken power-law:
\be
\frac{dN_{p}}{dR} \propto \left(\frac{R}{R_{0,j}^p} \right)^{-\gamma^{p}_{j}},
\label{eq:Spectra}
\ee
with two breaks at $R=R_{0,1}^p \sim 10-30~\GV$ and $R=R_{0,2}^p \sim 300$ GV
with the spectral index $\gamma^{p}_{1}$ lying in the range $1.85-2.1$ at low rigidities, $\gamma^{p}_{2}$ in the range $2.3-2.5$ at intermediate rigidities
 and $\gamma^{p}_{3}$ ranging in $2.18-2.35$ at high rigidities. 
This choice is motivated {\it a posteriori} by fitting the local proton data and
in particular from the combined fit of the recent
\textit{PAMELA} proton 
spectrum \cite{Adriani:2011cu}, and CREAM spectrum above $2.5$ TV \cite{Yoon:2011zz}.
  
Electrons and positrons accelerated between a pulsar and the termination shock of the wind nebula, may also contribute to the high energy $e^{\pm}$ spectrum \cite{Kobayashi:2003kp, Hooper:2008kg, Profumo:2008ms, Malyshev:2009tw, DiBernardo:2010is}, and then to the $\gamma$-ray flux \cite{Zhang:2008tb, FaucherGiguere:2009df}. In particular, middle aged pulsars were found to be particularly well suited \cite{Grasso:2009ma,DiBernardo:2010is}.
Each pulsar contribution to the $e^{\pm}$ fluxes can be described by an injection
spectrum $\sim E^{-n}$ with a high energy break $E_{b}$ which is estimated at the time the surrounding PWN is disrupted leading to the $e^{\pm}$ escaping into the ISM \cite{Malyshev:2009tw}.~\footnote{Even though it is clear that the higher energy $e^{\pm}$ escape
earlier into the ISM, the differences in the estimated time scales for 
$e^{\pm}$ of energies between 1-$10^{4}$ GeV to escape into the ISM are
negligible compared to the propagation time from the PWN to us 
\cite{Gelfand:2009aa, Malyshev:2009tw}.} 
Furthermore, each pulsar has an initial rotational energy $W_{0}$ of which only a portion $\eta$ is injected into 
the ISM as CR $e^{\pm}$. The ranges for those parameters within different pulsars are very broad. Indeed,  
$n$ can range between 1 and 2, $W_{0} \sim 10^{49 - 50}$ erg,
$\eta \sim 0.1$ \cite{Gelfand:2009aa, Malyshev:2009tw} and $E_{b} \sim 10$ TeV. However, the actual observed 
flux  of $e^{\pm}$ in our position from a pulsar has a break that is related to the cooling time (from ICS and synchrotron radiation) 
of the $e^{\pm}$ during their propagation in the ISM \cite{Kobayashi:2003kp, Malyshev:2009tw}. 
To account for these effects, we choose to fit the properties of a pulsar distribution following the parametrization of \cite{Malyshev:2009tw}:  
\be
Q_{p}(r,z,t,E) = J_{0} E^{-n} e^{-E/M} f_{p}(r,z). 
\label{eq:PulsarSource}
\ee
M is a ``statistical" cut-off, $n$ the injection index for the distribution of pulsars,\footnote{Due to energy losses, the fluxes of  CR $e^{\pm}$ with $E\sim$ TeV, at any position of the Galaxy, are typically dominated by the contribution from the sources within $\sim O(10^2)$ pc from each position. Thus the CR $e^{\pm}$ spectra at $\sim$ TeV energies will differ significantly between different positions of the Galaxy (even at the same galacto-centric distances). Since the IC $\gamma$-ray spectra are studied in wide regions of the sky (and thus the Galaxy), we care for the averaged $e^{\pm}$ flux, as measured by many different observers in the Galaxy.
For such a (statistically) averaged $e^{\pm}$ flux from pulsars, we follow \cite{Malyshev:2009tw} where the statistical cut-off $M$ and injection index $n$ refer to the statistically averaged values.}

\be
J_{0} = \frac{\eta W_{0} N_{b}}{\Gamma(2-n) M^{2-n} V_{gal}}, 
\ee 
(see Eq.~24 of \cite{Malyshev:2009tw}) with $N_{b}$ the pulsar birth rate in the Galaxy and
\be
V_{gal} = \int_{-z_{max}}^{z_{max}}\int_{0}^{rmax}dz \, dr \, 2 \pi r \, f_{p}(r,z).    
\ee
$f_{p}(r,z)$ describes the spatial distribution of young and middle aged pulsars.
Since pulsars have typical kick 3D speeds of 200-400 km/s \cite{2009ASSL..357.....B, 2004MNRAS.352.1439H}  
, a $10^5$ yr old (middle aged) pulsar would move away from its original position
by $\sim 30$ pc, and thus the spatial distribution of middle aged pulsars is
practically identical to that of their birth distribution  
in the Galaxy as given in \cite{FaucherGiguere:2005ny}:
\be
 f_{p}(r,z) \propto \left(\frac{r+R_{1}}{r{\odot}+R_{1}}\right)^{a} \exp{\left[ -b\left(\frac{r-r_{\odot}}{r_{\odot}+R_{1}}\right) \right]} \exp{\left[-\frac{\mid z \mid}{z_{1}}\right]},
\label{eq:PulsarSpatDistr}
\ee 
with $R_{1} =0.55$ kpc, $z_{1}=0.1$ kpc, $a = 1.64$ and $b=4$.  

%------------------------------------------------------------------------------
\subsection{Diffusion and Magnetic Fields}
\label{sec:Diffusion_Bfields}

Our galaxy is permeated by a large scale, so called regular, magnetic field, and by a randomly varying, so called turbulent, magnetic field with comparable strength on the disk. 
The large scale galactic magnetic field is generally assumed to be a bi-symmetrical spiral with a small pitch angle \cite{Jansson:2009ip}. Here we assume that the regular magnetic field is purely azimuthal, $ \vec{B_0}=B_0 \hat{\phi}$, and has the form
\be 
B_0=B_h\exp{\left(-\frac{r-r_{\astrosun}}{r_h}\right)}\exp{\left(-\frac{|z|}{z_h}\right)}. 
\label{eq:Bfield}
\ee 
Based on the analysis of WMAP synchrotron intensity and polarization data 
in \cite{MivilleDeschenes:2008hn}, 
as well as works including extragalactic rotation 
measures \cite{Sun:2007mx, beck:2009mm, Jansson:2009ip, 2010MNRAS.401.1013J}, 
we choose $B_h=3 \, \mu$G and $r_h=11$ kpc, with vertical scale $z_h=2$ kpc. 
Although these values are affected by large uncertainties, they have little impact
on our analysis, since the magnetic field enters only in the electron energy losses, 
which are anyway dominated by the ICS losses above few GeV, as we show in Appendix~\ref{sec:SynchLoss}.

The diffusion tensor can be in general decomposed in a component parallel to the direction of the regular magnetic field, $D_\parallel$, and a component describing diffusion perpendicular to the regular magnetic field, $D_\perp$. It can be shown \cite{1993A&A...268..726P} that, assuming cylindrical symmetry and that the regular magnetic field is azimuthally symmetric, parallel diffusion is irrelevant and only $D_\perp$ has an effect.
We consider then for simplicity that the diffusion is described by just one quantity, 
the diffusion coefficient. 

The diffusion coefficient is in general expected to depend on the position, because turbulence is not uniformly distributed in the Galaxy.

In a phenomenological approach, we choose $D$ to be described by:
\be
D(r,z,R)=D_0 \beta ^{\eta}\left(\frac{R}{R_0}\right)^{\delta}\exp{\left(\frac{r-r_{\astrosun}}{r_d}\right)}
\exp{\left(\frac{|z|}{z_d}\right)}
\label{eq:DiffCoef}
\ee
with the radial and vertical scales, $r_d$ and $z_d$, defining the diffusion profile in the Galaxy. 
Such a parametrization of the diffusion coefficient can be motivated on large scales since the Galactic magnetic field decreases away from the galactic center (both in $r$ and $z$). Particles gyrating along ordered field lines, may scatter due to magnetic irregularities. As the magnetic fields become weaker, assuming that the particles gyroradius remains small enough that the particles probe the ordered field component, the diffusion length (and coefficient) will increase. 
We choose $R_0=3$ GV as the reference rigidity, while $\delta$ is the diffusion spectral index which is related to the ISM turbulence power-spectrum. 
The dependence of diffusion on the particle velocity, $\beta=v_p/c$, is naturally expected to be linear ($\eta=1$), 
however the analysis by \cite{Ptuskin:2005ax} shows an increase in diffusion at low energies. 
To account for such a possibility, the parameter $\eta$ has been introduced (see e.g. \cite{Maurin:2010zp, DiBernardo:2010is, Putze:2010fr}). We will also consider the case where there is a break in the diffusion spectral index $\delta$.

In addition to spatial diffusion, the scattering of CRs on randomly moving 
magneto-hydro-dynamical
(MHD) waves leads to diffusion in momentum space which results in stochastic acceleration of CRs. 
The corresponding diffusion coefficient in momentum space is related to the 
diffusion coefficient in physical space by $D_{pp}\propto p^2 v_A /D_{xx}$, 
where $v_A$ is the Alfv\'en velocity, associated to the propagation of MHD waves \cite{1990acr..book.....B}. 

%------------------------------------------------------------------------------
\subsection{Energy Losses and Diffuse Gamma-Rays}
\label{sec:ElossDiffGammas}
CR nuclei and CR electrons and positrons lose energy during propagation in the ISM.
Depending on the energy, electrons and positrons energy losses are 
dominated by inverse Compton scattering of low energy photons of the 
interstellar radiation field (ISRF), for which we use the model of \cite{Porter:2005qx}, and, to a less extent, by synchrotron radiation. At energies $\lesssim 1~\GeV$ bremsstrahlung, ionization and Coulomb losses become relevant. 
For protons and nuclei ionization and Coulomb losses are the dominant continuous energy loss mechanisms.

In addition to ionization and Coulomb losses, collisions of heavy nuclei with 
hydrogen or helium of the ISM gas can lead to inelastic scattering that can also cause the fragmentation of the parent CRs.

CRs can also undergo radioactive decays, with the radioactive isotopes
created both by fragmentation of heavier nuclei (e.g.~$^{10}$Be is created from B, C, N, O)
and directly in CR sources (e.g.~$^{26}$Al).
All these processes are incorporated in DRAGON \footnote{Given similar 
assumptions, for the propagation of CRs in the Galaxy, we have checked that
the output of DRAGON and Galprop\cite{Strong:2007nh} agree within an accuracy 
of $5-10\%$ in the total galactic diffuse $\gamma$-ray flux; and by an 
accuracy of no more than $5\%$ in the total $\gamma$-ray flux, at energies 
between 100 MeV and 300 GeV in the three windows that we study 
($0^{\circ}<l<360^{\circ}$, $10^{\circ} < \mid b\mid < 20^{\circ}$ / 
$20^{\circ} < \mid b\mid < 60^{\circ}$ / $\mid b\mid > 60^{\circ}$).}.

For $E_{\gamma} > 10$ MeV, there are three processes that contribute mainly to the diffuse galactic component.
Inelastic $pp$ collisions producing  $\pi^{0}$s which subsequently decay to 2 photons,
constitute the main contribution to the diffuse gamma 
ray flux from the Milky Way in the intermediate GeV range, and trace the ISM target distribution dominated by the HI and H2 gasses.
The spectral shape of these gamma rays is essentially determined by the spectral shape of the proton spectrum along the line of sight. 
CR electrons may also produce $\gamma$-rays via bremsstrahlung in the ISM gas, or by up-scattering low energy photons \cite{Blumenthal:1970gc, Longair, Koch:1959zz, 1969PhRv..185...72G, Jones:1968zza, Strong:1998fr}.
Since at distances far from the galactic disk the optical and the IR (mainly emitted by dust) photon densities are less than those close to the disk, IC $\gamma$-rays at high latitudes are mainly due to up-scattered CMB, and thus due to their isotropy, they set a good probe to study the CR $e^{-}$ far from the disk. 
Apart from galactic diffuse $\gamma$-rays the observed fluxes include the extragalactic flux modeled in \cite{Abdo:2010nz} and  
galactic point sources \cite{Abdo:2010nz}. The CRs misidentified as $\gamma$-ray events have instead already been subtracted in the spectral data that we use, with the remaining CR contamination being negligible up to $E_{\gamma} = 100$ GeV. 
%------------------------------------------------------------------------------
\subsection{Interstellar Gas}\label{sec:GasDistr}
The interstellar matter is made up of gas and dust with an average mass ratio of 100:1 \cite{Schlickeiser}. 
Interstellar gas is composed of hydrogen, helium and small contributions from heavier elements, 
with hydrogen observed in atomic (HI), molecular (H2) and ionized (HII) states.

The three dimensional distribution of HI gas can be derived from Lyman-$\alpha$, 
from 21-cm spectra information and from rotation curves 
\cite{Watson:2010hf, 2001ApJ...555..301M, 2006ApJ...636..721B}, 
with the 21-cm line emission being due to the transition between the atomic 
hydrogen $S^{2}$ ground state levels split by the hyperfine structure. 

We will use as a reference HI gas model the recent result obtained by 
\cite{Nakanishi:2003eb}, but also refer to \cite{1976ApJ...208..346G, 1990ARA&A..28..215D}
which has been widely used in the literature.

Molecular hydrogen can exist only in dark cool clouds where it is protected 
against the ionizing stellar ultraviolet radiation. It can be traced with 
the $\lambda$ = 2.6 mm (J = 1 $\to$ 0) emission line of CO, since collisions 
between the CO and H2 molecules in the clouds are responsible for the 
excitation of CO. The CO to H2 conversion factor, $X_{CO}$ which relates the 
H2 column density, $N_{H2}$ , to the velocity-integrated intensity of the CO 
line, has considerable uncertainties. 

We use as our reference H2 model the map provided by \cite{Nakanishi:2006zf},
assuming the conversion factor to vary exponentially with galacto-centric radius as:
\be 
X_{CO}(r)=1.4\exp(r/11 \, \textrm{kpc}) \times 10^{20} H_2 \cm^{-2} \textrm{K}^{-1} \km^{-1} \s, 
\label{eq:XCO}
\ee
comparing it with the models developed by~\cite{1988ApJ...324..248B}.

The ionized hydrogen is concentrated in the vicinity of young O and B stars, with the ultraviolet radiation from these stars ionizing the ISM. 
It is known that the contribution of the HII gas to the total mass in the ISM is negligible \cite{1982MitAG..57..207D}, while its distribution is very similar to that of the free electrons in the galaxy. Thus we choose not to vary the averaged large scale distribution of this gas 
component, for which we use the parameterization of \cite{1991Natur.354..121C},
that was calculated for the Galactic distribution of free electrons:
\be 
n_{HII}(r,z)= n_{e}(r,z) =\langle n_{e} \rangle_{1} \exp{\left[{-\frac{|z|}{z_{1}}-\left(\frac{r}{r_{1}}\right)^2} \right]} + \langle n_{e} \rangle_{2} \exp{\left[{-\frac{|z|}{z_{2}}-\left(\frac{r-r'}{r_{2}}\right)^2}\right]}. 
\label{eq:HIIdistr}
\ee
We used the mean values for $z_{1} = 1$~kpc, $z_{2} = 0.15$~kpc,
$r_{2} = 2$~kpc, $r' = 4$~kpc, $\langle n_{e} \rangle_{1} = 0.025 \; \cm^{-3}$, 
$\langle n_{e} \rangle_{2} = 0.2 \; \cm^{-3}$ and the minimum value for $r_{1} = 20$~kpc.

In general, the use of 2D, spatially smoothed gas distributions would not be accurate enough to interpret $\gamma$-ray sky maps with the angular resolution of the Fermi instrument. However, we will compare $\gamma$-ray spectra with observed spectra in very wide, longitudinally averaged, intermediate and high latitude portions of the sky, where the small scale features of the gas present in detailed 3D models are washed out on average mainly within equal latitude regions. Indeed, using our reference model, we checked that passing from a 2D to a 3D gas model changes our $\pi^{0}$ and bremsstrahlung results by no more than 10\% in the regions of interest, as we show in Appendix~\ref{sec:2Dvs3D}; with the difference on the predicted total galactic diffuse model being at the 5$\%$ level. 
Given that in deriving our physical conclusions we use the $\chi^{2}$ analysis carried out between the total $\gamma$-ray \textit{Fermi} fluxes and the total $\gamma$-ray predicted fluxes which include the extra galactic background (EGB) and point sources, the impact of using a 2D ISM gas model is minimal.

Finally, Helium appears to follow the hydrogen distribution with a factor He/H $= 0.10 \pm 0.08$. Following \cite{Asplund:2004eu} 
we adopt a value of He/H = 0.11, which is widely used in the literature and neglect heavier nuclear species.

%------------------------------------------------------------------------------
\section{Methodology}
\label{sec:astro_analysis}

As just illustrated, there are many unknowns involved in the modeling of 
both the ISM and the propagation of CRs. We are 
then forced to focus our discussion introducing a few benchmark scenarios 
for both aspects of the problem.
 
One important parameter is the spectral index of diffusion, $\delta$, inferred from the spectral slope 
of the secondary to primary ratios at high energy. It ranges between about $\delta = 0.3$ up to about 0.7.  
We will discuss mainly the theoretically motivated 
frameworks of the Kraichnan turbulence spectrum 
\cite{1980RPPh...43..547K, 1967PhFl...10..859K} corresponding to $\delta = 0.5$, and the 
Kolmogorov \cite{1941DoSSR..30..301K} corresponding to $\delta = 0.33$.

We choose a range for the 
vertical and radial scales of the diffusion coefficient, $z_d$ and $r_d$ in Eq.~\ref{eq:DiffCoef}.
The discrete values that we choose in our analysis are  $z_d=(1,4,10)$~kpc and $r_d=(5,10,20)$~kpc, allowing for cases where the diffusion is highly homogeneous within our simulation volume
($D \sim e^{\mid z \mid /10~\kpc} e^{(r-r_{\odot})/20~\kpc}$) and cases such as $D \sim e^{\mid z \mid /1 \kpc} e^{(r-r_{\odot})/5 \kpc}$ where the diffusion properties vary significantly within the Galaxy. 

For most of the discussion we will neglect convective effects. However, we will also consider the effects of strong convective winds introducing in one benchmark model a convective velocity, directed only along the vertical direction outwards from the galactic plane $V_C(z) = dv_C/dz\cdot |z|$, with $dv_C/dz = 50~\km/\s/\kpc$.

In the first nine lines of Tab.~\ref{tab:Param} we summarize the models we consider for the diffusion properties of the Galaxy. In the other models we instead use our reference model for propagation, but we vary the properties of the ISM gas and SNRs.
\begin{sidewaystable}
\centering
\begin{tabular}{|c||c|c|c|c||c|c|c||c|c|c|c||c|c|c||c|c|c|}
   \hline
    & \multicolumn{4}{|c|}{Benchmark} & \multicolumn{3}{|c|}{Fitted} & \multicolumn{4}{|c|}{Fitted} & \multicolumn{3}{|c|}{Fitted} & \multicolumn{3}{|c|}{Predicted} \\
   \hline
   Name & $\delta$ & $z_d$ & $r_d$ & $dv_{C}/dz$ & $D_0 \times 10^{28}$ & $v_A$ & $\eta$  & $\gamma_1^p/\gamma_2^p/\gamma_3^p$ & $R_{0,1}^p$ & $\gamma_1^e/\gamma_2^e$ & $\overline{\eta W_{0}}$ & $\chi^2_{B/C}$ & $\chi^2_p$ & $\chi^2_{(e^- + e^+)}$ & $\chi^2_{\bar{p}}$ & $\chi^2_{\gamma} $ & $\chi^2_{\bar{p} \& \gamma}$\\
    &  & \kpc & \kpc & $\km \, \s^{-1}\,\kpc^{-1}$ & $ \cm^2 \s^{-1}$ & $\km\,\s^{-1}$ & & & \GV & & $\times 10^{49} \erg$ &  &  &  &  & & \\
   \hline\hline
    KRA4-20 & 0.5 & 4 & 20 & 0 & 2.49 & 19.5 & -0.363 & 2.06/2.35/2.18 & 14.9 & 1.6/2.62 & 0.77 & 0.34 & 0.6 & 0.4 & 0.73 & 1.02/0.42/0.60 & 0.71 \\ 
   \hline\hline
   KRA1-20 & 0.5 & 1 & 20 & 0 & 0.55 & 16.3 & -0.521 & 2.07/2.34/2.18 & 16.5 & 1.5/2.58 & 0.27 & 0.4 & 0.51 & 0.57 & 0.76 & 3.21/1.67/0.29 & 1.29 \\
   \hline
   KRA10-20 & 0.5 & 10 & 20 & 0 & 4.29 & 19.1 & -0.373  & 2.05/2.35/2.18 & 15.2 & 1.6/2.62 & 1.01 & 0.32 & 0.48 & 0.33 & 0.70 & 0.91/0.32/0.60 & 0.65 \\
   \hline
   KRA4-5  & 0.5 & 4 & 5 & 0 & 2.76 & 16.9 & 0.0  & 2.07/2.35/2.18 & 27 & 1.6/2.62 & 0.71 &0.64 & 0.54 & 0.4 & 1.45 & 1.06/0.46/0.45 & 1.01 \\
   \hline
   KRA4-10 & 0.5 & 4 & 10 & 0 & 2.58 & 19.1 & -0.247  & 2.05/2.35/2.18 & 17.5 & 1.6/2.62 & 0.78 & 0.36 & 0.52 & 0.41 & 0.93 & 1.02/0.42/0.55 & 0.78 \\
   \hline  
   RUN4-20 & 0.4 & 4 & 20 & 0 & 3.21 & 23.1 & 0.32  & 2.06/2.44/2.28 & 14 & 1.7/2.64 & 0.76 & 0.34 & 0.41 & 0.29 & 1.36 & 1.11/0.45/0.51 & 0.99 \\
   \hline   
   KOL4-20 & 0.33 & 4 & 20 & 0 & 3.85 & 24.8 & 0.765  & 2.03/2.49/2.35 & 10.7 & 1.7/2.64 & 0.70 & 0.5 & 0.3 & 0.45 & 2.86 & 1.33/0.54/0.40 & 1.70 \\
   \hline
   CON4-20 & 0.6 & 4 & 20 & 50 & 0.645 & 27.2 & 0.755 & 1.85/2.48/2.19 & 12.3 & 1.6/2.62 & 0.44 & 0.61 & 0.44 & 0.82 & 1.05 & 2.18/0.98/0.25 & 1.09 \\
   \hline
   Scenario B & 0.5/0.33 & 4 & 20 & 0 & 2.49 & 19.5 & -0.363 & 2.06/2.35/2.35 & 14.9 & 1.6/2.62 & 0.67 & 0.34 & 0.6 & 0.4 & 0.74 & 1.04/0.43/0.59 & 0.71 \\ 
   \hline
   NS low & 0.5 & 4 & 20 & 0 & 1.94 & 14.6 & -0.324 & 2.07/2.35/2.18 & 15.4 & 1.6/2.62 & 0.68 & 0.32 & 0.44 & 0.42 & 0.67 & 3.69/1.77/0.21 & 1.34 \\
   \hline
   NS high & 0.5 & 4 & 20 & 0 & 3.04 & 24.4 & -0.411 & 2.06/2.35/2.18 & 17 & 1.6/2.62 & 0.74 & 0.31 & 0.57 & 0.55 & 0.85 & 0.52/0.55/1.96 & 0.94 \\
   \hline
   Bronf & 0.5 & 4 & 20 & 0 & 3.39 & 26.5 & -0.526 & 2.08/2.35/2.18 & 17.6 & 1.6/2.62 & 0.79 & 0.38 & 0.58 & 0.52 & 0.69 & 1.45/1.42/3.44 & 1.47 \\
   \hline
   Source B & 0.5 & 4 & 20 & 0 & 2.49 & 19.6 & -0.355 & 2.04/2.34/2.18 & 15.9 & 1.6/2.62 & 0.77 & 0.3 & 0.36 & 0.39 & 0.86 & 1.26/0.50/0.43 & 0.79 \\
   \hline
   Source C & 0.5 & 4 & 20 & 0 & 2.33 & 19.2 & -0.44 & 2.05/2.34/2.18 & 16.6 & 1.6/2.62 & 0.84 & 0.33 & 0.58 & 0.53 & 0.73 & 0.73/0.34/1.44 & 0.79 \\ 
   \hline
\end{tabular}
\caption{The parameters for the benchmark models for propagation used for $\gamma$-ray predictions in 
Fig.~\ref{fig:RefAstroModelGammas},~\ref{fig:DiffIndexVary},~\ref{fig:rdVary},~\ref{fig:zdVary},~\ref{fig:Conv},~\ref{fig:GasVary} 
and~\ref{fig:SNRVary}. $\gamma^e$'s are the injection indices for primary electrons below and above a break at 5~GV. Our reference model 
``KRA4-20" is also referred to in the text as ``Scenario A", ``NS mean" and ``Source A". For ``CON4-20'' we assumed $dv_{C}/dz = 50$ $\km \, \s^{-1}\, \kpc^{-1}$. The ``Scenario B'' model accounts for a possible break in the spectral index. In ``NS low" and ``NS high" we use H2 gas densities, respectively, 1 $\sigma$ lower and 1 $\sigma$ higher than mean values of \cite{Nakanishi:2006zf}, however they share the same HI gas distribution of \cite{Nakanishi:2003eb}. In ``Bronf" we use the model of \cite{1988ApJ...324..248B} for H2 and the model of \cite{1976ApJ...208..346G, 1990ARA&A..28..215D} for HI gas distributions. 
The ``Source B" and ``Source C" models refer to different assumptions for the primary CR source distributions. See text for more details on the definition of the other parameters. $\chi^2$'s refer either to the goodness of our fits of CR nuclei, protons and leptons or show the level of agreement of our predictions with $\gamma$-ray and antiproton data.}\label{tab:Param}
\end{sidewaystable}

%%%%%%%%%%%%%%%%%%%%%%%%%%%%%%%%%%%%%%%%%%%%%%%%%%%%%%%%%%%%%%%%%%%%%
\begin{sidewaystable}
\centering
\begin{tabular}{|c||c|c|c||c|c|c||c|c|c|c||c|c|c||c|c|c|}
   \hline
    & \multicolumn{3}{|c|}{Benchmark} & \multicolumn{3}{|c|}{Fitted} & \multicolumn{4}{|c|}{Fitted} & \multicolumn{3}{|c|}{Fitted} & \multicolumn{3}{|c|}{Predicted} \\
   \hline
   Name&$\delta$&$z_d$&$r_d$&$D_0 \times 10^{28}$&$v_A$&$\eta$&$\gamma_1^p/\gamma_2^p/\gamma_3^p$&$R_{0,1}^p$&$\gamma_1^e/\gamma_2^e$&$\overline{\eta W_{0}}$&$\chi^2_{B/C}$&$\chi^2_p$&$\chi^2_{(e^-+e^+)}$&$\chi^2_{\bar{p}}$&$\chi^2_{\gamma}$&$\chi^2_{\bar{p} \& \gamma}$\\
    &  & kpc & kpc & cm$^2$/s & km/s & & & GV & & $\times 10^{49}$ erg &  &  &  &  & & \\
   \hline\hline
    KRA1-20 NS low& 0.5 & 1 & 20 & 0.414 & 11.9 & -0.454 & 2.07/2.35/2.18 & 16.6 & 1.6/2.62 & 0.31 & 0.44 & 0.56 & 0.67 & 0.74 & 6.77/3.89/0.77 & 2.43 \\ 
   \hline\hline
   KRA10-20 NS low & 0.5 & 10 & 20 & 3.35 & 14.4 & -0.331 & 2.06/2.35/2.18 & 15.2 & 1.6/2.62 & 0.95 & 0.36 & 0.85 & 0.31 & 0.68 & 2.84/1.28/0.14 & 1.09 \\ 
   \hline\hline
   KRA1-20 Bronf & 0.5 & 1 & 20 & 0.831 & 25.4 & -0.563 & 2.07/2.35/2.18 & 18.3 & 1.6/2.62 & 0.33 & 0.29 & 0.9 & 0.47 & 0.9 & 0.66/0.47/1.11 & 0.81 \\ 
   \hline\hline
   KRA10-20 Bronf & 0.5 & 10 & 20 & 5.6 & 24.6 & -0.576 & 2.09/2.36/2.18 & 17.6 & 1.6/2.62 & 1.15 & 0.35 & 0.75 & 0.58 & 0.68 & 1.43/1.29/3.23 & 1.4 \\ 
   \hline

\end{tabular}
\caption{Parameters for models (not shown in any Figure), that represent two extreme cases of either a thin diffusion halo with a low ISM gas density assumption: "KRA1-20 NS low", a thick diffusion halo with a high ISM gas density assumption: "KRA10-20 Bronf"; and two intermediate cases: "KRA10-20 NS low" (thick diffusion halo with low ISM gas), "KRA1-20 Bronf"(thin diffusion halo with high ISM gas). As can be seen from the "predicted" $\chi^{2}$ columns the $\gamma$-ray spectra are more sensitive than the antiproton spectra in discriminating among some of these cases.}\label{tab:Param2}
\end{sidewaystable}

Models are labeled in the following way: models with ``KRA$z_d$-$r_d$" correspond to Kraichnan-like turbulence ($\delta=0.5$) and fixed values of $z_d$ and $r_d$. 
In the same way, the ``KOL4-20" model corresponds to $\delta=0.33$, $z_d=4$ kpc and $r_d=20~\kpc$ and the ``CON4-20" has a significant convective velocity.
The models labeled by ``NS low" and ``NS high" have the same ($\delta$, $z_d$, $r_d$) as our reference model, 
but use different H2 gas distributions than our reference to probe the uncertainties derived from 
\cite{Nakanishi:2006zf}. The ``Bronf" model assumes an HI (H2) gas distributions modeled by 
\cite{1976ApJ...208..346G, 1990ARA&A..28..215D,1988ApJ...324..248B}. 
The ``Source B" and ``Source C" scenarios are used to study the effects of different source distributions in the Galactic disk.

\begin{figure}[tbp]
\begin{center}
\includegraphics[scale=.43]{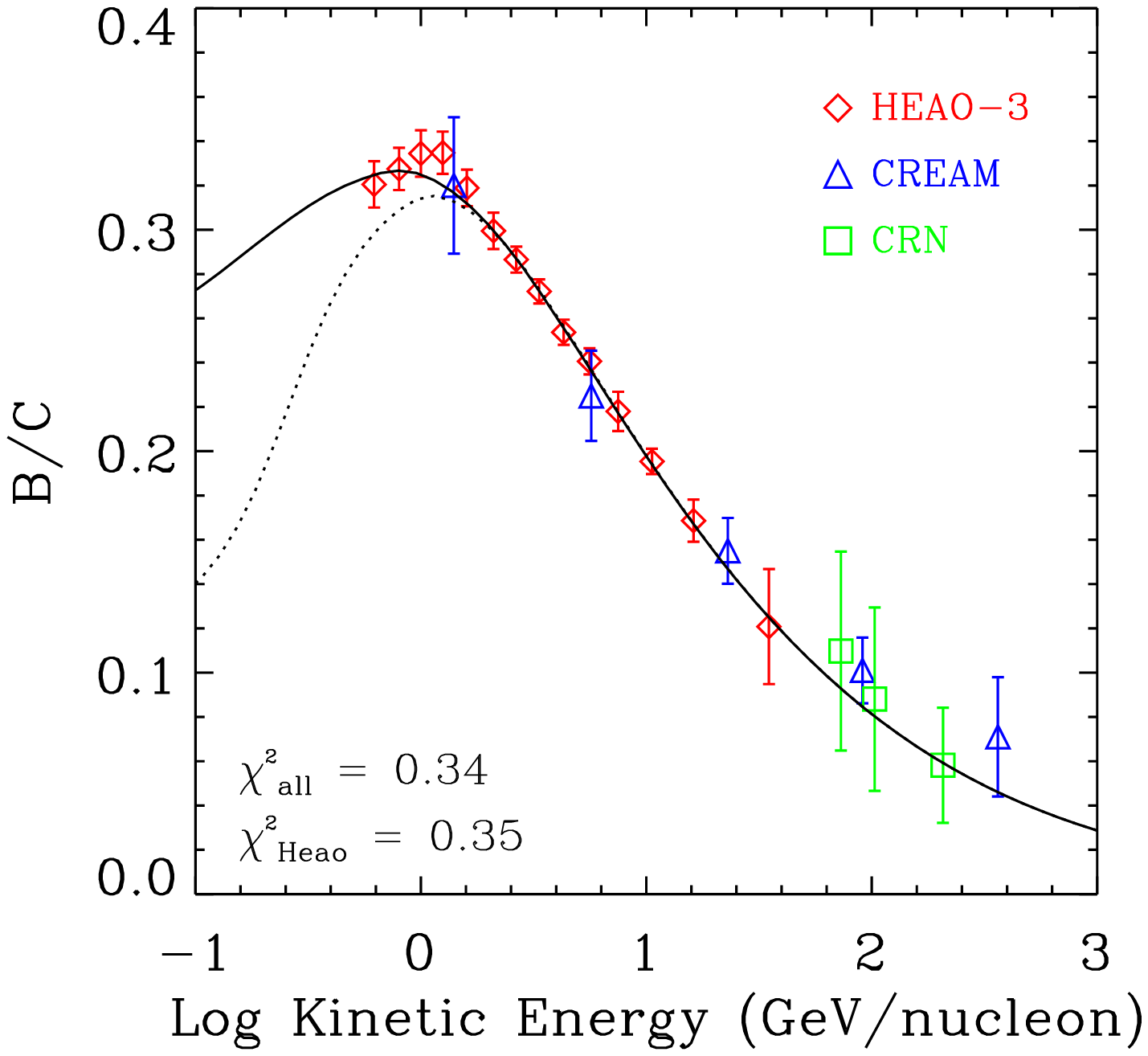}
\hspace{-0.1cm}
\includegraphics[scale=.43]{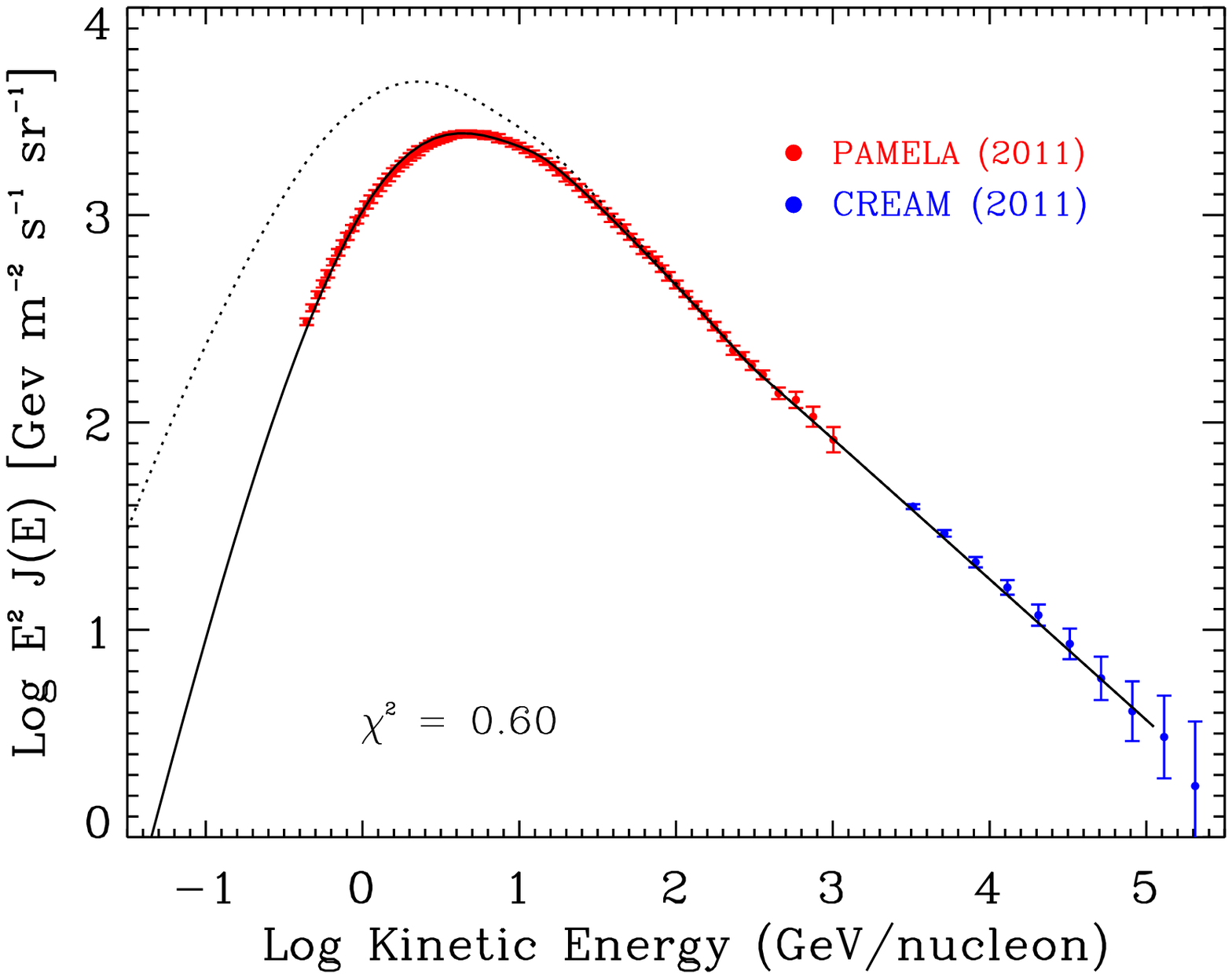}
\\
\includegraphics[scale=.43]{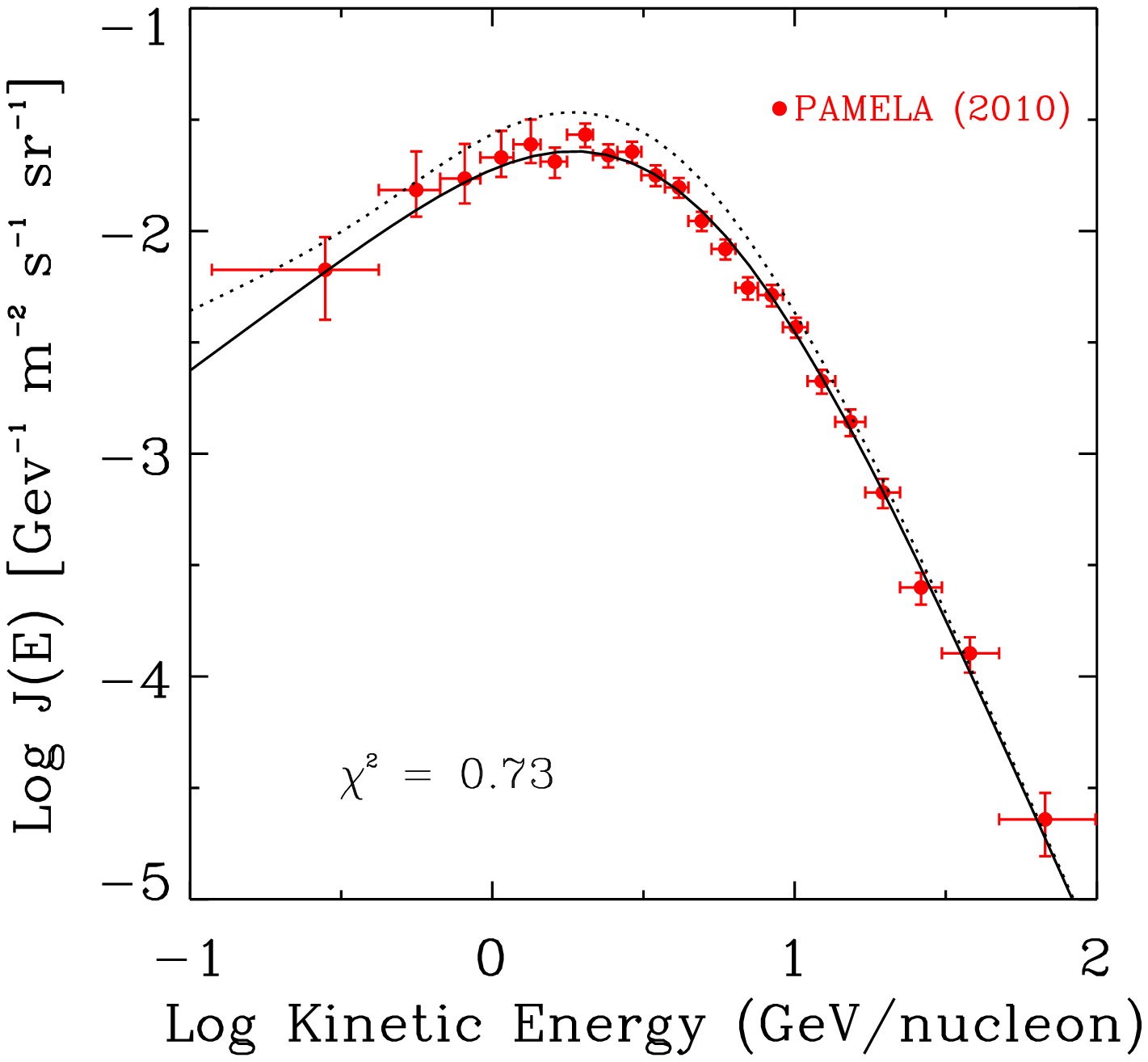}
\hspace{-0.1cm}
\includegraphics[scale=.43]{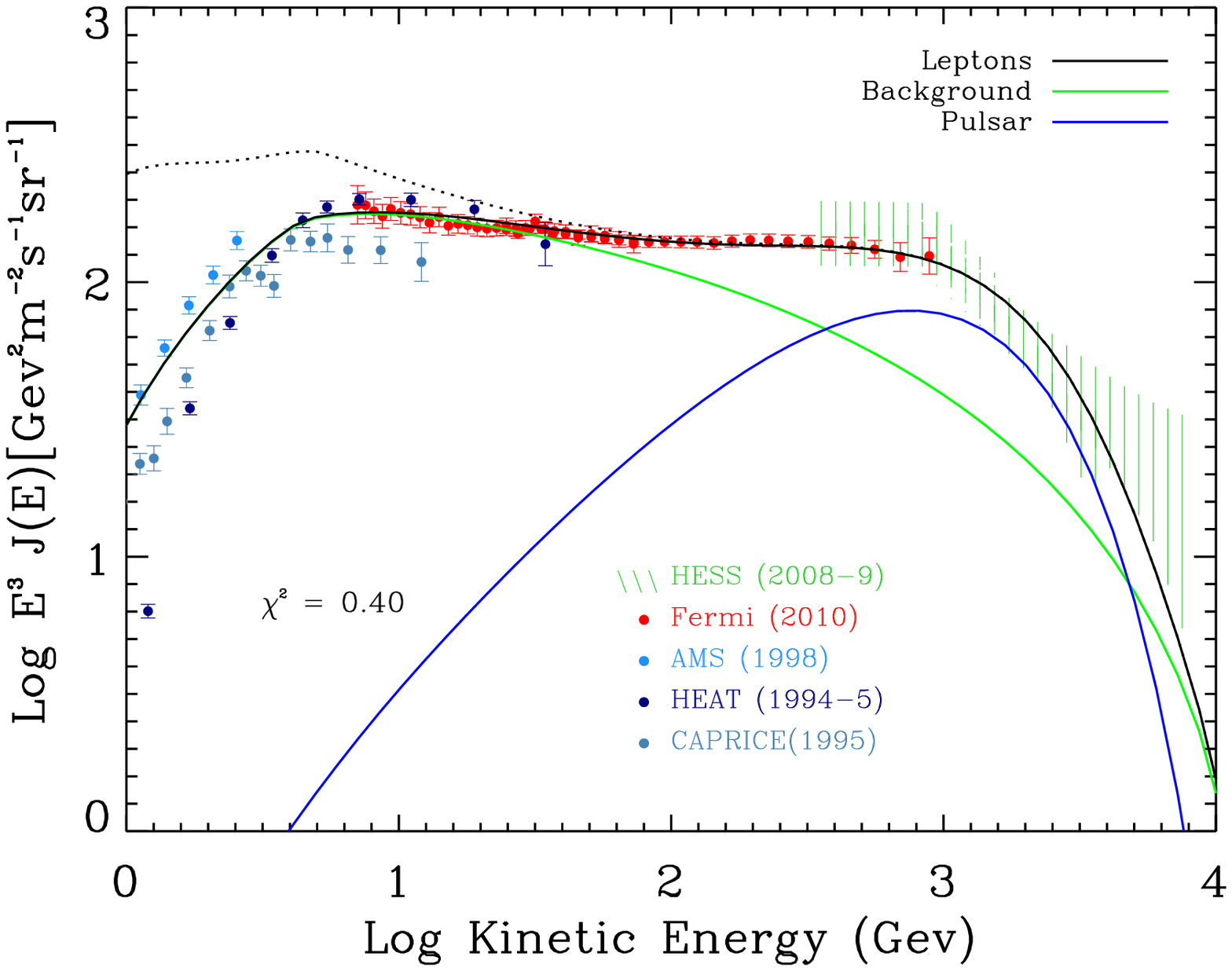}
\end{center}
\caption{Reference Astrophysical model. We assume that the diffusion coefficient $D = 2.49\times10^{28}~\cm^{2}/\s~\beta^{-0.363} (R/3~\GV)^{0.5} e^{\mid z \mid/4~\kpc}e^{(r-r_{\odot})/20~\kpc}$ 
(where $R$ is the rigidity) as in Eq.~\ref{eq:DiffCoef}. Values of $\chi^2$ of the reference model for each observable are also shown.
\emph{Upper left}: fit to the B/C data. 
\emph{Upper right}: proton spectrum, where we fit the injection spectrum using \textit{PAMELA} \cite{Adriani:2011cu} and CREAM data \cite{Yoon:2011zz} (see section~\ref{sec:CRSources}). 
\emph{Lower left}: the predicted antiproton spectrum, which provides a good fit to the \textit{PAMELA} data \cite{Adriani:2010rc}.
\emph{Lower left}: $e^{-}+e^{+}$ flux, including \textit{Fermi} observations \cite{Ackermann:2010ij}. Constraints on the primary, secondary and pulsar fluxes properties are obtained by fitting to the data. Dotted lines refer to unmodulated CR fluxes.}
\label{fig:RefModelCRs}
\end{figure}

For each model with a set of values of $\delta$, $z_d$, $r_d$ and $dv_C/dz$ we derive the other propagation parameters ($D_0, \eta, v_A$) by minimizing the $\chi^2$ of $B/C$ data, thus fitting our galactic (global) models to the local data, as we show in Fig.~\ref{fig:RefModelCRs} (upper left panel).  
We use the \textit{HEAO-3} \cite{1990A&A...233...96E}, \textit{CRN} \cite{1990ApJ...349..625S} and CREAM \cite{Ahn:2008my} data points. 
We then fix the spectral indices for protons $\gamma^{p}_{1}, \gamma^{p}_{2}, \gamma^{p}_{3}$ below and above the 
rigidity breaks $R_{0,1}^p, R_{0,2}^p$ by fitting to the recently released \textit{PAMELA} \cite{Adriani:2011cu} and CREAM data \cite{Yoon:2011zz} (see Fig.~\ref{fig:RefModelCRs} upper right panel). 
We then refit the normalization for the proton spectral data and modulation potential (in the force field approximation \cite{Gleeson:1968zz}) checking also for consistency with the BESS data of years 1997-1999 \cite{2007APh....28..154S}, where the only free parameter in fitting the entire spectra of different years is the modulation potential. 
We fit in the same way the injection parameters of He nuclei (not shown in Tab.~\ref{tab:Param}) by fitting to the most recent data \cite{Adriani:2011cu, Yoon:2011zz} up to the highest energies.
Having reproduced primary protons and He, we check also whether the predicted antiproton flux is consistent with local data (Fig.~\ref{fig:RefModelCRs} lower left panel). The corresponding $\chi^2$'s are listed in Tab.~\ref{tab:Param}.

Having chosen the ISM gas models and fixed the diffusion and re-acceleration properties in the ISM, as well as the CR nuclei spectra, the remaining task before calculating $\gamma$-ray spectra is to fix CR electrons (and positrons) source properties. Since the $e^{+} + e^{-}$ spectrum below $E \sim$ 30 GeV
is dominated by shock accelerated electrons in SNRs (primaries), and by secondary electrons (and positrons)
from inelastic collisions of CR nuclei with the ISM, we fit the primary and secondary electron spectral properties to the low energy $e^{+} + e^{-}$ spectrum between 7-30 GeV as measured by \textit{Fermi} \cite{Ackermann:2010ij}.
Pulsars within $\sim 3~\kpc$ can contribute to the $e^{+} + e^{-}$ spectrum up to $O(0.1)$ at $E \approx 50$ GeV
 and up to $O(1)$ at $E \approx 500$ GeV (\cite{Malyshev:2009tw} and also  
similar works \cite{Hooper:2008kg, Yuksel:2008rf, Profumo:2008ms, DiBernardo:2011wm}). 
Thus, if we assume pulsars contribute maximally we find from the \textit{Fermi} data the 
injection index $n$ for the distribution of pulsars of Eq.~\ref{eq:PulsarSource} 
and the averaged total energy injected into the ISM through CR $e^{\pm}$ per pulsar 
$\eta W_{0}$. Best fit values are found to be $n\sim1.4$ and $M\sim1.2$ TeV. Using 
them, and assuming a constant birth rate of $N_{b} = 30~\yr^{-1}$, we find 
that on average $\overline{\eta W_{0}} \simeq 10^{49}$ erg for our various 
propagation models (see Table~\ref{tab:Param}), which is well within the 
allowed range of values \cite{2003A&A...404..939V, 2006ARA&A..44...17G, 2006A&A...451L..51H, Malyshev:2009tw, Gelfand:2009aa}.

Having fixed all the properties of the CR electrons from SNRs, pulsars and inelastic collisions, we can then compute the $\gamma$-ray diffuse spectra via up-scattering of ISRF and CMB photons by the CR $e^{\pm}$, bremsstrahlung in the ISM gas by both CR electrons and protons and by decays of $\pi^0$ produced in $pp$ collisions in the ISM gas.
We note that while it is clear that the high energy part of the $e^{-}+e^{+}$ 
flux is dominated by local sources, in our approach we fit the properties of 
the statistically averaged flux ($n$, $M$ and $\overline{\eta W_{0}}$) to the 
$e^{-}+e^{+}$ data, and extend those properties to the entire distribution 
of \cite{FaucherGiguere:2005ny} of pulsars in the Galaxy. If for some reason the local $e^{-}+e^{+}$ flux above 100 GeV is severely enhanced (suppressed) by the presence (absence) of strong local sources, this should have an effect in our model over-predicting (under-predicting) the ICS components of the spectra mainly at lower latitudes.
     
We also note that we have checked for consistency with the PAMELA positron fraction \cite{Adriani:2010ib, Adriani:2008zr} and the recently released electron spectrum \cite{PAMELA:2011xv}.      

We bring attention to the fact that for each model we use exactly the same gas
distribution model to first fit propagation and injection properties against
primary CRs, then to predict secondary antiprotons and leptons and finally
to produce $\gamma$-ray maps.
%------------------------------------------------------------------------------
\section{Results}
\label{sec:astro_results}

\subsection{Reference Model}\label{sec:astro_results_RefModel}
\begin{figure}[tbp]
\begin{center} 
\includegraphics[width=0.45\textwidth]{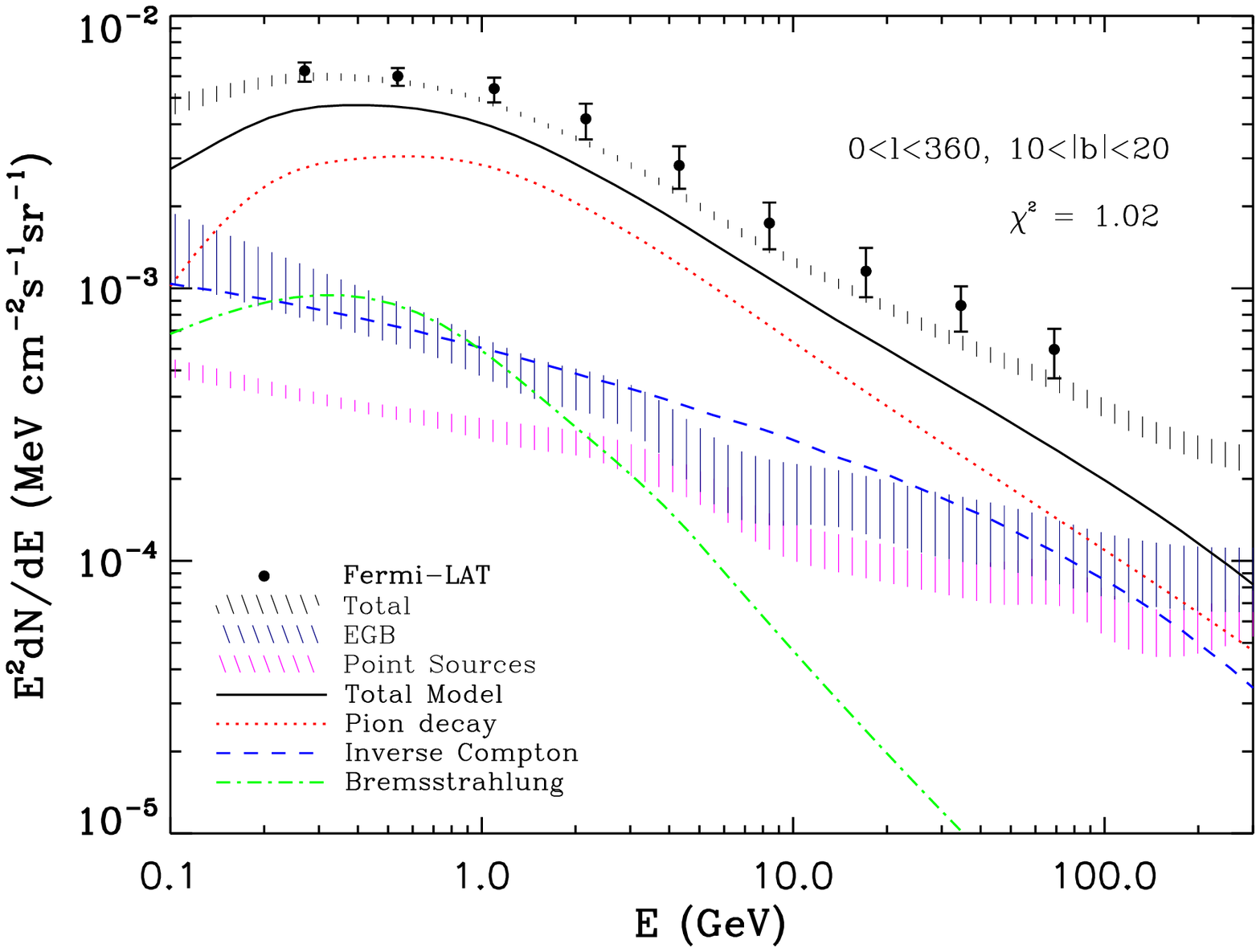} 
\includegraphics[width=0.45\textwidth]{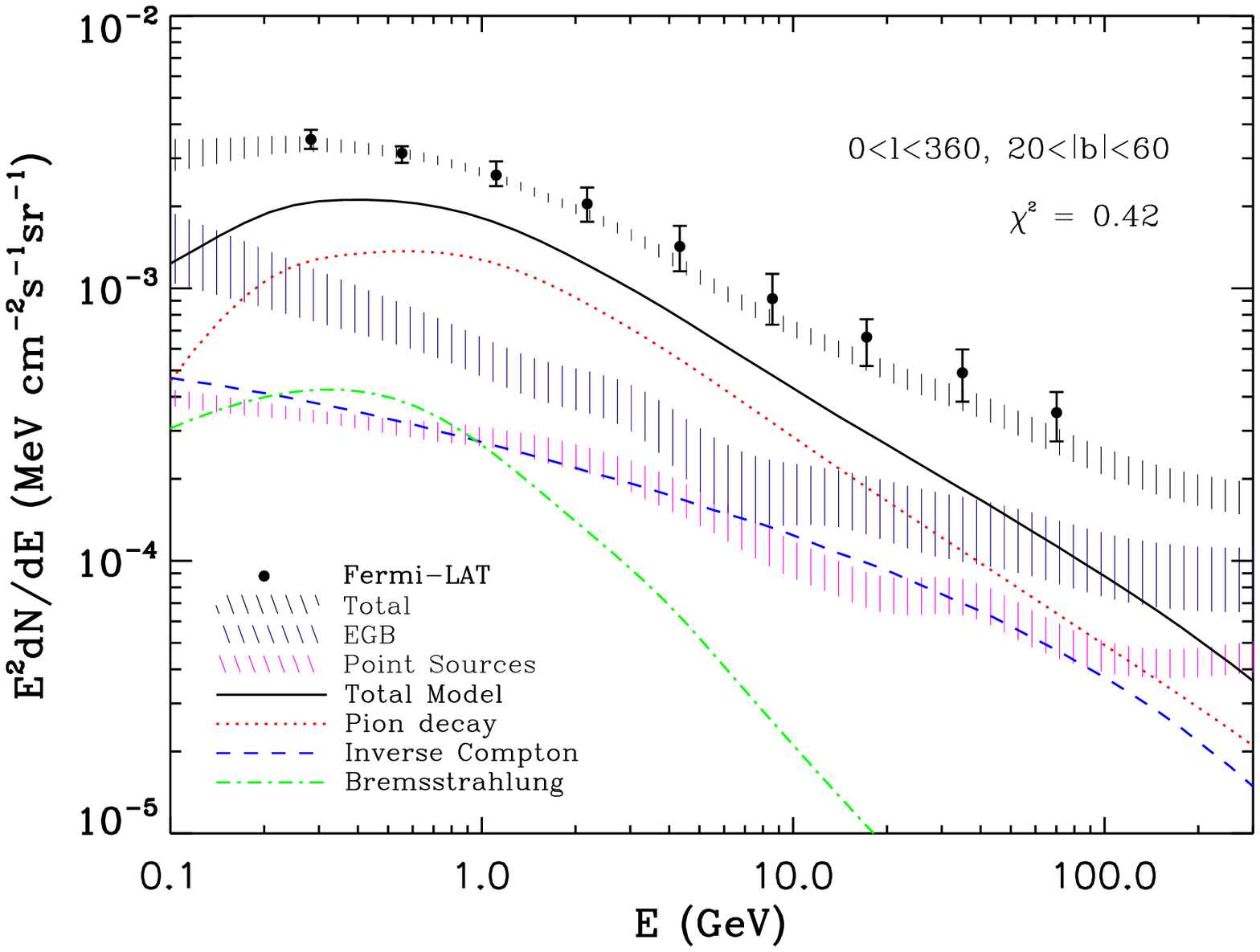}
\includegraphics[width=0.45\textwidth]{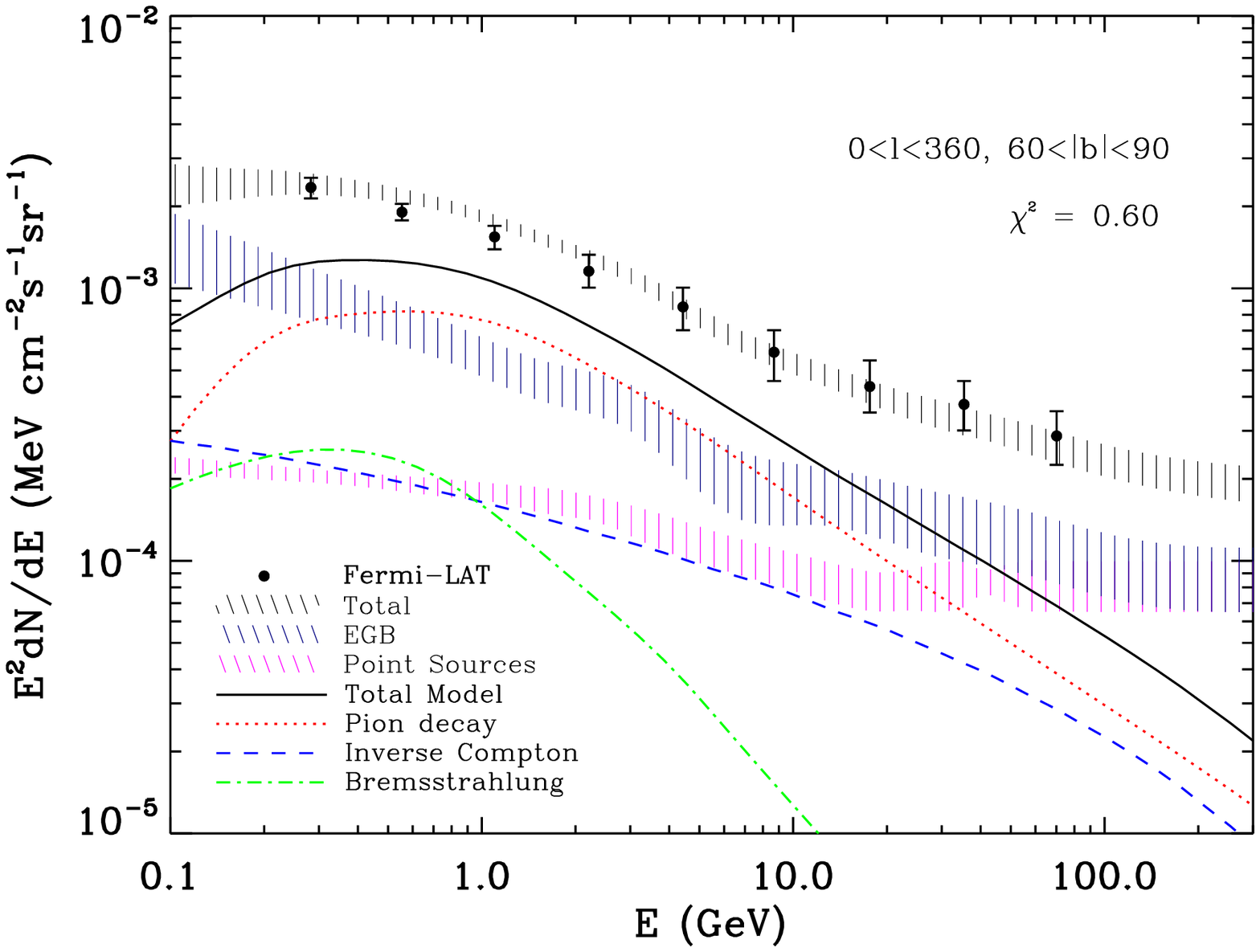}
\end{center}
\caption{Gamma-ray spectra for given sky regions predicted in our Reference Astrophysical model ``KRA4-20".
\emph{Upper left}: $10^\circ< \mid b \mid <20^\circ$ and 
$0^\circ< l < 360^\circ$, \emph{Upper right}: $20^\circ < 
\mid b \mid < 60^\circ$ and $0^\circ< l < 360^\circ$,
\emph{Lower}: $60^\circ < \mid b \mid < 90^\circ$ and $0^\circ< l < 360^\circ$.}
\label{fig:RefAstroModelGammas}
\end{figure}

\begin{figure}[tbp]
\begin{center}
\includegraphics[width=0.45\textwidth]{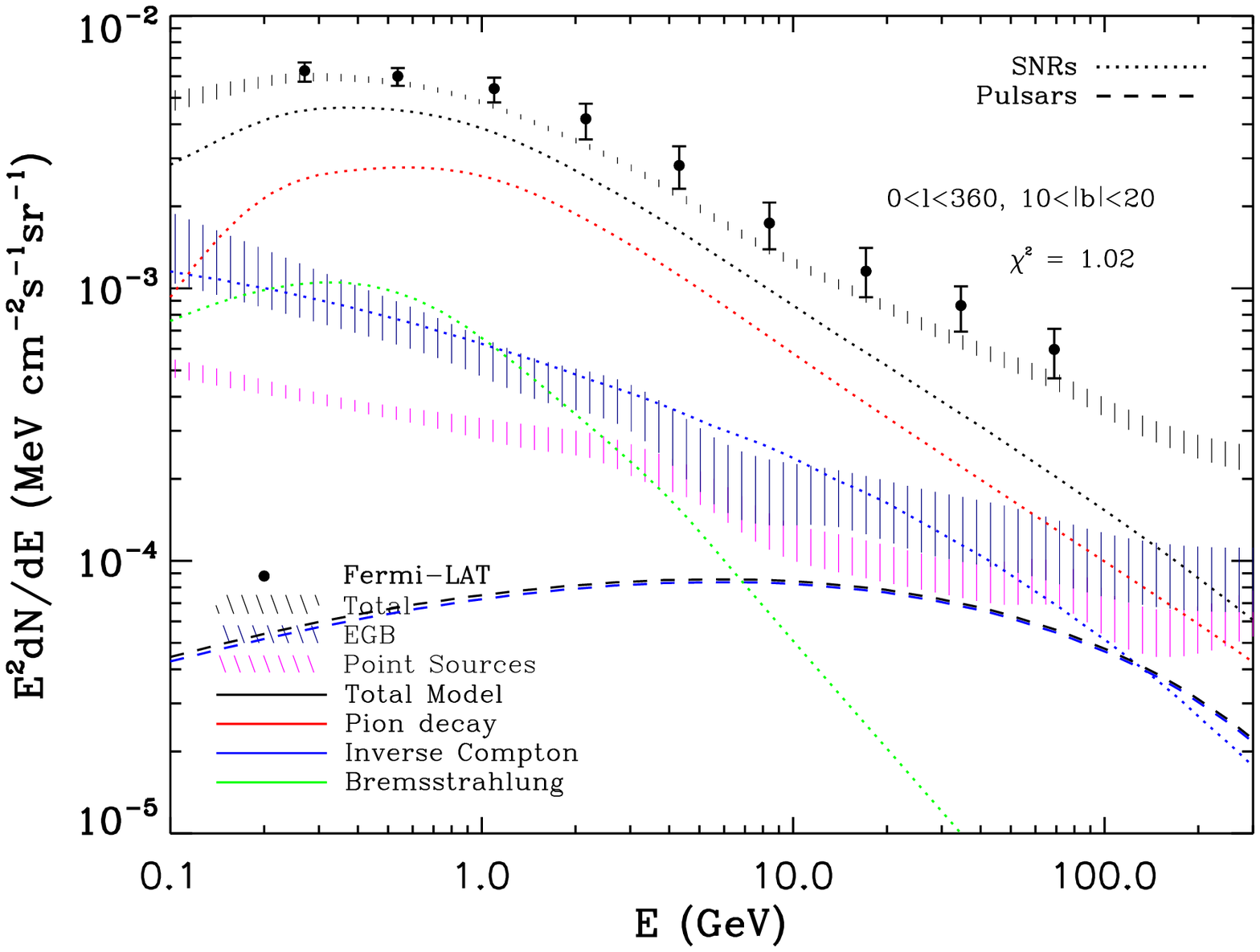}
\includegraphics[width=0.45\textwidth]{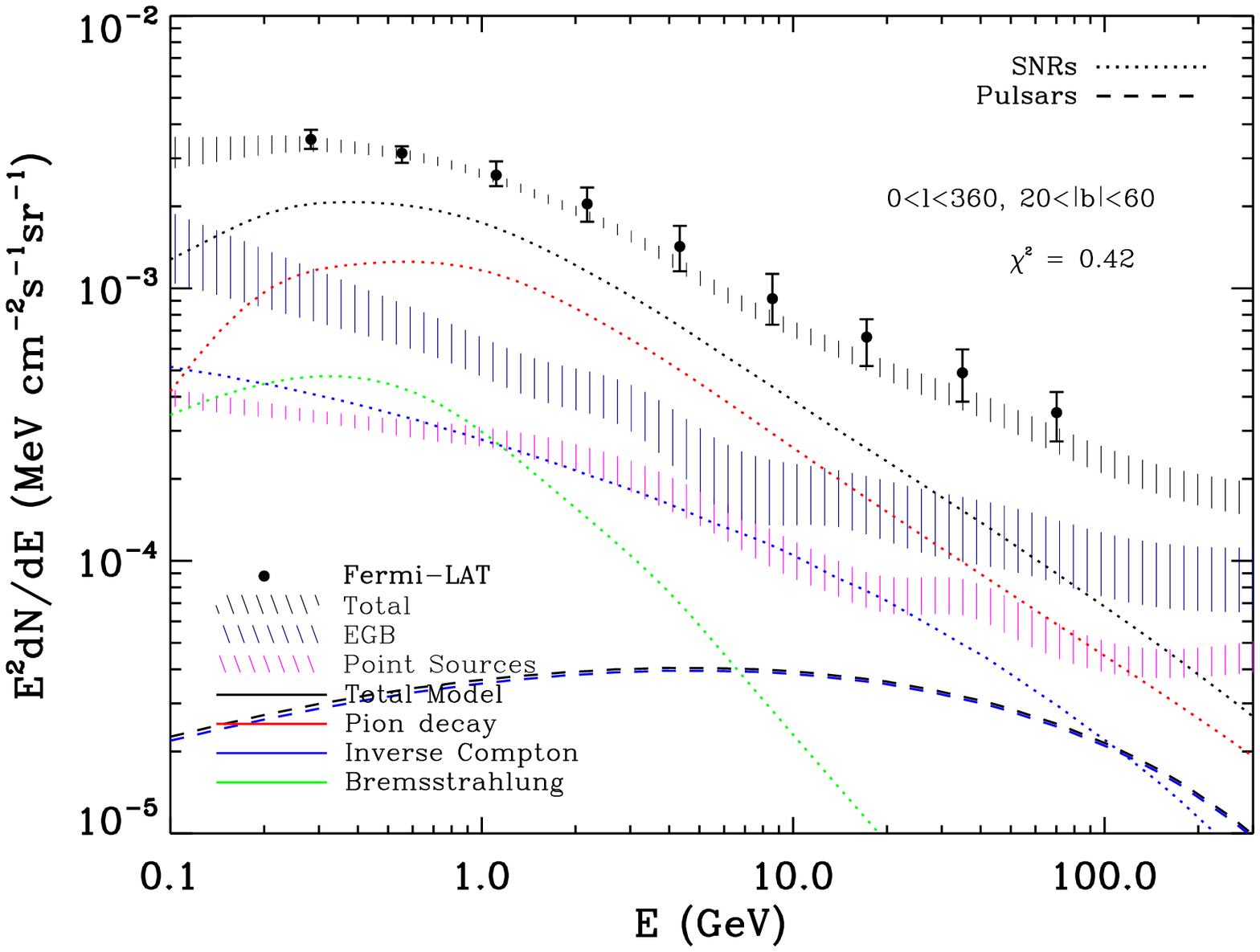}
\includegraphics[width=0.45\textwidth]{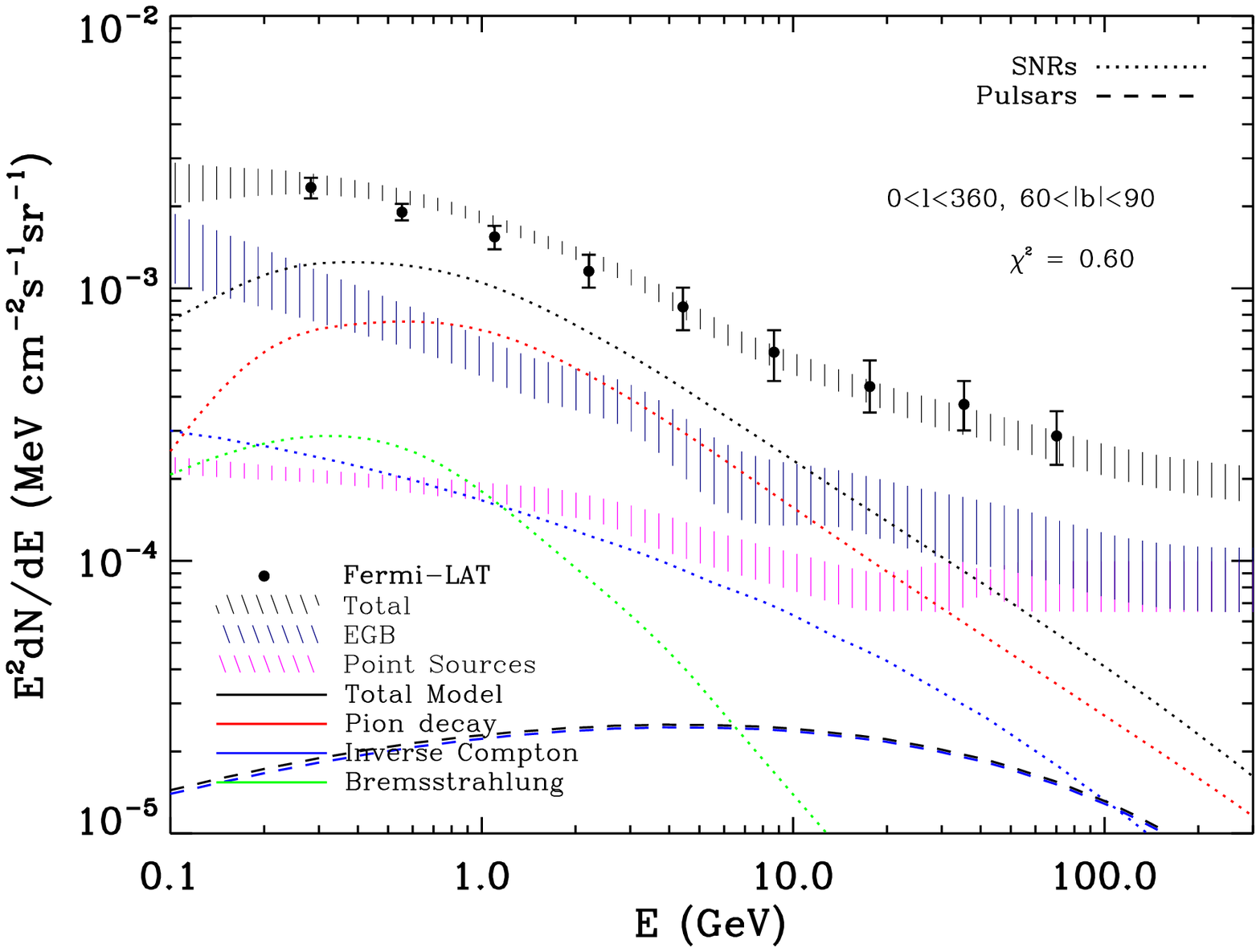}
\end{center}
\caption{Seperate SNR \emph{dotted} and pulsar \emph{dashed} sources contributions to the diffuse $\gamma$-ray spectra for the Reference Astrophysical model ``KRA4-20''. Sky windows are as in Fig.~\ref{fig:RefAstroModelGammas}.}
\label{fig:RefModGamma_SNRandPulsars}
\end{figure}

As it is clear from Fig.~\ref{fig:RefAstroModelGammas}, our reference model ``KRA4-20" (see Table~\ref{tab:Param}) provides a very good combined fit of the local CRs (see Fig.~\ref{fig:RefModelCRs}) and the $\gamma$-rays at intermediate and high latitudes. In Fig.~\ref{fig:RefModGamma_SNRandPulsars} we also show the SNRs and pulsars contributions to the total spectra separately. 

The best fit to the $\gamma$-ray spectra is achieved at the higher latitudes ($\mid b \mid > 20^{\circ}$) that are also less affected by uncertainties in the sources distributions.
This can also be seen by comparing in Fig.~\ref{fig:RefModGamma_SNRandPulsars} our predictions between $10^{\circ} < \mid b \mid < 20^{\circ}$ and $\mid b \mid > 60^{\circ}$ where we show separately the contribution from SNRs and pulsars that have different source functions, given in eq.~\ref{eq:injNucl} and~\ref{eq:PulsarSource}. 
At intermediate latitudes it slightly under-predicts the observed flux while still giving a fairly good fit. We note that the ``source" component that we show is composed by the sources detected with at least 14$\sigma$ and also weaker sources that have been catalogued by LAT \cite{Abdo:2010nz}. 
Yet, very dim $\gamma$-ray sources that would be contributing, per energy bin and pixel, less photons than the uncertainty of the true diffuse background 
are not included in the ``sources'' component. 
Such a class of sources could be millisecond pulsars (MSPs) in the Galactic 
Ridge and halo that are not accounted for. 

MSPs that are not in globular 
clusters, can contribute in the lower latitudes and could possibly compensate 
for our under-prediction of the total gamma-ray flux at $\sim$ few GeV.
The uncertainties in the contribution of dim MSPs to the diffuse 
galactic flux have been shown to be very significant \cite{Malyshev:2010xc} 
due to the great energy loss time scale ($\sim$10 Gyr 
\cite{1993ApJ...402..264C, 2005hpa..book.....L, 2009ASSL..357.....B}) of MSPs 
which results in their total population being greatly affected by the 
uncertainties in the evolution of the Galactic halo \cite{Malyshev:2010xc}. 
Recently, \cite{Calore:2011bt, SiegalGaskins:2010mp} have suggested that MSPs could be contributing to the isotropic diffuse $\gamma$-ray flux. Since MSPs occur in regions of high stellar densities such as the Galactic Ridge, and  possibly the Galactic halo \cite{Malyshev:2010xc} at its earlier stages, it is 
unlikely that the main part of their diffuse contribution is going to be isotropic. 
While individual MSP spectra may vary significantly, based on the measured spectra of 8 MSPs \cite{2009Sci...325..848A} their distribution spectrum could be described by
\be
\frac{dN_{\gamma}}{dE} \sim E^{-\Gamma} e^{-E/E_{c}},
\ee
with $\Gamma = 1.5 \pm 0.4$ and $E_{c} = 2.8 \pm 1.9$ GeV and luminosity in $\gamma$-rays of $L = 10^{33.9 \pm 0.6}$~erg/s. By comparing to the data, we find that to fit to 
the $10^\circ < \mid b \mid < 20^\circ$ data we need a flux of 
\be
\frac{dN_{\gamma}}{dE} \approx 10^{-3}~\MeV \,\cm^{-2}\s^{-1} \sr^{-1},\; \textrm{between 1 and 10 GeV,}
\ee
which would lead to a population of $\sim 10^4$ MSPs in that region following the 
assumptions of \cite{Malyshev:2010xc}.

\subsection{Varying the Diffusion of CRs in the ISM}
\label{subsec:VaryDiffuse}
\begin{figure}[tbp]
\begin{center}
\includegraphics[width=0.45\textwidth]{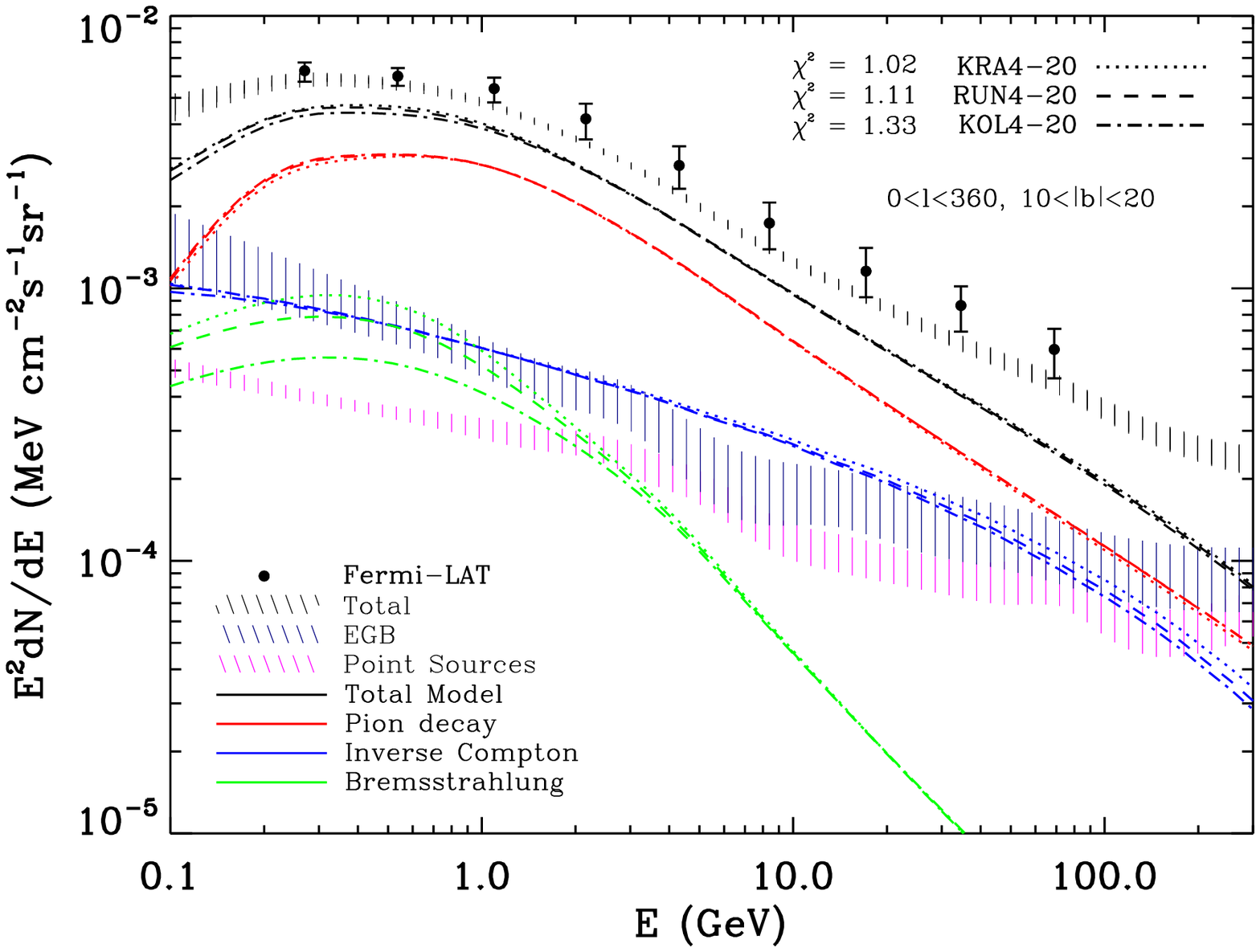}
\includegraphics[width=0.45\textwidth]{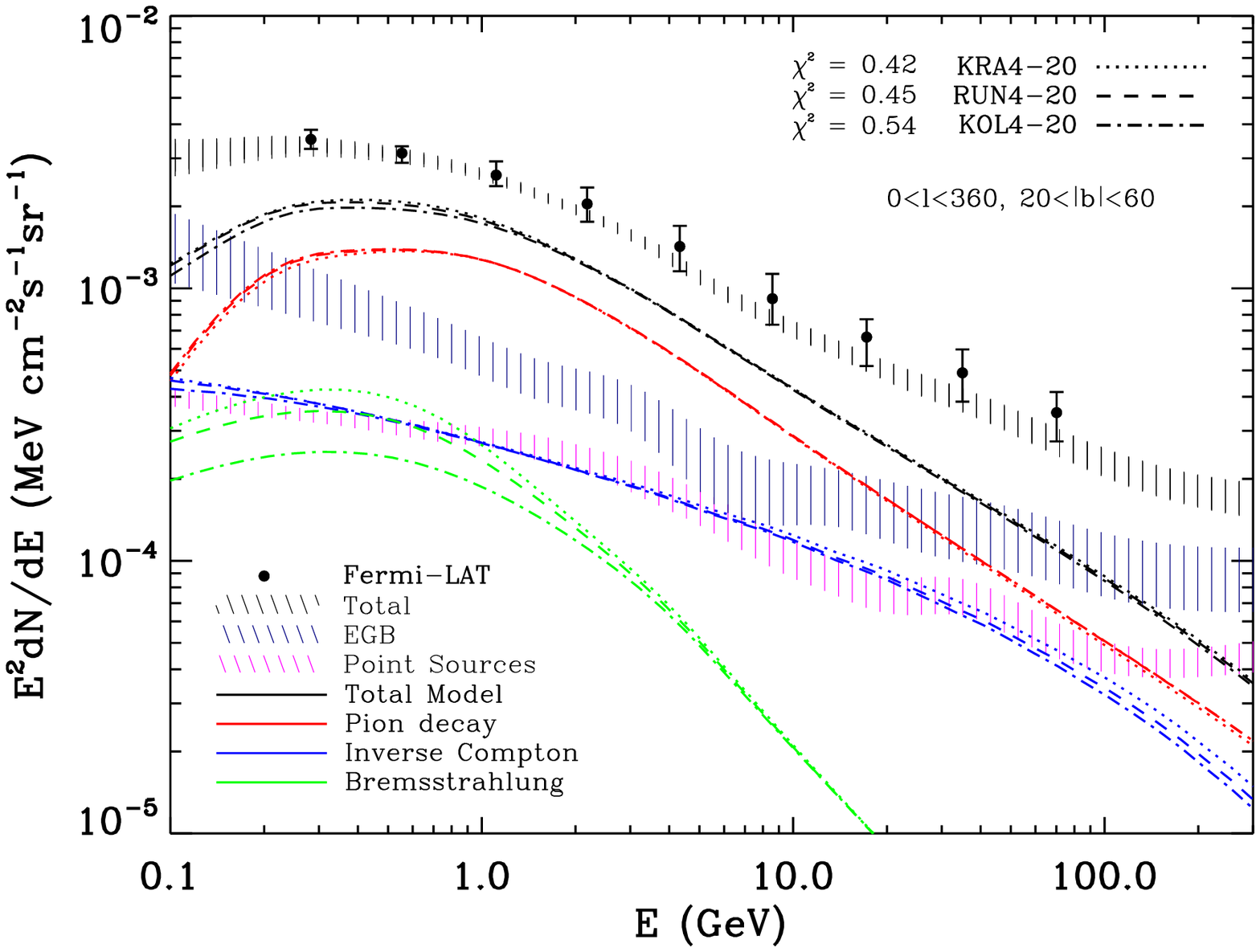}
\includegraphics[width=0.45\textwidth]{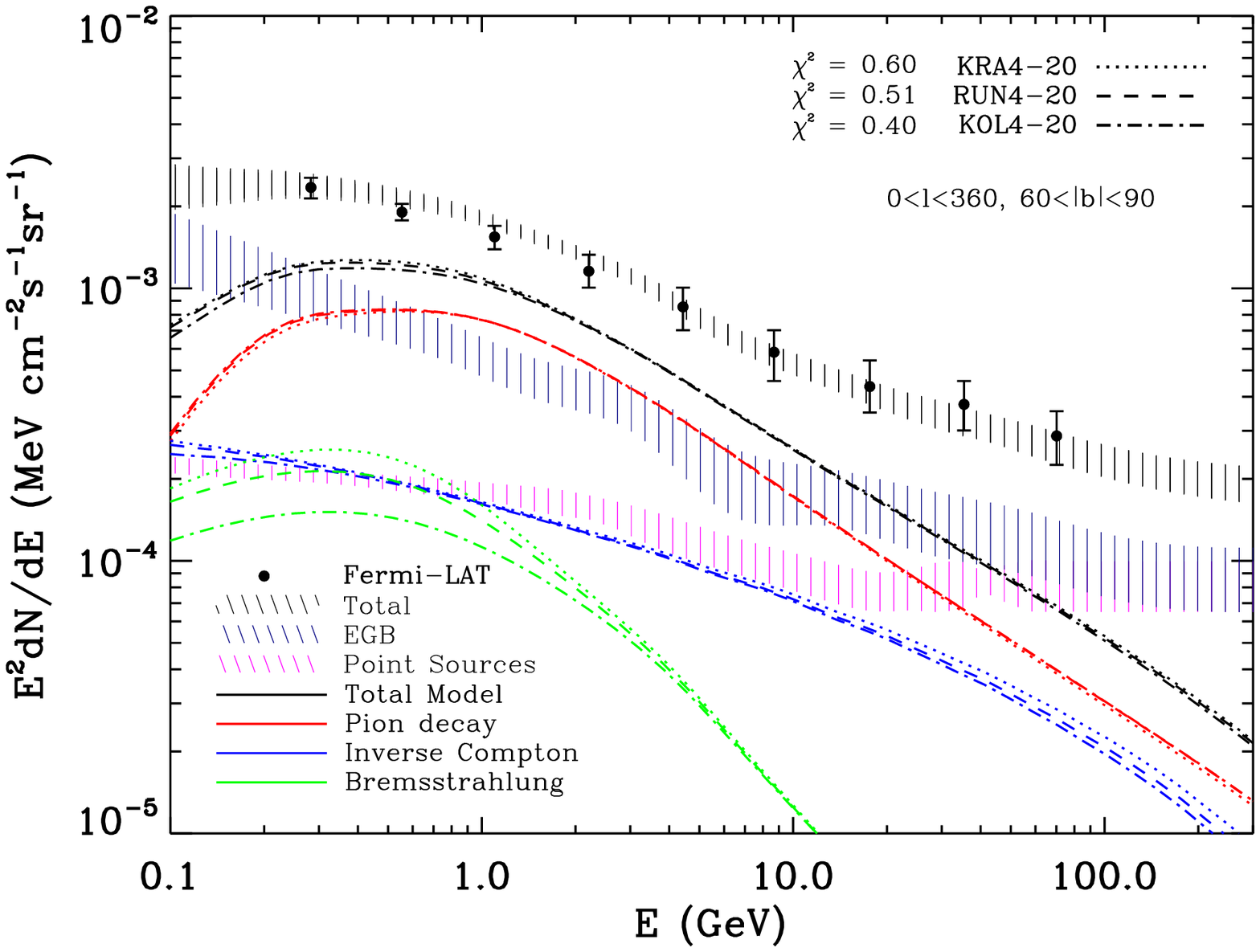}
\end{center}
\caption{Gamma-ray spectra predicted in models with different $\delta$. Plots 
refer to the different sky regions of our study. 
\emph{dotted lines}: $\delta = 0.5$, \emph{dashed lines}: $\delta = 0.4$, 
\emph{dashed-dotted lines}: $\delta = 0.33$. 
For all cases $z_d=4$~kpc and $r_d=20$~kpc.}
\label{fig:DiffIndexVary}
\end{figure}

In Fig.~\ref{fig:DiffIndexVary} we compare the $\gamma$-ray spectra predicted by models with different diffusion spectral index $\delta$ in the three sky regions under study. 
Changing $\delta$ affects the proton spectra whose propagation 
timescale depends on the diffusion timescale.
Lower values of diffusion index $\delta$ make the protons propagated spectra 
be harder for the same injection properties, \textit{resulting in the need for a 
softer proton injection index at high energies} to reproduce the data, as shown in Table~\ref{tab:Param}, where we show our results for $\delta=0.33$ (``KOL4-20"),  $\delta=0.5$ (``KRA4-20") and the intermediate value $\delta=0.4$ (``RUN4-20"). This in turn produces differences in the $\pi^{0}$ fluxes at the highest energies.

Unlike protons, electron propagation at energies above 5~GeV is significantly affected by the energy loss time-scale and, since the ISRF and B-field model are kept fixed, the ICS and the higher part of the bremsstrahlung spectrum
are not largely affected. In the very high energy part of the ICS spectrum
we see a hardening for the models with greater $\delta$.
That hardening is due to the fact that for greater $\delta$ the higher energy $e^{\pm}$ diffuse faster out of the Galactic disk compared to lower energy $e^{\pm}$, reaching the higher latitudes where we observe them through their ICS. 
In the lower energy part of the spectrum, bremsstrahlung varies
significantly among the models since very different Alfv\'en velocities and $\eta$ values are used in those models in order to fit the CR data. 
While the overall fit of the $\gamma$-ray spectra is not affected much due to
opposite effects of changing the value of $\delta$ on the bremsstrahlung and the $\pi^{0}$ spectra below a few GeV, the relative ratio of bremsstrahlung to $\pi^{0}$ flux among the models with different $\delta$ changes by up to a factor of two (at $E_{\gamma} \simeq 0.5$~GeV). Since both $\pi^{0}$ and bremsstrahlung are morphologically correlated to the gasses, discriminating among those components is very difficult. On the other hand, the predicted $\bar{p}$ spectrum favors larger values of $\delta$ within our parameter search region.

\begin{figure}[tbp]
\begin{center}
\includegraphics[width=0.45\textwidth]{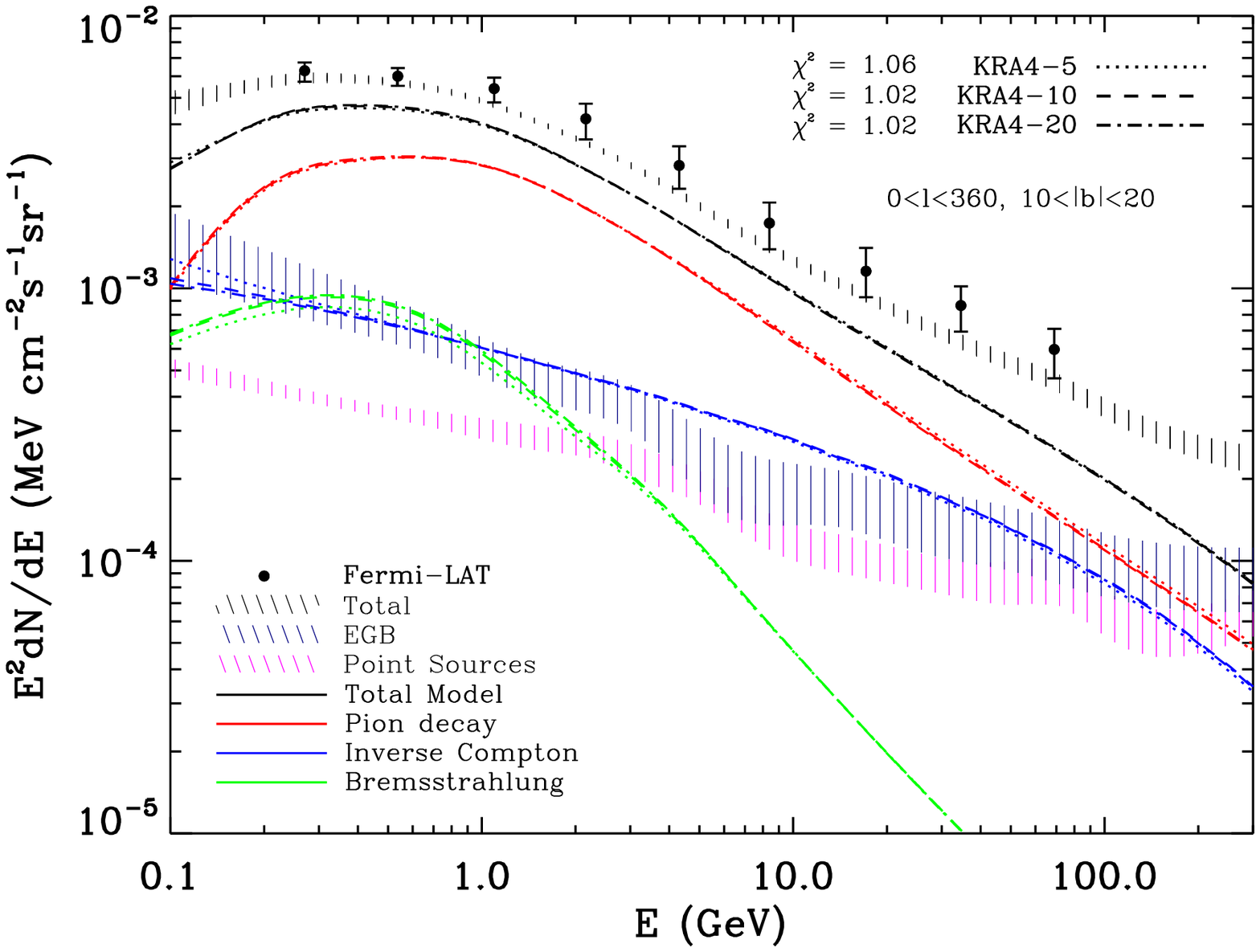}
\includegraphics[width=0.45\textwidth]{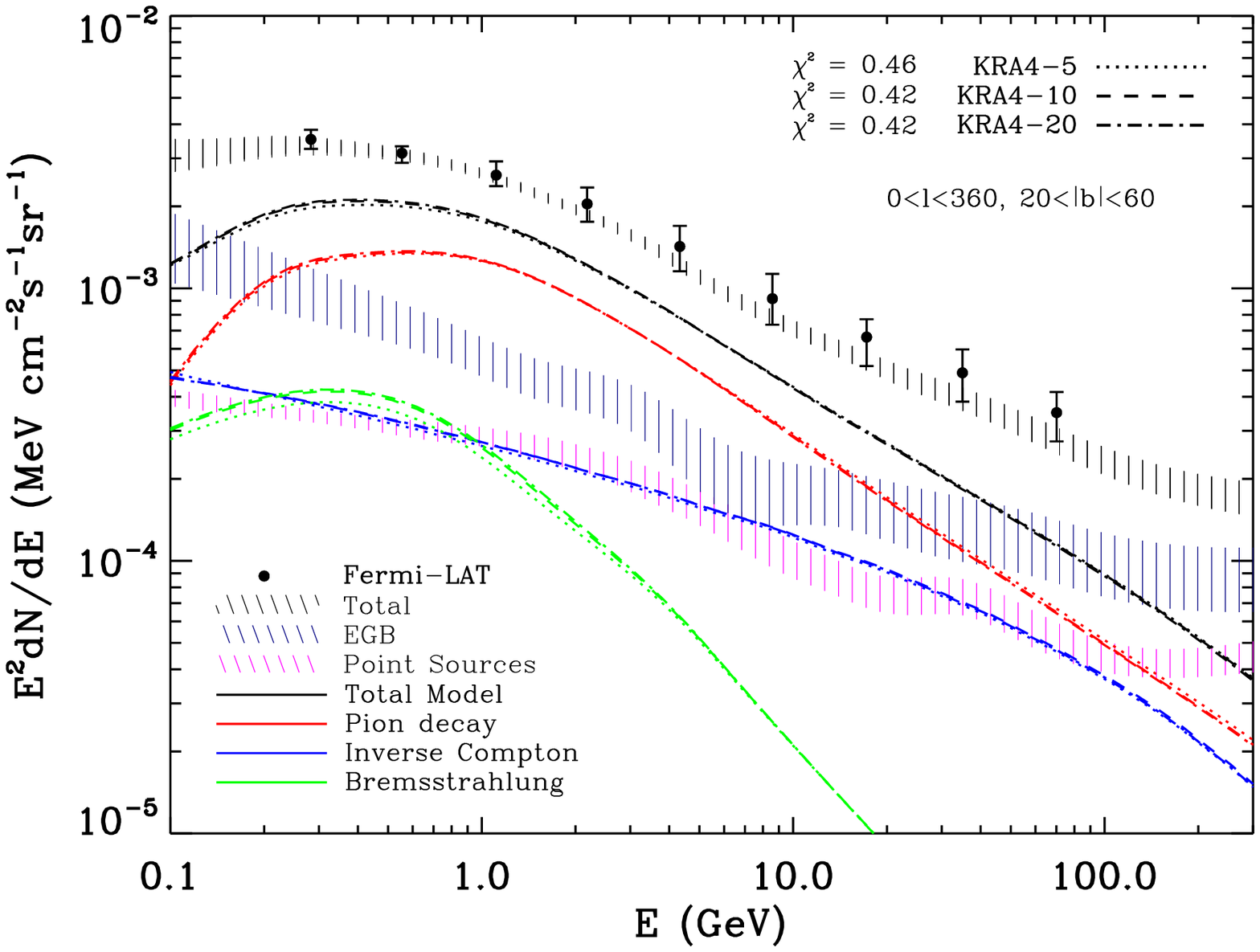}
\includegraphics[width=0.45\textwidth]{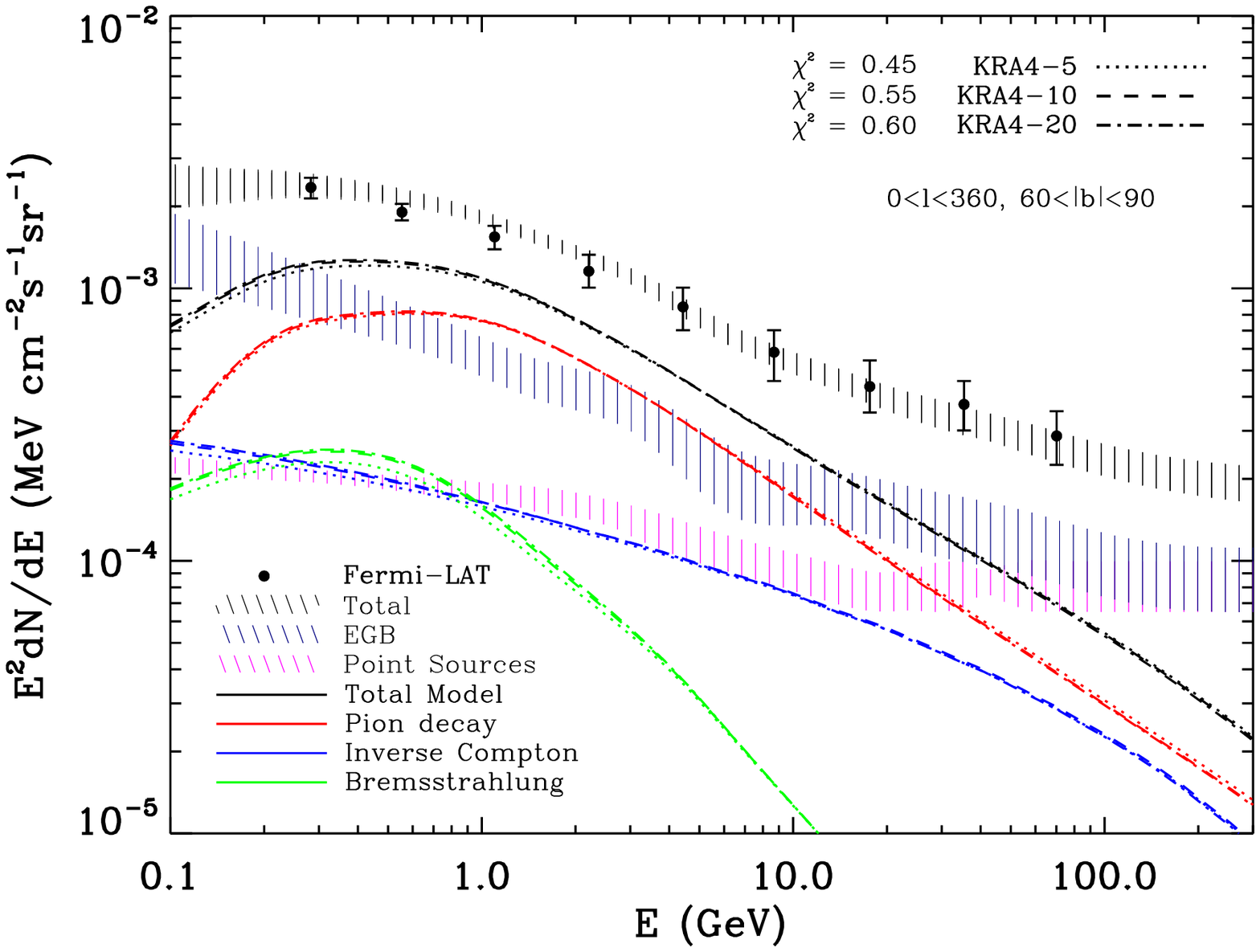}
\end{center}
\caption{Gamma-ray fluxes for models in which we vary the diffusion radial scale $r_{d}$. Plots refer to the different sky regions of our study. \emph{dotted lines}: $r_{d} = 5$~kpc, \emph{dashed lines}: $r_{d} = 10$~kpc, 
\emph{dashed-dotted lines}: $r_{d} = 20$~kpc. For all cases $\delta=0.5$ and 
$z_d=4$~kpc.}
\label{fig:rdVary}
\end{figure}

In Fig.~\ref{fig:rdVary} we show the effect of varying the radial scale for the diffusion coefficient $r_{d}$.
Decreasing the value of $r_{d}$ results in lower values for the diffusion coefficient towards the Galactic center relative to the Sun's position, which forces the $e^{\pm}$ and $p$ produced by sources closer to the Galactic 
center to spend a greater time close to the disk than those produced by sources close to the Sun.
After refitting the diffusion coefficient normalization $D_{0}$ (see Table~\ref{tab:Param}) to the CR nuclear data, the net change in the fluxes is negligible. We also find that the difference between the radial independent case and the case with $r_d = 20~\kpc$ is negligible. 
Yet the quality of the fit of the predicted $\bar{p}$ to the \textit{PAMELA} $\bar{p}$ is affected by changing the scale $r_{d}$ and disfavors the smaller values of $r_{d}$ as is shown in Table~\ref{tab:Param}.

\begin{figure}[tbp]
\begin{center}
\includegraphics[width=0.45\textwidth]{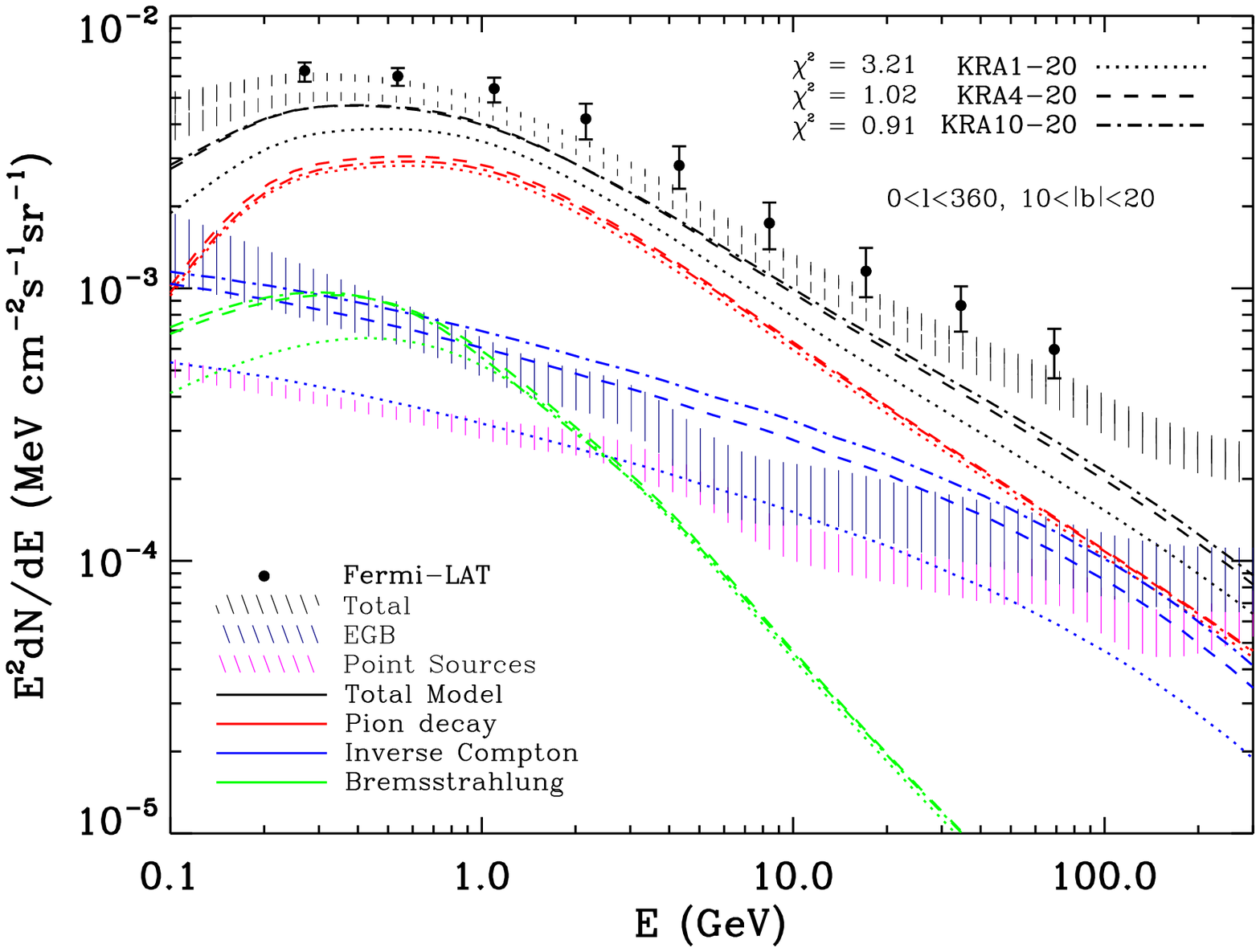}
\includegraphics[width=0.45\textwidth]{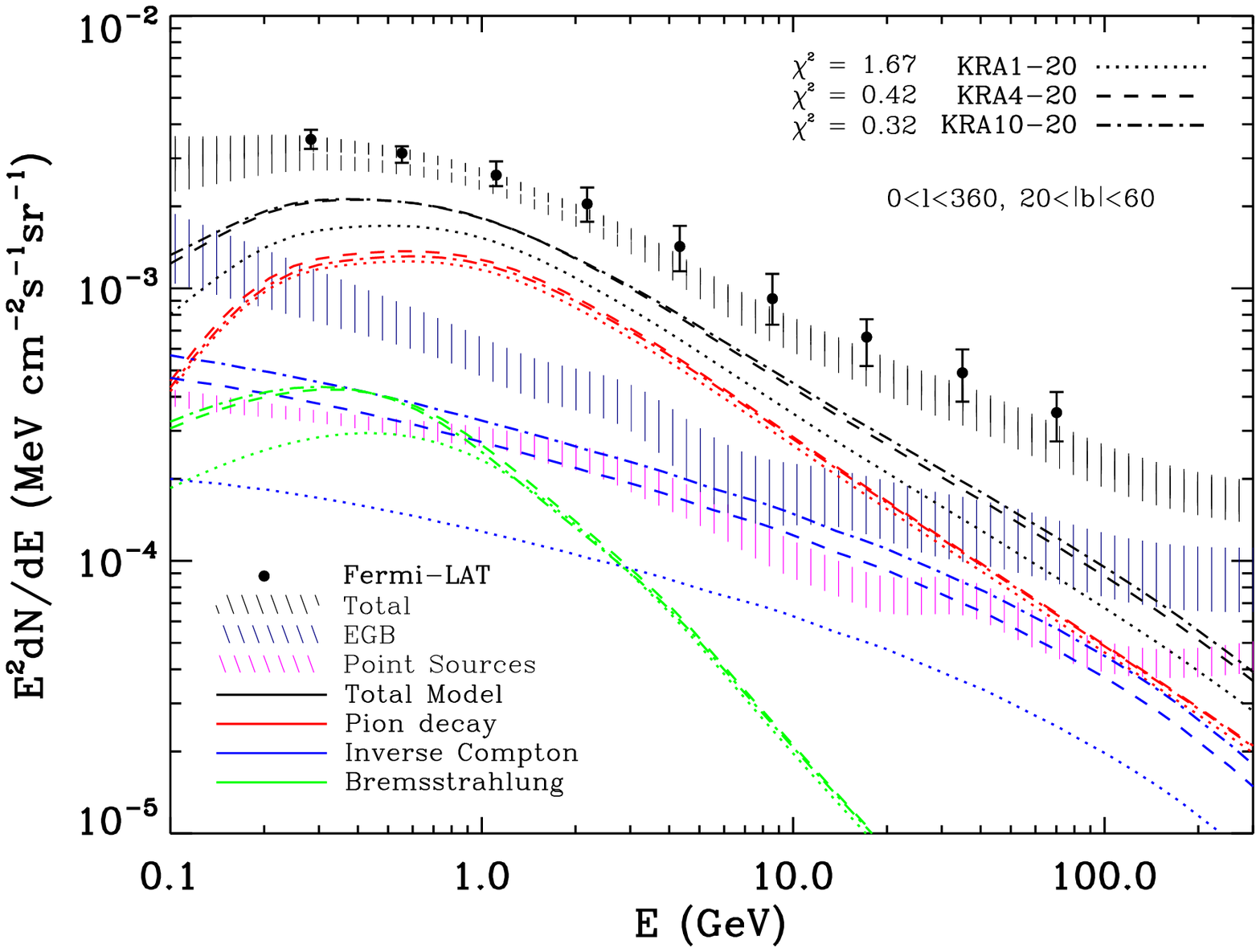}
\includegraphics[width=0.45\textwidth]{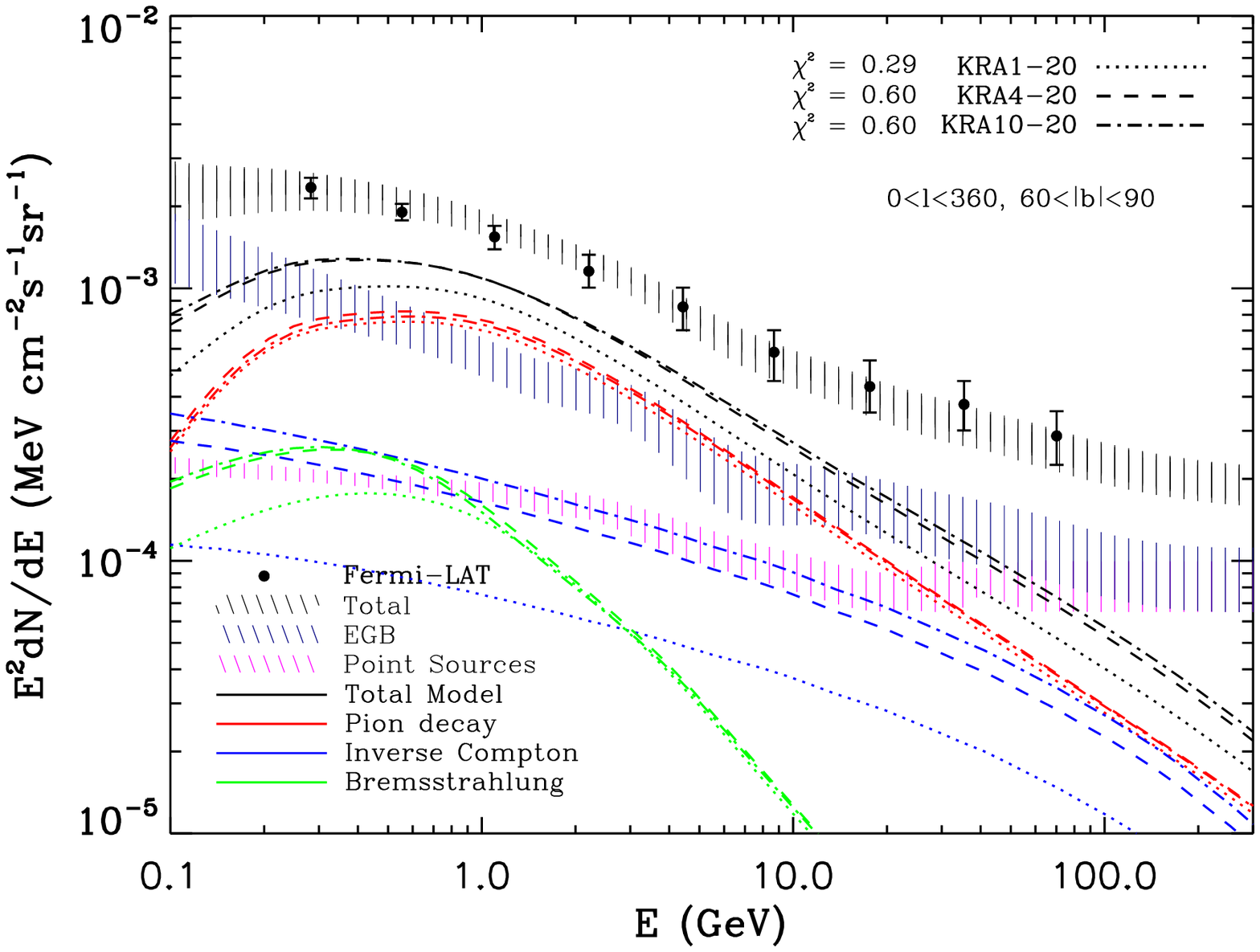}
\end{center}
\caption{Gamma-ray fluxes for models with different diffusion scale $z_{d}$. Plots refer to the different 
sky regions of our study. \emph{dotted lines}: $z_{d} = 1$~kpc, 
\emph{dashed lines}: $z_{d} = 4$~kpc, \emph{dashed-dotted lines}: 
$z_{d} = 10$~kpc. For all cases $\delta=0.5$ and $r_d=20~\kpc$.}\label{fig:zdVary}
\end{figure}

In Fig.~\ref{fig:zdVary} we show the effect of varying the diffusion vertical scale $z_{d}$, that is correlated to the height of the diffusion zone \cite{Evoli:2008dv}. 
Since smaller values of $z_{d}$ yield a greater diffusion coefficient above the galactic plane, an overall rescaling of the diffusion normalization $D_0$ is necessary to fix the secondary to primary ratio (see Table~\ref{tab:Param}). Concerning protons, for which energy losses are less significant, changing $z_{d}$ does not affect much their spectrum while also keeps the $\pi^{0}$ spectrum and flux unchanged. 
Bremsstrahlung emission is also weakly affected by the changes in $z_{d}$, because it is correlated morphologically to the gas distribution which is concentrated close to the galactic disk. 
Finally the ICS spectrum is mainly affected by the actual distribution of 
electrons being confined within thinner (thicker) diffusion zones resulting
in lower (higher) total IC flux. For the case $z_{d} = 1$ this results in 
a poor fit to the gamma-ray spectra. Thus thin diffusion zone models such as 
those that have been suggested by~\cite{Hooper:2009fj}, in order to give low 
$\bar{p}$ fluxes from Kaluza-Klein DM annihilation models, while simultaneously explaining the leptonic excesses observed by ATIC and \textit{PAMELA}, are in tension with the combination of CR \textit{and} $\gamma$-ray spectra. 

\begin{figure}[tbp]
\begin{center}
\includegraphics[width=0.45\textwidth]{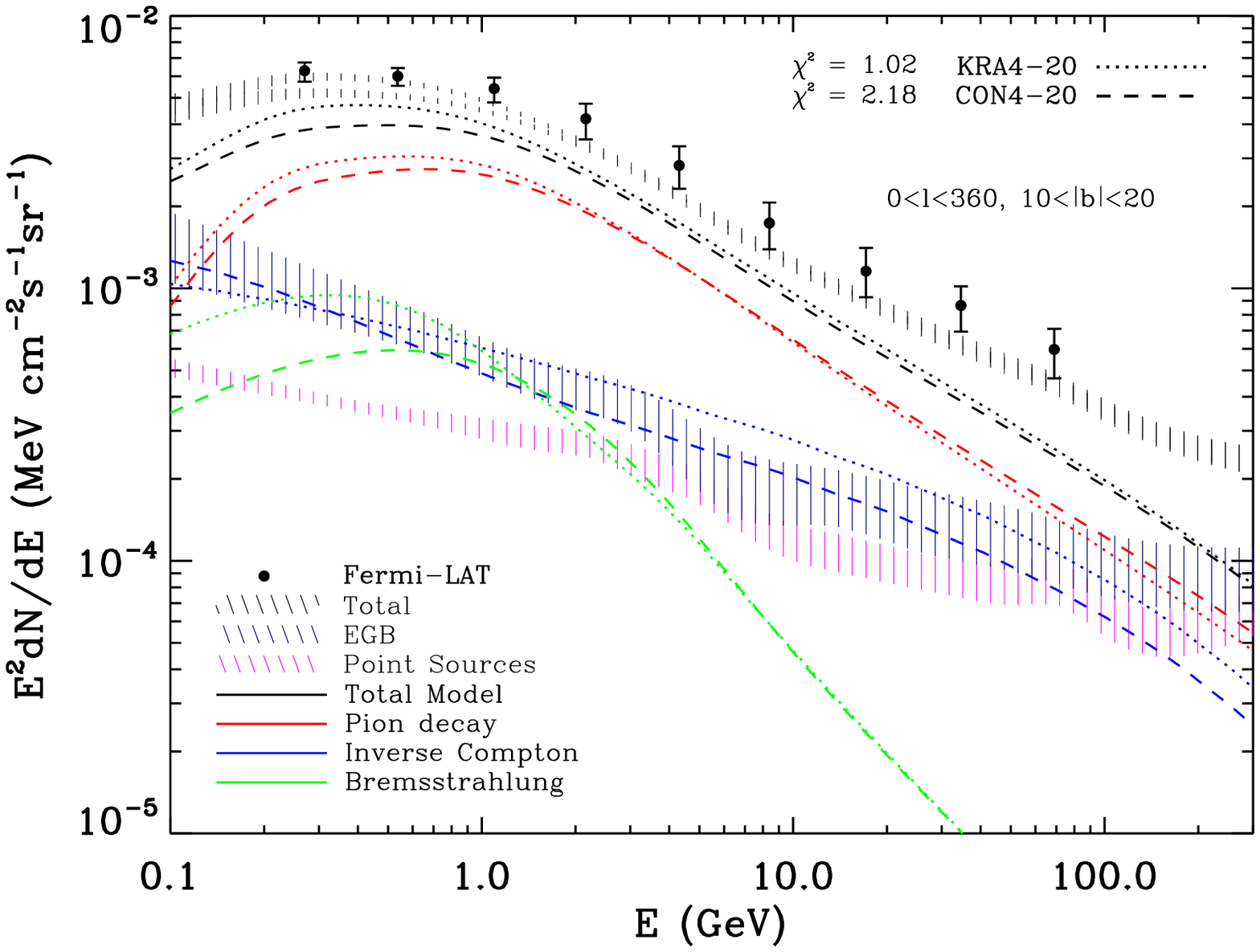}
\includegraphics[width=0.45\textwidth]{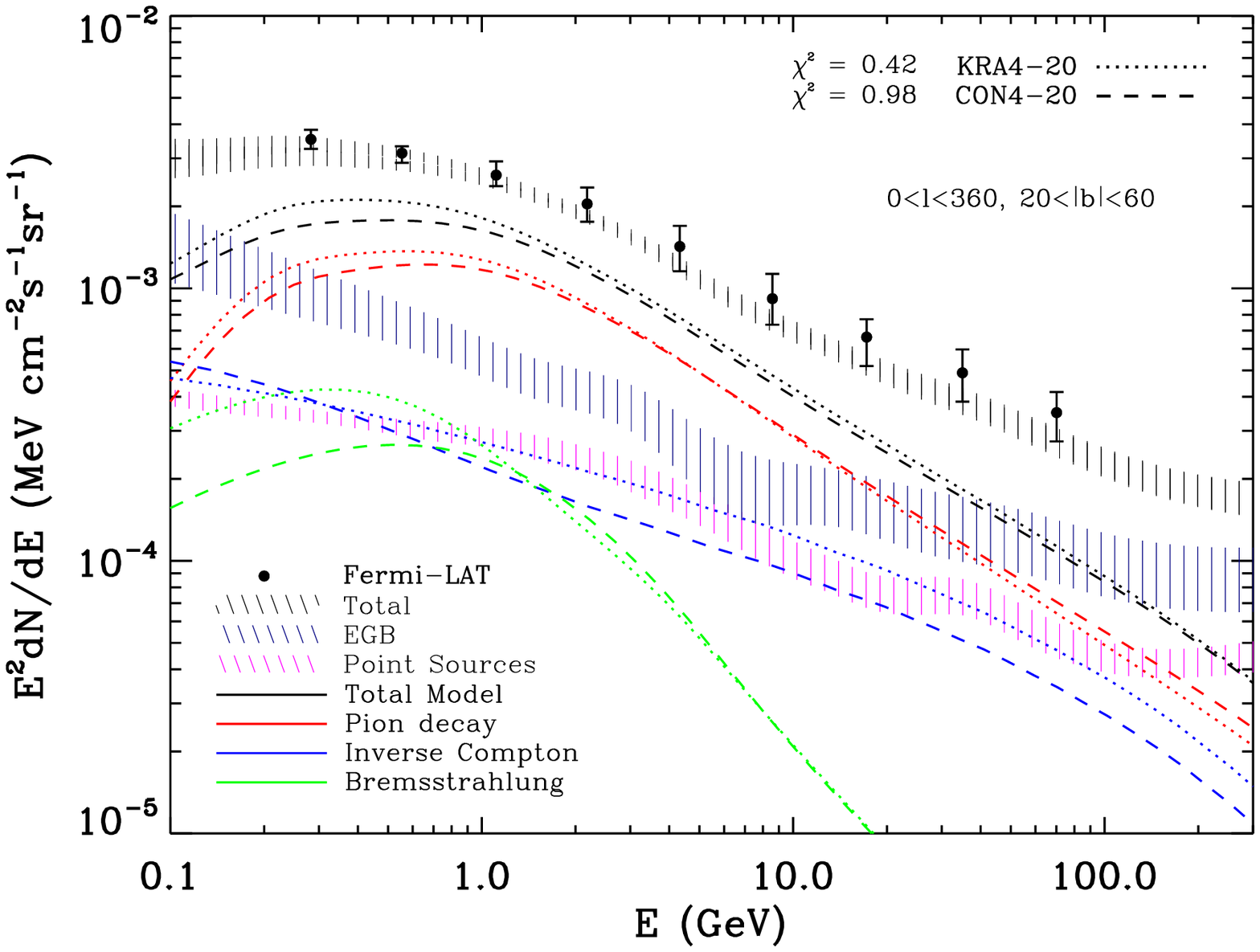}
\includegraphics[width=0.45\textwidth]{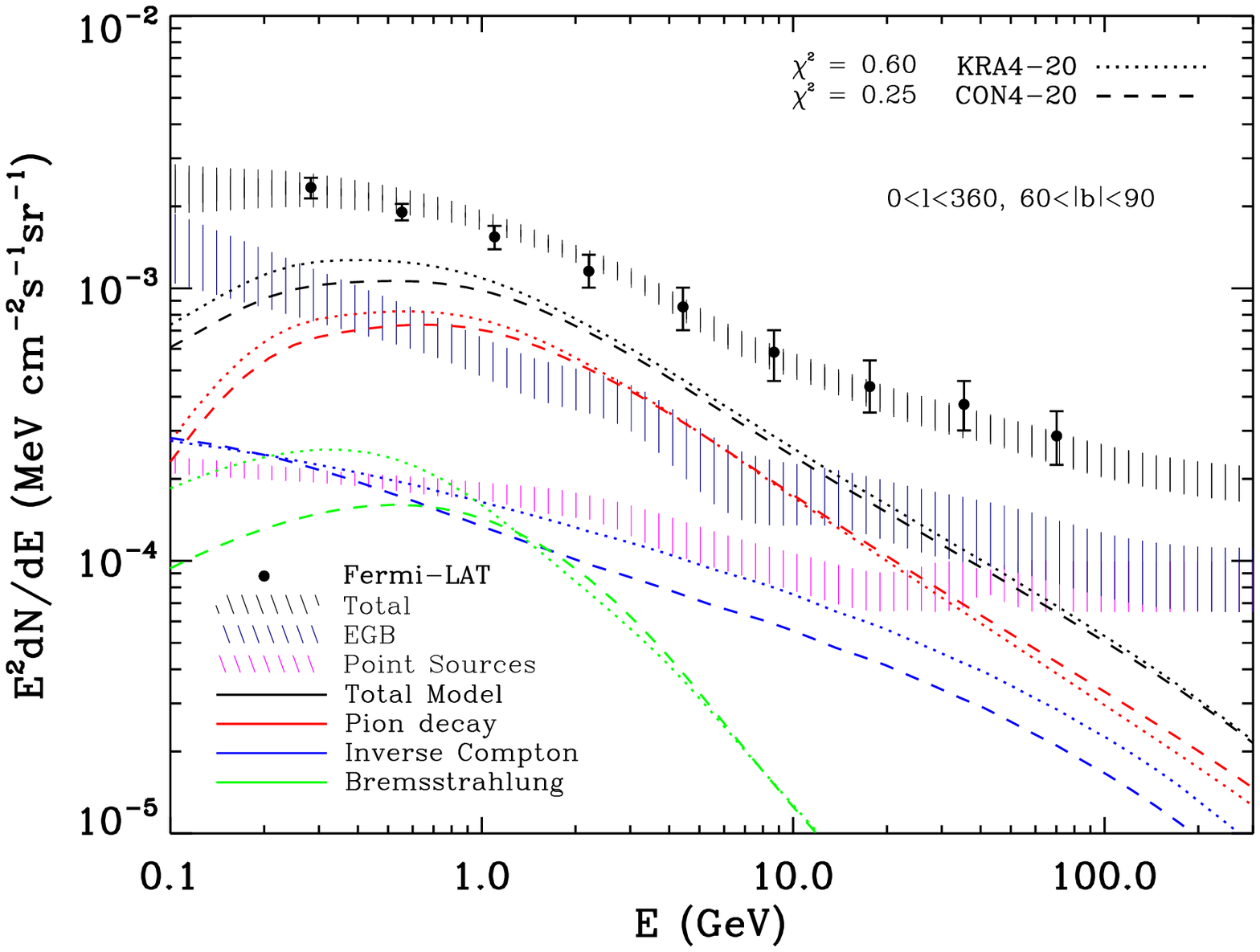}
\end{center}
\caption{Effects of convection on $\gamma$-ray spectra compared to the reference model. Plots refer to the different 
sky regions of our study. \emph{dotted lines}: reference model, 
\emph{dashed lines}: ``CON4-20" model.}
\label{fig:Conv}
\end{figure}

In Fig.~\ref{fig:Conv} we consider the effects of convective winds in the Galaxy. Convection introduces a new time scale into the propagation of CRs which mainly affects the protons, since energy losses still dominate electron propagation. After we refit to the B/C and proton fluxes, the diffusion properties are strongly affected (see Tab.~\ref{tab:Param}) resulting in a significantly altered ICS and bremsstrahlung components. This is most evident at the low energy part of the spectra, where convection is more important relative to ICS and synchrotron losses. We find that high convection models are not favored by $\gamma$-ray data in the middle latitude region. In fact, we also find that low energy positron and electron fluxes are in tension with PAMELA positron fraction data.
\subsection{Rigidity break in injection or diffusion}

Recently \textit{PAMELA} \cite{Picozza:2006nm} has observed a break at rigidity $R \simeq 230$~GV, at both the proton and He spectra \cite{Adriani:2011cu}, suggesting a hardening of the CR spectra at high energies.
The harder spectral power-law at high rigidities is confirmed by the CREAM data as well \cite{Yoon:2011zz}. Our combined fit of \textit{PAMELA} and CREAM data leads to a break rigidity in our models at $R_{0,2}^p \sim 300~\GV$.

One possible explanation for the observed rigidity break, is that the same break originates at the CR acceleration sites, at the SNRs shocks.
Such a scenario  has been suggested by studies of SNRs \cite{Damiano:2011tp} and
from diffusive shock acceleration semi-analytical calculations
\cite{2000ApJ...533L.171M, 2002APh....16..429B, 2005MNRAS.364L..76A, Blasi:2006wj, Caprioli:2010uj}.
The pressure on accelerated
particles around the shock leads to the formation of a precursor \cite{Damiano:2011tp} where the upstream
fluid is slowed down and compressed \cite{Damiano:2011tp}. 
For diffusively accelerated particles moving with respect to the shock, and thus between regions of different pressure, their gained energy depends on the "compression ratio". 
On average, the higher energy particles which have larger diffusion lengths will probe the entire (or a greater part of the) precursor than the lower energy particles, leading to a concave shape spectrum. Thus the highest energy particles will "feel the total compression ratio" \cite{Damiano:2011tp} which (from first order Fermi acceleration) will result in the spectrum being harder than $E^{-2}$ at high energies and softer at low energies\cite{Damiano:2011tp, Caprioli:2010uj, Caprioli:2010ne, 2002APh....16..429B}.
 
Another possible explanation is that at $\sim 230$ GeV  we observe the 
emergence of a population of galactic sources (SNRs) that accelerate CRs with a resulting harder injection index. 
As long as this second SNR population is common enough in the Galaxy, and with a similar distribution as that injecting the softer CR spectra in the ISM, both possibilities can be modeled in the same way with the DRAGON code, i.e.~with the injection of CRs given by Eq.~\ref{eq:injNucl}-\ref{eq:Spectra}. We will refer to this scenario as ``scenario A", under which the $\gamma$-ray spectra of Fig.~\ref{fig:RefAstroModelGammas}-\ref{fig:zdVary} were produced.

A third possibility is that instead we observe a change in the turbulence power spectrum of the ISM. 
The properties of the interstellar magnetic turbulence can be indirectly inferred from CR measurements. Before the \textit{PAMELA} data, CR spectra were not measured accurately enough to exclude any break in the diffusion coefficient rigidity index.  
Coincidentally the needed change in the diffusion 
index $\Delta\delta \simeq 0.17$ is the same as when considering the transition from a Kraichnan type turbulence at low $R$ to a Kolmogorov type at high $R$.
We pursue this scenario (``scenario B") and after refitting the pulsar normalization flux to the total \textit{Fermi} $e^{+} + e^{-}$ spectra, we calculate the $\gamma$-ray spectra, for $\mid b \mid > 10^\circ$.
\begin{figure}[tbp]
\begin{center}
\includegraphics[scale=.42]{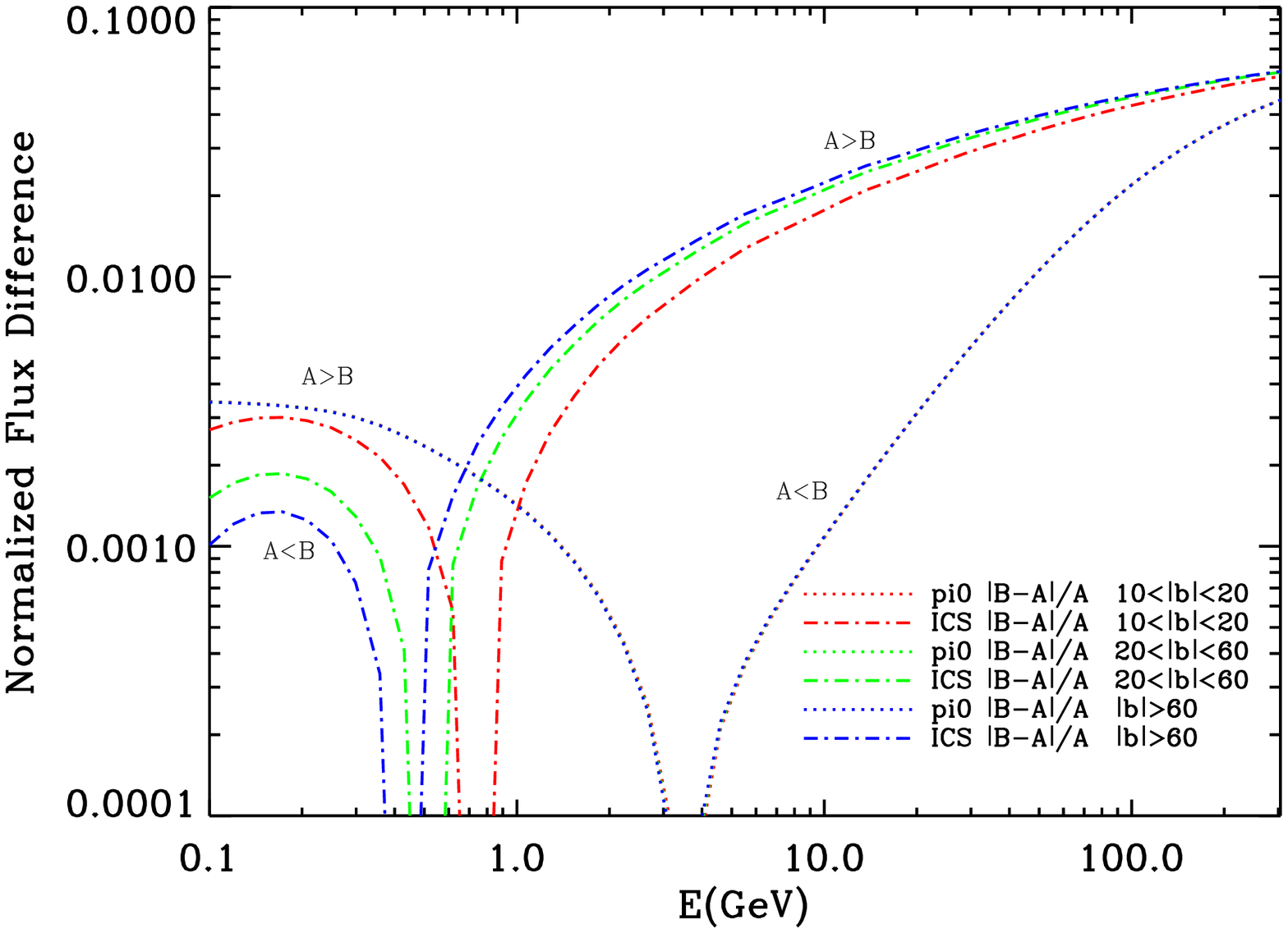}
\includegraphics[scale=.42]{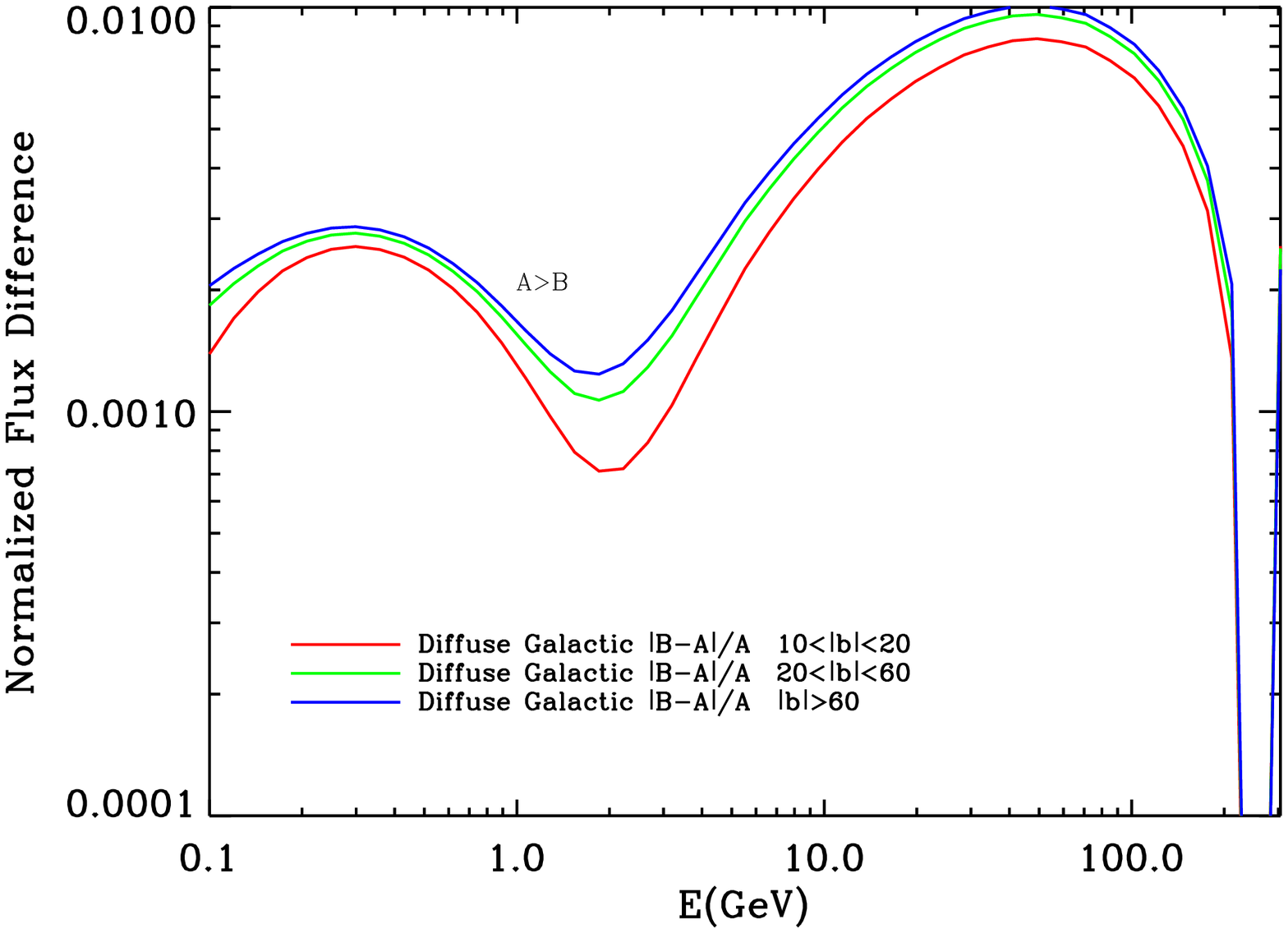}
\end{center}
\caption{Absolute difference in fluxes between scenarios A and B (see text) normalized to the flux of scenario A. \emph{Left panel}: $\pi^0$ (\emph{dotted lines}) and ICS (\emph{dashed-dotted lines}). \emph{Red}: $10^\circ<\mid b \mid <20^\circ$,  \emph{Green}: 
$20^\circ<\mid b \mid <60^\circ$ and  \emph{Blue}: $\mid b \mid >60^\circ$. Notice that the normalized differences for the $\pi^0$ diffuse component have almost identical spectra. 
\emph{Right panel}: Absolute difference in the total diffuse fluxes between scenarios A and B (including bremsstrahlung). Colors as in left-hand panel plot. The difference in the total diffuse are up to 2\% (for $E_{\gamma} < 300$~GeV), due to cancellation between $\pi^0$ and ICS differences.}
\label{fig:A_VS_B_scenarios}
\end{figure}
In Fig.~\ref{fig:A_VS_B_scenarios} we present the difference in the $\gamma$-ray flux between scenarios A and B, normalized to the flux from scenario A, where we used the global diffusion model ``KRA4-20" for our calculations.
As it can be seen in Fig.~\ref{fig:A_VS_B_scenarios} (left) the maximal difference between scenarios A and B, is up to $O(0.1)$ in both ICS and $\pi^{0}$ spectra at energies $E_{\gamma} \sim 100$~GeV, but of opposite sign, resulting in a difference in the total diffuse galactic component to be less than $O(10^{-2})$ (Fig.~\ref{fig:A_VS_B_scenarios}, right) over the whole considered spectrum, including $E_{\gamma} \sim 100$~GeV, in all three regions of interest. Such differences are too small to be probed by $\gamma$-rays at such high energies, because the extragalactic background flux becomes more important, while also CR contamination modeling uncertainties, low statistics and the possible contribution from Dark Matter add to the total uncertainty. 

Thus as also suggested by \cite{Evoli:2011id} the best way to discriminate between scenarios A and B is through the $\bar{p}$ flux, where scenario A would give a soft break only at $R \sim 10$ GV due to the break of the p and He spectra at $\sim230$ GV, while scenario B would also give a harder break at $\sim 230$ GV (with the same spectral index
change as in the p and He fluxes) from the diffusion of the secondary $\bar{p}$ in the ISM (see discussion in \cite{Evoli:2011id}).

Finally we note that \cite{Donato:2010vm}, have discussed the impacts of a
\textit{smooth hardening} in the power-law of CRs and its impact on high 
energy $\gamma$-rays and $\bar{p}$s. 
After refitting the propagation parameters
to the current wealth of data (which includes the most recent \textit{PAMELA} 
data \cite{Adriani:2011cu}), and including the ICS and bremsstrahlung 
components, our differences in the total $\gamma$-ray spectra (shown in~\ref{fig:A_VS_B_scenarios} left) turn out to be significantly smaller than those of
\cite{Donato:2010vm}. 

\subsection{Significance of the ISM Gasses}

\begin{figure}[tbp]
\begin{center}
\includegraphics[scale=.45]{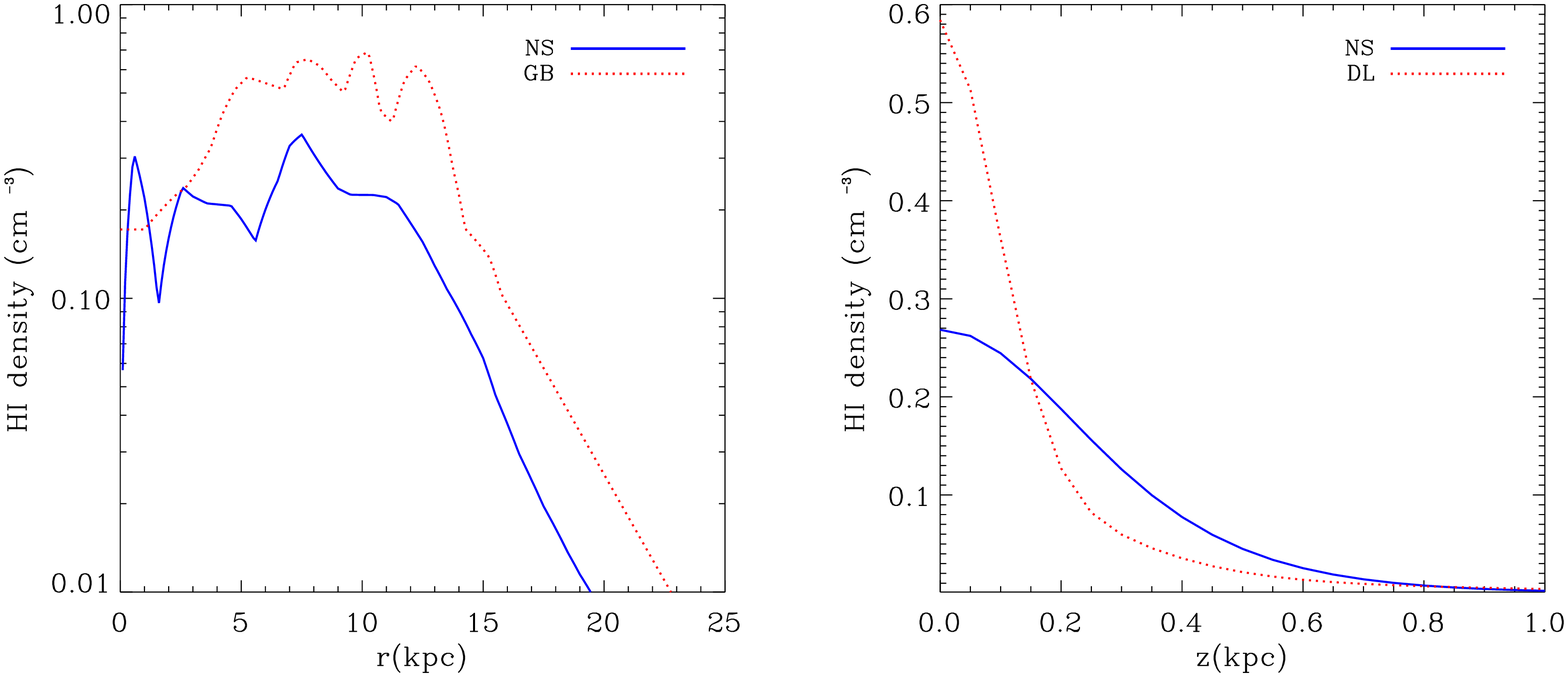}
\includegraphics[scale=.45]{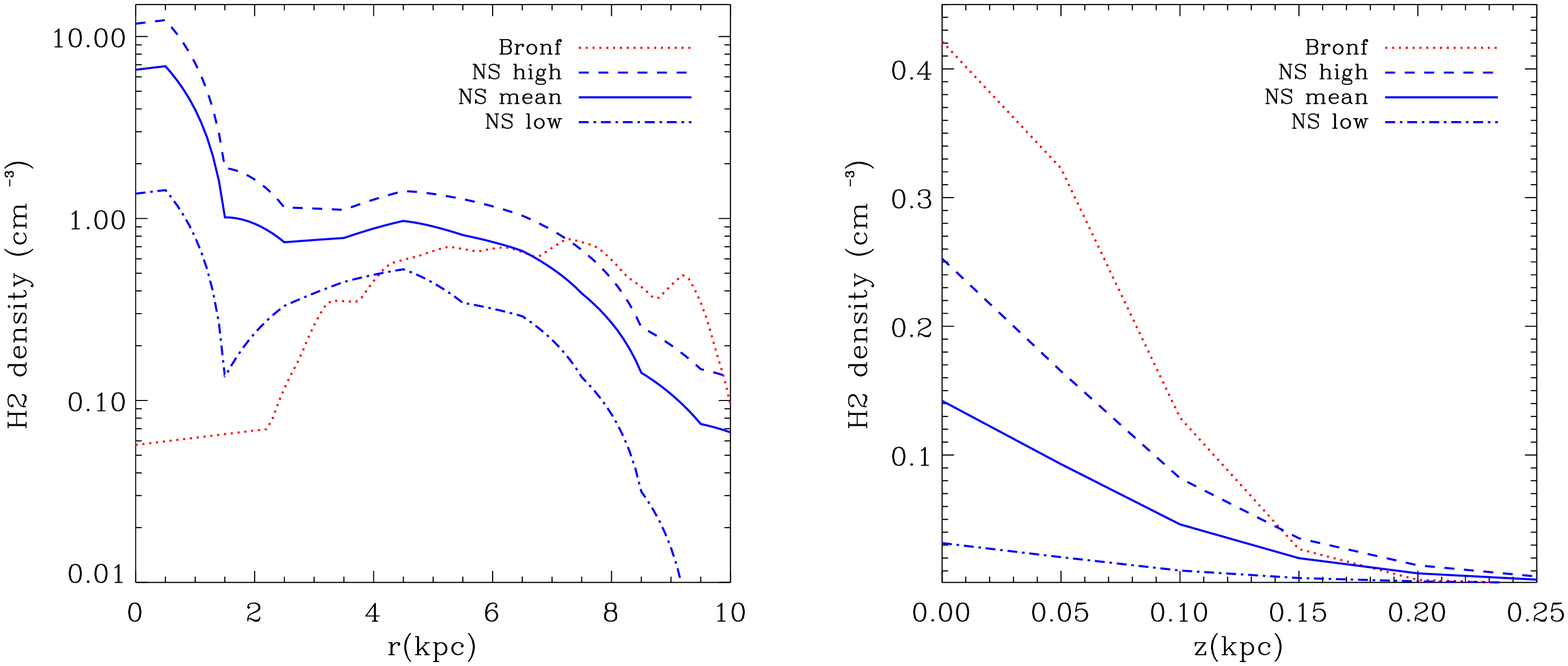}
\end{center}
\caption{Large scale density distributions of atomic (\emph{top}) and molecular (\emph{bottom}) hydrogen in the Galaxy vs $r$ for $z=0$ (\emph{left}); vs $z$ for $r=r_{\odot}$ (\emph{right}). For HI, ``NS" refers to \cite{Nakanishi:2003eb}, ``GB" refers to \cite{1976ApJ...208..346G} and ``DL" to \cite{1990ARA&A..28..215D}. For H2, ``Bronf" refers to \cite{1988ApJ...324..248B}, ``NS high", ``NS mean" and ``NS low" refer to \cite{Nakanishi:2006zf} using, respectively, high, mean and low values of the midplane density.}
\label{fig:HI_H2_profiles}
\end{figure}

\begin{figure}[tbp]
\begin{center}
\includegraphics[width=0.45\textwidth]{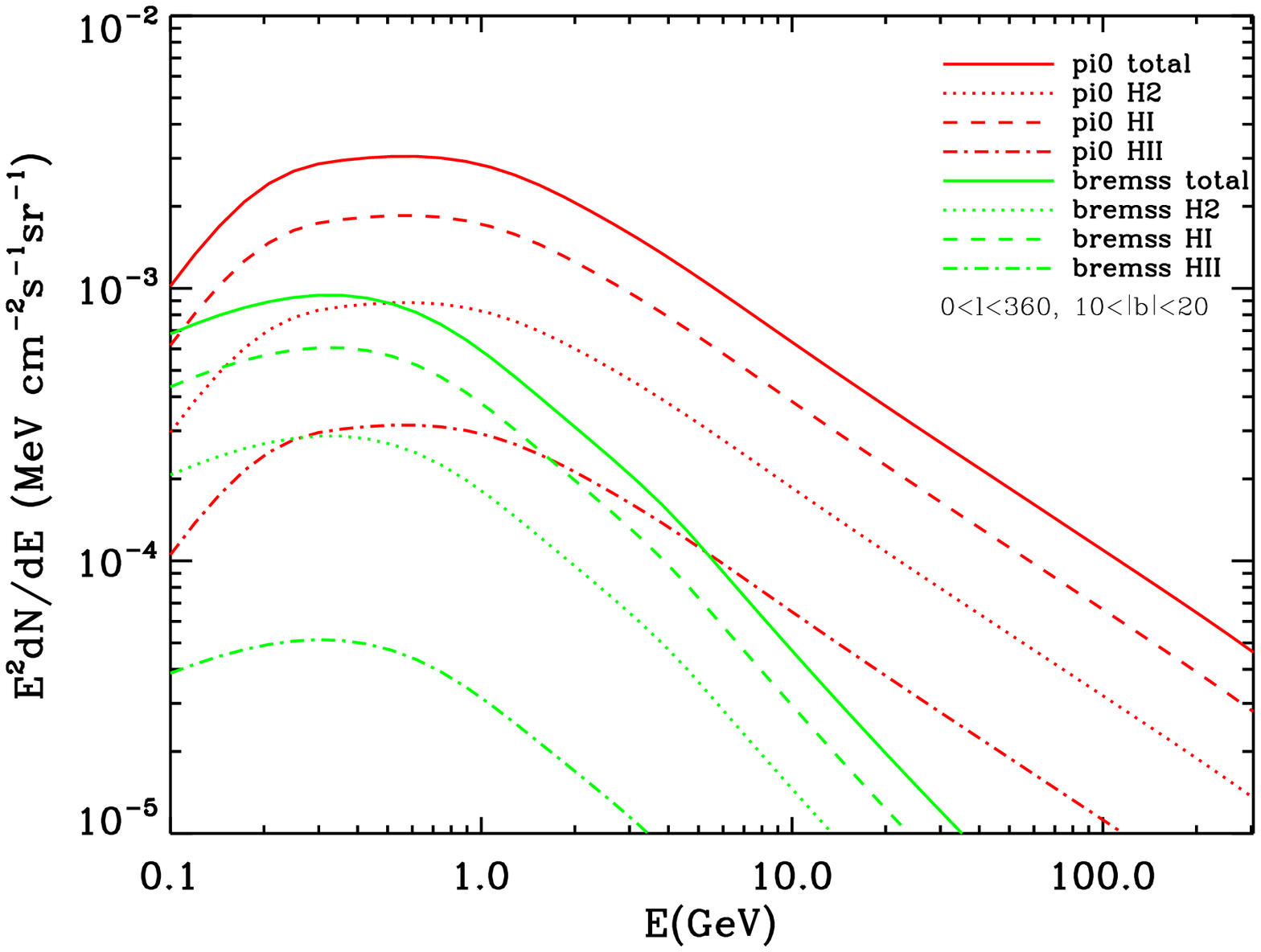}
\includegraphics[width=0.45\textwidth]{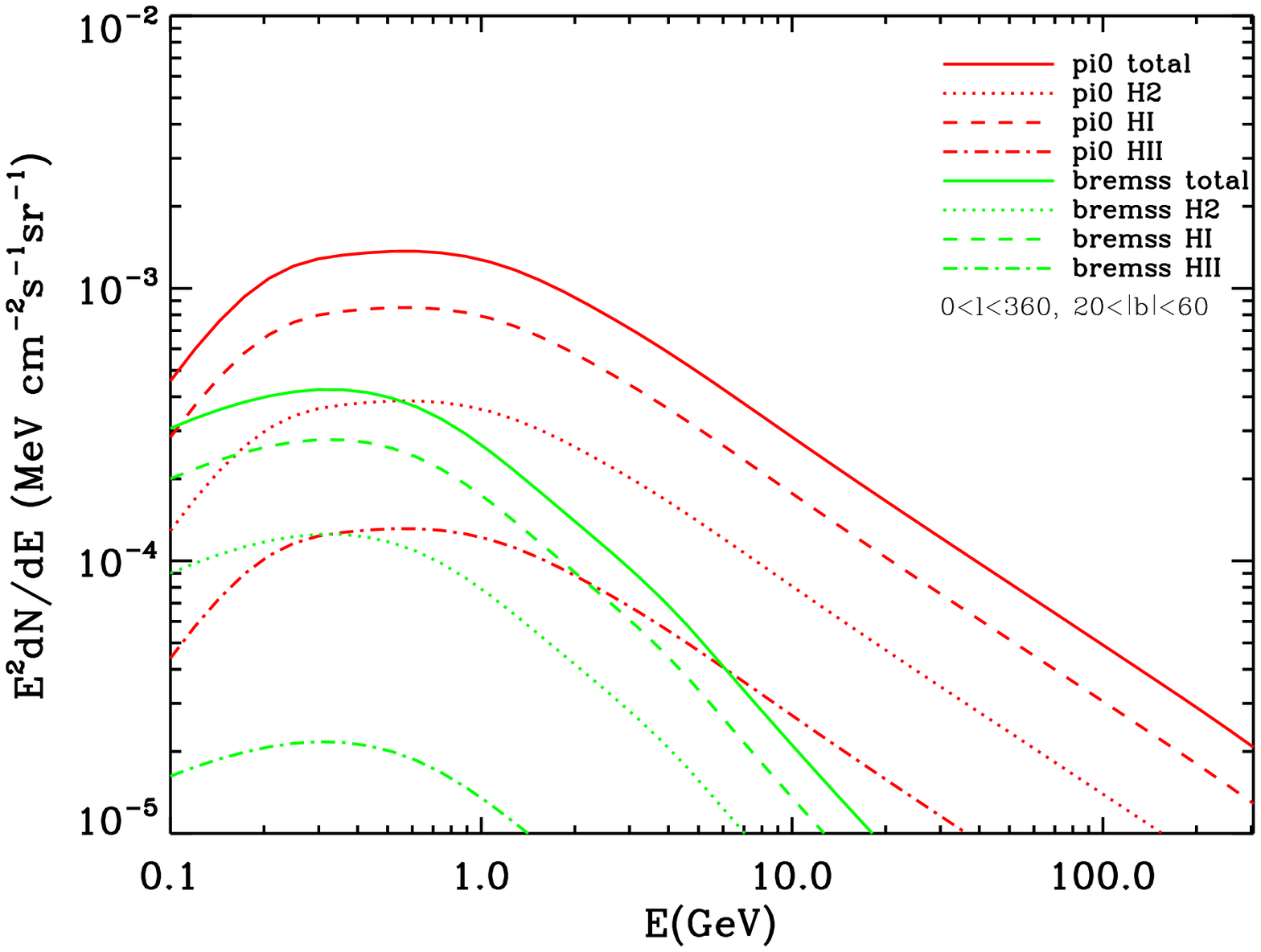}\\
\includegraphics[width=0.45\textwidth]{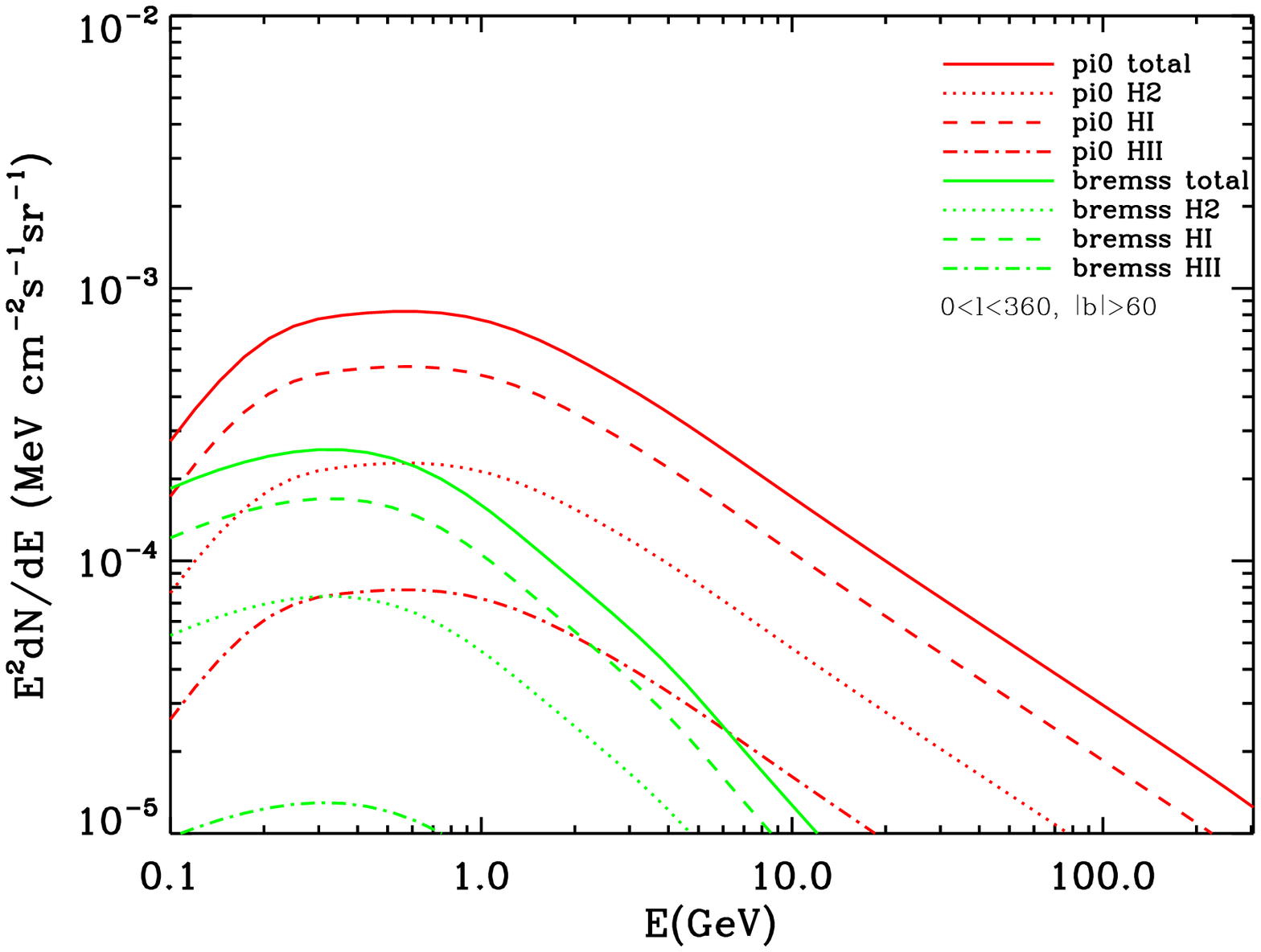}
\includegraphics[width=0.45\textwidth]{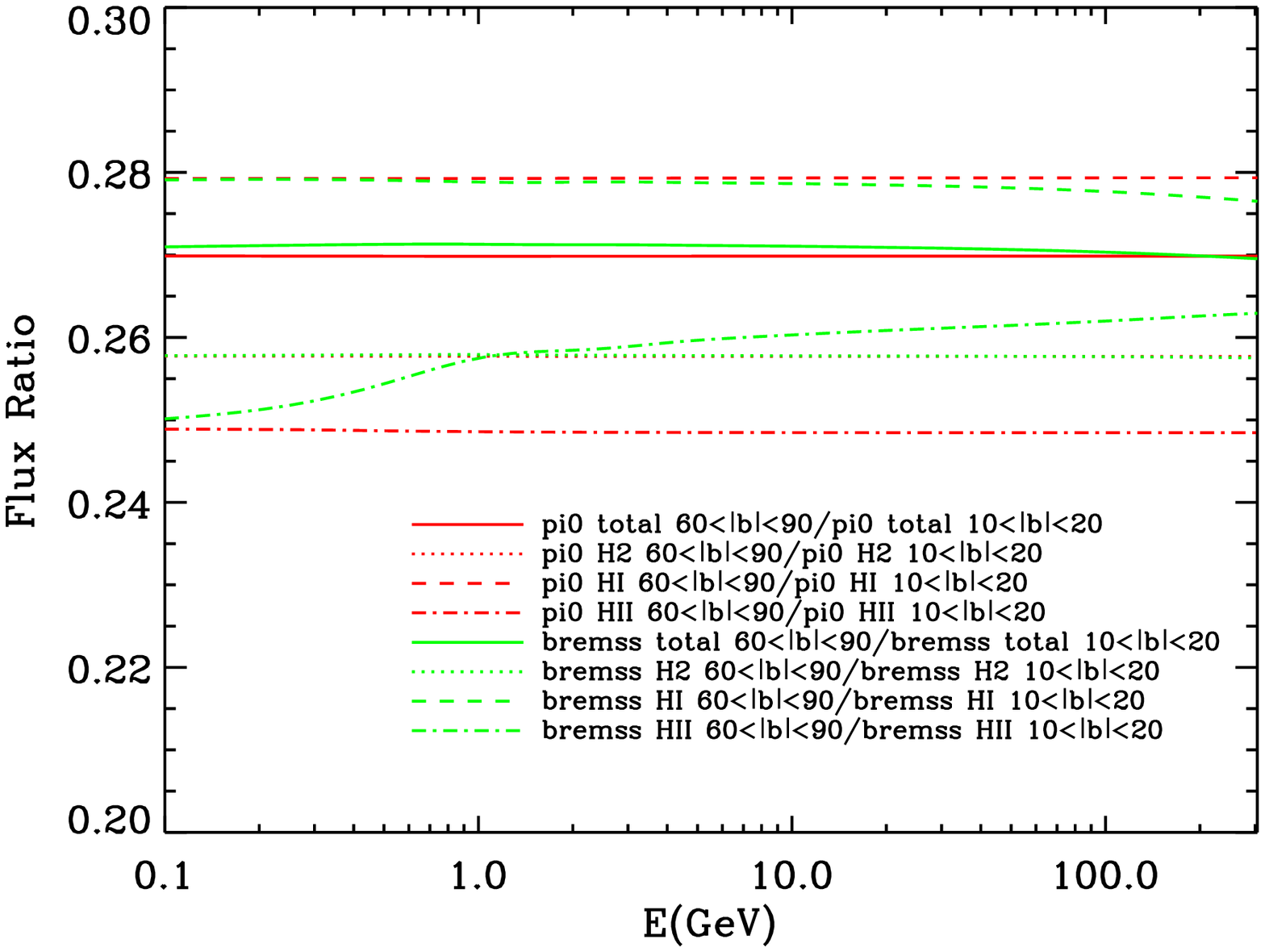}
\end{center}
\caption{The observed $\pi^{0}$(\emph{red}) and bremsstrahlung (\emph{green}) components of the diffuse $\gamma$-ray flux, from the three major ISM gas components.
\emph{Dotted lines}: from H2 following \cite{Nakanishi:2006zf} (mean values),
\emph{dashed lines}: from HI following \cite{Nakanishi:2003eb},
\emph{dashed-dotted lines}: from HII following \cite{1991Natur.354..121C},
\emph{solid lines}: total flux. \emph{Upper left}: $10^\circ< \mid b \mid <20^\circ$,
\emph{upper right}: $20^\circ< \mid b \mid <60^\circ$, \emph{lower left}: 
$\mid b \mid >60^\circ$ averaging over all latitudes. 
\emph{Lower right}:
flux ratios between the components for $\mid b \mid >60^\circ$ to $10^\circ< \mid b \mid <20^\circ$.}
\label{fig:GasComponents}
\end{figure}

\begin{figure}[tbp]
\begin{center}
\includegraphics[scale=.73]{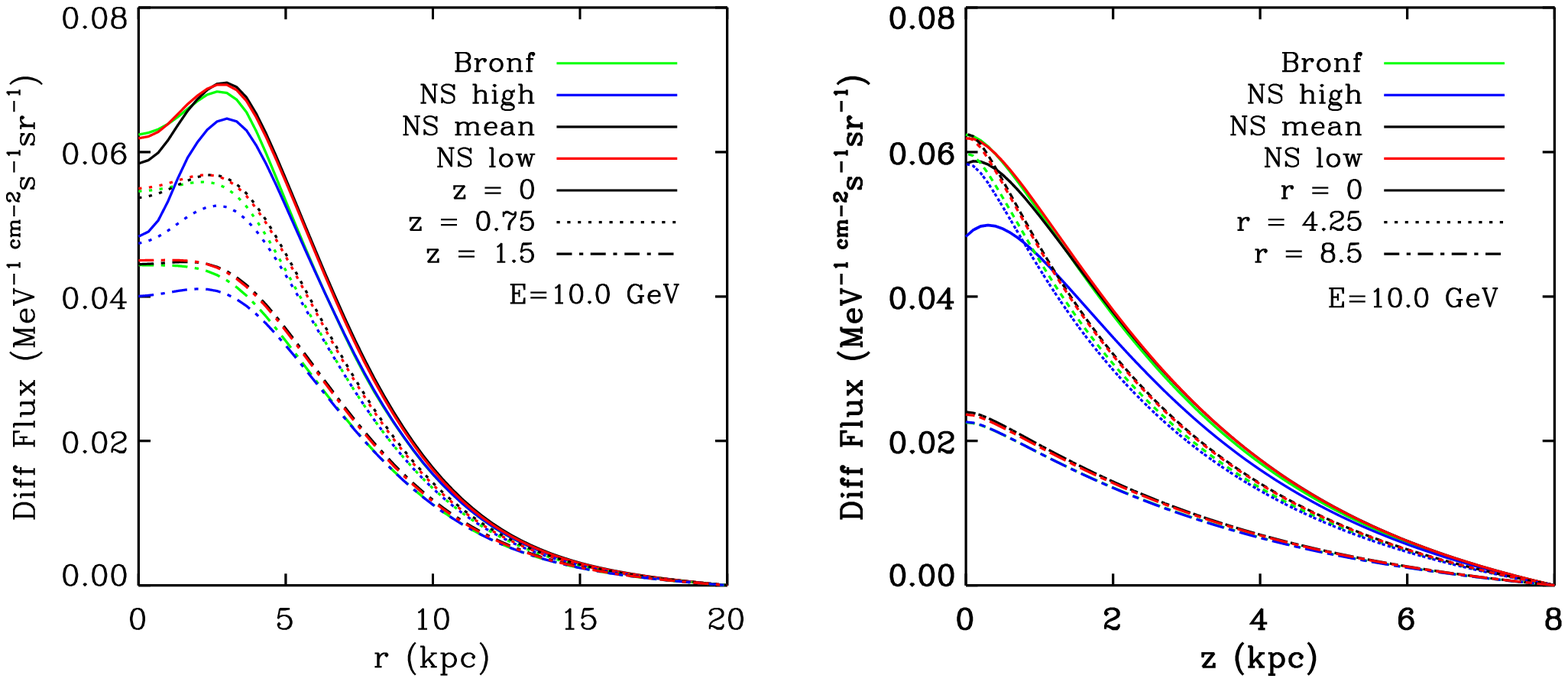}
\includegraphics[scale=.73]{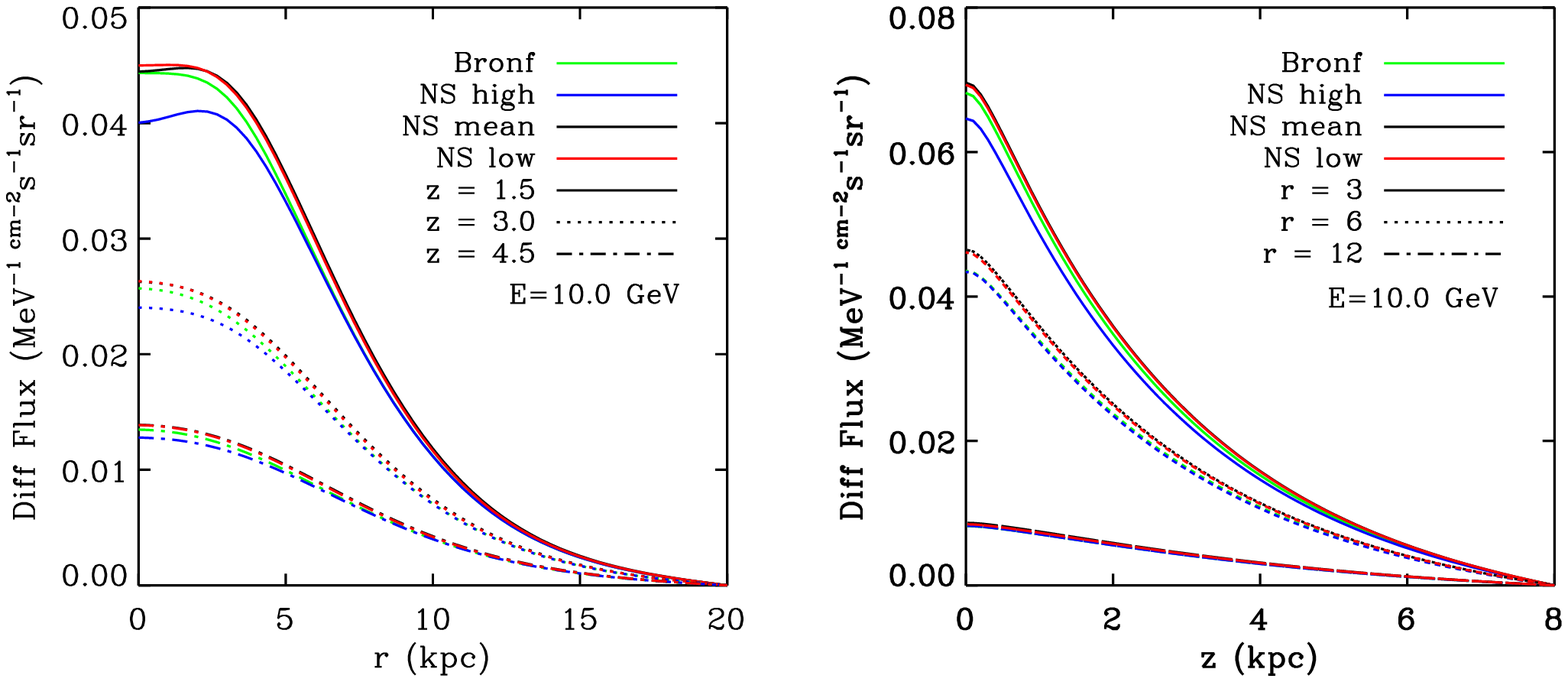}
\end{center}
\caption{CR proton flux profile at $E=10~\GeV$ (at steady state) as a function of $r$ for given $z$ (\emph{left}) and as a function of $z$ for given $r$ (\emph{right}). In ``Bronf" we use the model of \cite{1988ApJ...324..248B} for H2 and the model of \cite{1976ApJ...208..346G, 1990ARA&A..28..215D} for HI gas distributions. In ``NS high", ``NS mean" and ``NS low" we use for H2 gas distribution, respectively, high, mean and low values of midplane density of \cite{Nakanishi:2006zf}, however they share the same HI gas distribution of \cite{Nakanishi:2003eb}.}
\label{fig:Proton_profiles}
\end{figure}

As discussed in section~\ref{sec:GasDistr}, conventional models describing the HI and H2 interstellar gas distributions have been updated by the work of \cite{Nakanishi:2003eb, Nakanishi:2006zf}, of which we show the profiles in Fig.~\ref{fig:HI_H2_profiles}. 
In Fig.~\ref{fig:GasComponents}, using the assumptions of \cite{Nakanishi:2003eb} for the HI, of \cite{Nakanishi:2006zf} for the H2 and of \cite{1991Natur.354..121C} for the HII ISM gas, we show what are the separate contributions to the $\pi^{0}$ and bremsstrahlung $\gamma$-ray fluxes from each of those gasses. As it is clear, the HII contributes on average up to $\approx 10\%$ to the component fluxes at all energies and latitude regions that we show. Thus, changing the assumptions on the HI and H2 models can have a significant effect on both the propagation parameters shown in Tables~\ref{tab:Param} and~\ref{tab:Param2} as has been suggested also in \cite{Maurin:2010zp}, as well as the $\gamma$-ray spectra, 
On the contrary the uncertainties on assumptions on HII (such as \cite{1991ApJ...372L..17R}) could at most result in a few \% change in the $\gamma$-ray flux, well below the uncertainties from either one of the other two ISM gasses. 
Also note that since the protons suffer small energy losses their equilibrium spectrum 
is almost the same in the entire propagation region. Thus after changing the 
ISM assumptions and refitting the injection and propagation properties we also 
have similar CR protons density profiles in the Galaxy. Thus the changes that 
we observe in the $\gamma$-ray $\pi^{0}$ fluxes are very tightly correlated 
to the gas (target) distributions. 
In Fig.~\ref{fig:Proton_profiles} we plot the steady state CR proton differential flux 
profiles (density profiles) at $E = 10$ GeV (note that 
for higher energies the differences between the profiles, for given $z$ or $r$, 
are even smaller).

\begin{figure}[tbp]
\begin{center}
\includegraphics[width=0.45\textwidth]{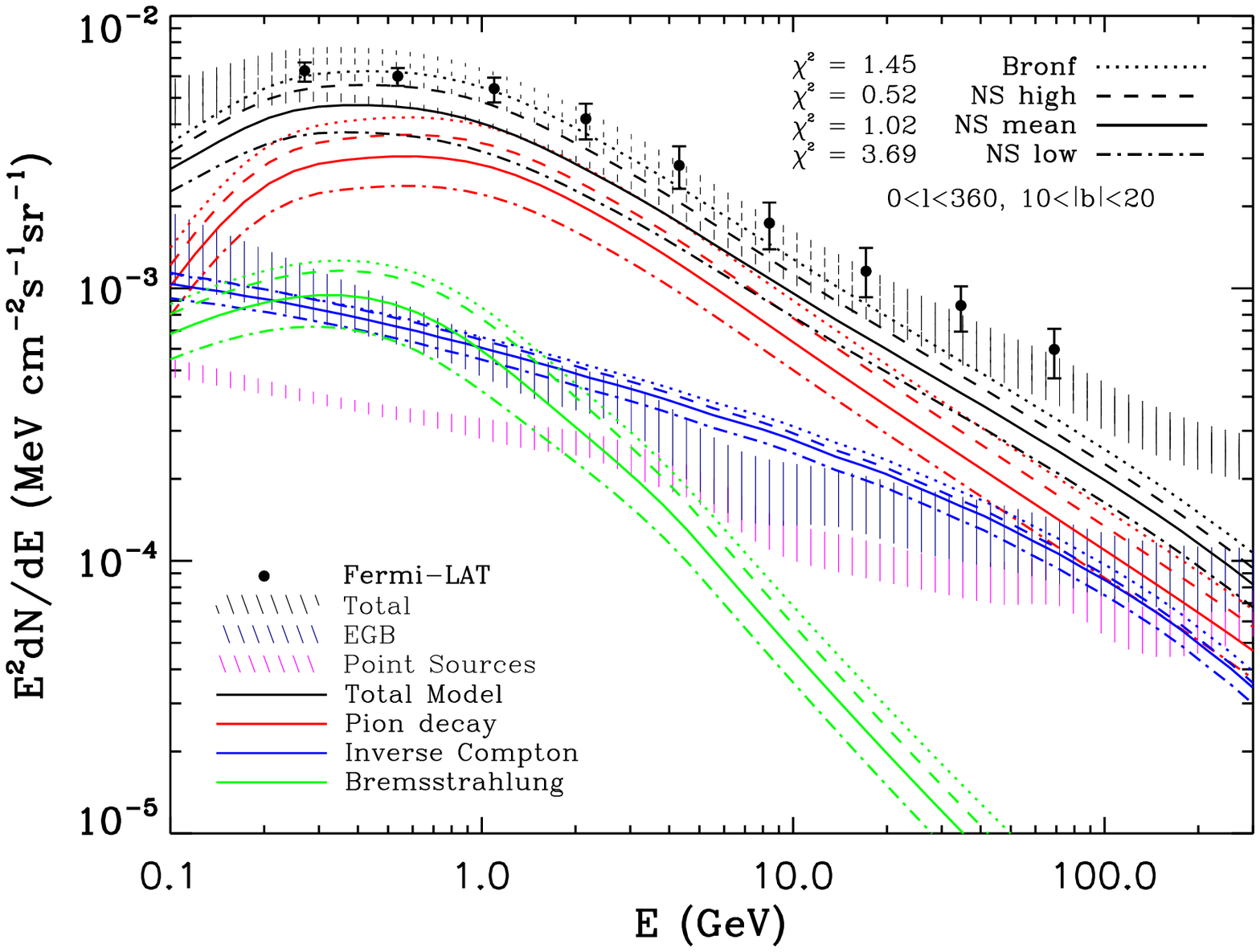}
\includegraphics[width=0.45\textwidth]{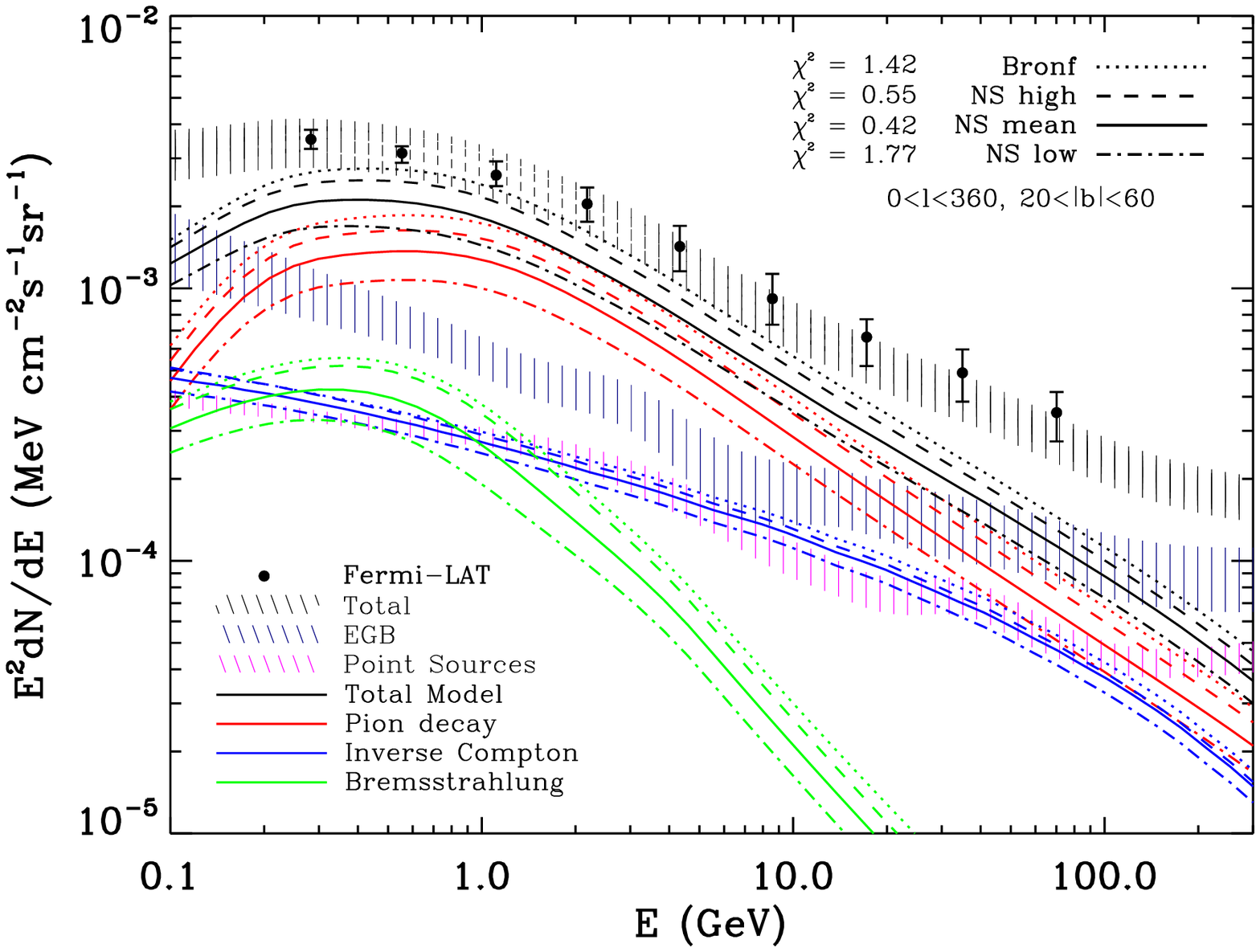} \\
\includegraphics[width=0.45\textwidth]{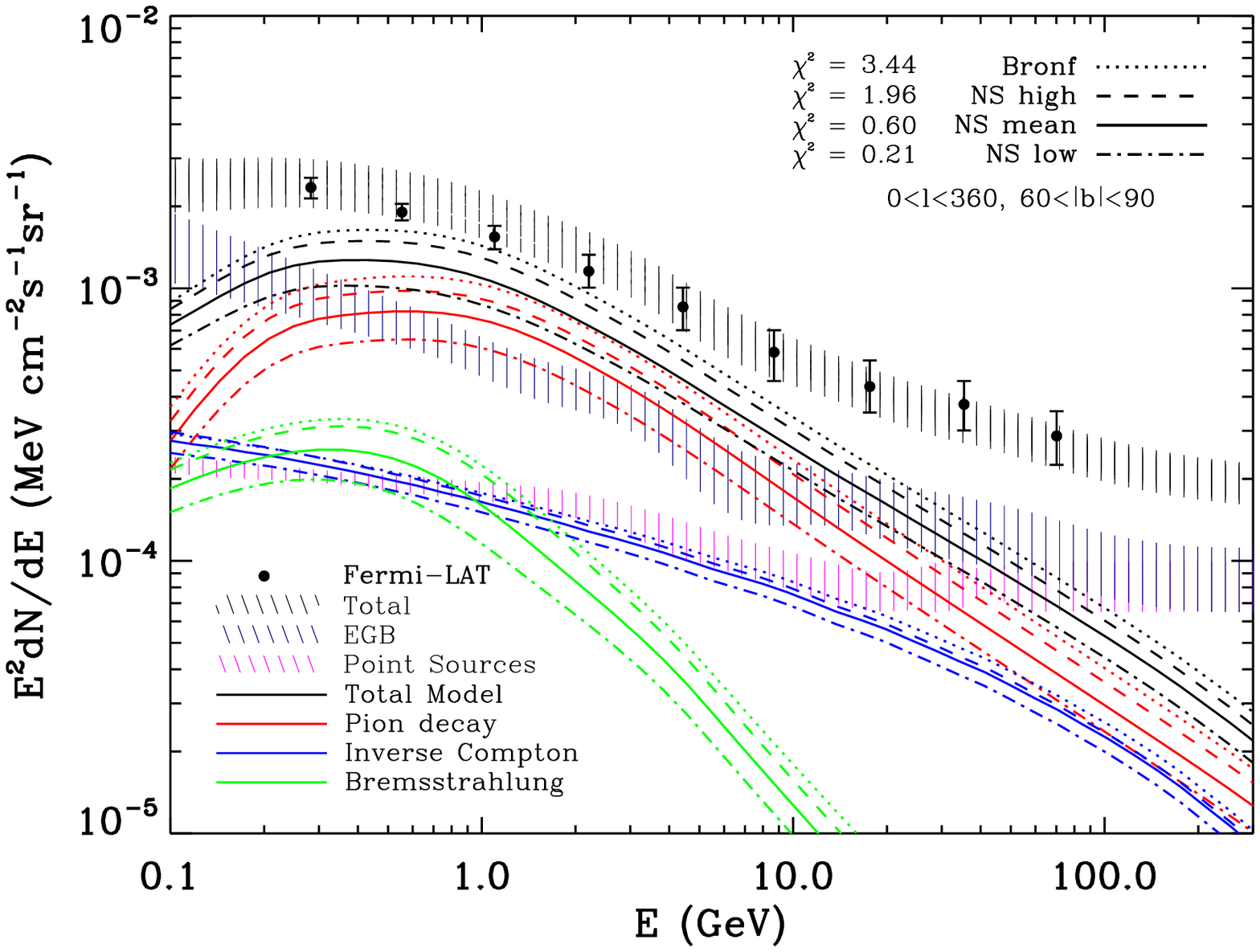}
\includegraphics[width=0.45\textwidth]{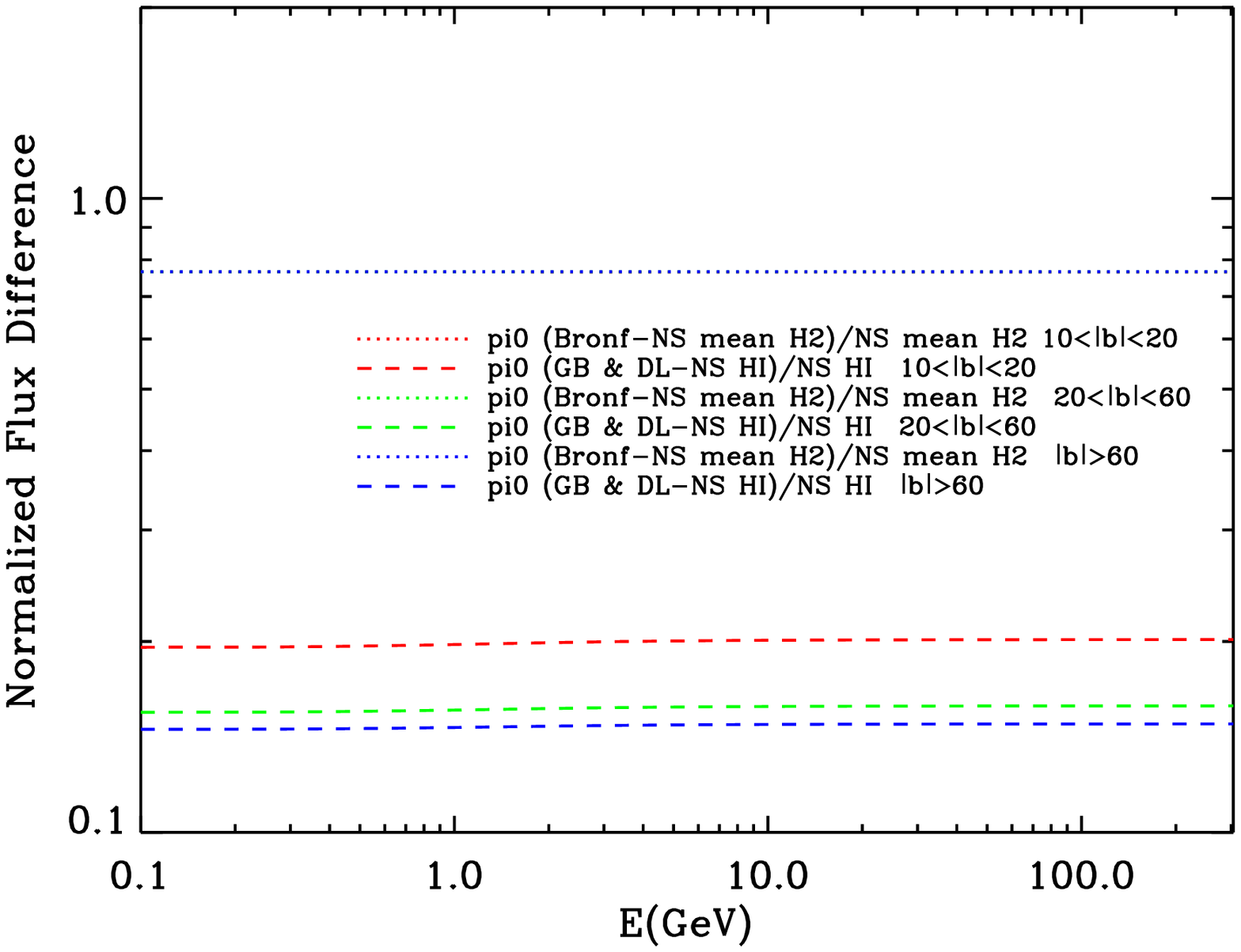}
\end{center}
\caption{Effects of changing the models that describe HI and H2
gas distributions on the $\gamma$-ray spectra in the sky regions of our study. 
\emph{dotted lines}: HI from \cite{1976ApJ...208..346G, 1990ARA&A..28..215D} and H2 from \cite{1988ApJ...324..248B}. 
\emph{solid lines}: HI from \cite{Nakanishi:2003eb} and H2 from \cite{Nakanishi:2006zf},  
\emph{dashed lines}: HI from \cite{Nakanishi:2003eb} and H2 from \cite{Nakanishi:2006zf} increased by 1 $\sigma$, 
\emph{dashed-dotted lines}: HI from \cite{Nakanishi:2003eb} and H2 from \cite{Nakanishi:2006zf}, diminished by 1 $\sigma$.
In the lower right panel we show the relative difference between the $\pi^0$ components predicted by various gas assumptions and the one predicted by our reference model. 
\emph{Red}: $10^\circ < \mid b\mid < 20^\circ$, \emph{green}: $20^\circ < \mid b \mid < 60^\circ$ and \emph{blue}: 
$\mid b\mid > 60^\circ$. The dotted lines are practically overlapping due to the H2 models of \cite{1988ApJ...324..248B} and \cite{Nakanishi:2006zf} having a similar vertical scaling.}
\label{fig:GasVary}
\end{figure}

For the $\gamma$-ray spectra, the models of \cite{1976ApJ...208..346G}
(\cite{1988ApJ...324..248B}) for HI (H2) give -compared to the mean values of 
\cite{Nakanishi:2003eb,Nakanishi:2006zf}- an increased $\pi^{0}$ and 
bremsstrahlung flux by $\simeq 50\%$ in the entire range of the spectra and at latitudes above $10^{\circ}$ as shown in Fig.~\ref{fig:GasVary}.
We note that $\gamma$-ray data favor the models \cite{Nakanishi:2003eb} and 
\cite{Nakanishi:2006zf} for the HI and H2 gas distributions, while the models of \cite{1976ApJ...208..346G, 1990ARA&A..28..215D} and \cite{1988ApJ...324..248B} are marginally disfavored, from observations at the highest $\mid b \mid > 60^{\circ}$, or intermediate $10^{\circ} < \mid b \mid < 20^{\circ}$ latitudes (see Table~\ref{tab:Param}). This is mainly due to the higher density of the local H2 local, as shown in Fig.~\ref{fig:HI_H2_profiles} (lower row).

That can also be understood from Fig.~\ref{fig:GasVary} (lower right panel) where we plot the relative difference in the $\pi^{0}$ $\gamma$-ray fluxes between the H2 gas model of \cite{1988ApJ...324..248B} and the reference H2 model. The differences in the $\pi^{0}$ fluxes between these models are of $O(1)$. We also note that in the original parametrization of \cite{1988ApJ...324..248B}, the assumed $X_{CO}$ factor $\equiv N(H2)/W(CO)$ was taken to be constant, equal to $(2.8 \pm 0.4) \times 10^{20}~\cm^{-2}{\rm K}^{-1}\km^{-1} s$. This is a factor of 2 higher than the $X_{CO}$ factor of Eq.~\ref{eq:XCO}, that is in better agreement with recent $\gamma$-ray analysis of the ISM in the outer part of the Galaxy \cite{Tibaldo:2010yu}. Had we used the constant value for the $X_{CO}$ factor, our $\pi^{0}$ and bremsstrahlung fluxes would be enhanced, bringing them in more tension with the $\gamma$-ray data. Also we alternatively use the 1$\sigma$ higher and 1$\sigma$ lower values for the H2 density profile in $r$ of \cite{Nakanishi:2006zf}, which we plot also in Fig.~\ref{fig:GasVary} and~\ref{fig:HI_H2_profiles}(bottom), using however the mean values of \cite{Nakanishi:2003eb} for the HI gas. We note that both the 1$\sigma$ higher and the 1$\sigma$ lower cases are in tension with the data as well. 

From Fig.~\ref{fig:GasVary} (lower right panel) it can be also seen that switching from the older parametrization \cite{1976ApJ...208..346G, 1990ARA&A..28..215D}, to the newer one \cite{Nakanishi:2003eb} for the HI gas has much smaller effect compared to switching between H2 models. The reason for these much smaller differences in HI is that the steeper decrease with distance from the Galactic plane present in \cite{1990ARA&A..28..215D} (see Fig~\ref{fig:HI_H2_profiles} top right panel) as opposed to \cite{Nakanishi:2003eb}, compensates for its suggested higher density on the disk 
\cite{1976ApJ...208..346G} (see Fig.~\ref{fig:HI_H2_profiles} top left panel). 
Recently \cite{Timur:2011vv} has also shown the importance of the uncertainties
of the gas models and especially that of the $X_{CO}$ factor, suggesting as we do that the uncertainties in the H2 gas distribution are the greatest (see relevant discussion of \cite{Timur:2011vv}).

The increased number of target nuclei in the gas models \cite{1976ApJ...208..346G, 1990ARA&A..28..215D,
1988ApJ...324..248B}, results in the need of a faster escape of CR nuclei from the Galaxy, in order not to overproduce secondaries. This will then increase the diffusion coefficient normalization to keep the same B/C flux ratio. Therefore $e^{\pm}$ would propagate to larger distances from the Galactic disk, resulting in a 10\% increase of the observed IC flux, as well as to an effect on the $\pi^{0}$ and bremsstrahlung components.
 
A possible way to reconcile an increased number of target nuclei in the ISM gas would be to decrease the thickness of the diffusion halo.
That is shown in Table~\ref{tab:Param2} where we show the effects of interplaying the gas distribution and the thickness of the diffusion halo.
Our model ``KRA1-20 Bronf" represents our thinnest diffusion halo with the highest ISM gas assumption, which can not be considered in tension with any data. Yet the reverse case, of a thick halo with a low ISM gas (``KRA10-20 NS low") tends to under-predict the $\gamma$-ray spectra at low latitudes. Finally, the extreme cases of a very thin diffusion halo and a low ISM gas assumption (``KRA1-20 NS low"), or a high ISM gas in a thick diffusion halo (``KRA10-20 Bronf"), while can still predict the $\bar{p}$ flux in good agreement to the data, are in systematic tension with the $\gamma$-ray fluxes by either systematically under-predicting them or systematically over-predicting them.
Thus a combined analysis of CRs and $\gamma$-rays as is ours can probe the uncertainties in the large scale gas distributions, which when using only CR information, are large, especially for the H2 gas.

%We have also checked the effect of interplaying of gas distribution and the thickness of the halo 
%~\ref{tab:Param2}. As we described in~\ref{subsec:VaryDiffuse} and it is clear from Fig.~\ref{fig:zdVary}, the %predicted gamma ray spectrum for a thin halo ("KRA1-20") is below the data. This under-prediction can be %compensated with over prediction of models with more gas ("Bronf"). On the opposite side, the combination of 
%thin halo and lower gas densities ("NS low") leads to even less gamma ray spectrum.  

\subsection{SNR Distribution}
\label{sec:SRN}

\begin{figure}[tbp]
\begin{center}
\includegraphics[width=0.45\textwidth]{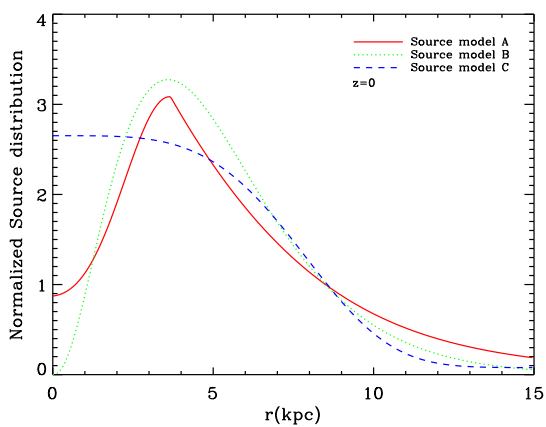}
\end{center}
\caption{SNR radial profiles.
\emph{Red}: parametrization of \cite{Ferriere:2001rg} given in Eq.~\ref{eq:FerriereDistr},
\emph{green}: discribed by Eq.~\ref{eq:GalpropSource} \cite{Strong:1998pw} and \emph{blue}: 
given by Eq.~\ref{eq:UlternativeSource}.}
\label{fig:SNRdistr}
\end{figure}

Since in our study we use the $\gamma$-ray data above $\mid b \mid > 10^\circ$ we are not very sensitive to the SNR source distribution. In particular, we are poorly sensitive to the inner few kpc, that are also hardly probed by direct observations of single sources. To study the significance of the source distribution in the inner part, we use three different source profiles as shown in Fig.~\ref{fig:SNRdistr}.
The Ferriere {\it et al.} \cite{Ferriere:2001rg} profile (``Source A"), our reference assumption, is recovered from Type I and Type II supernovae distribution models and is defined by
\begin{eqnarray}
f_{s}(r,z)&=& 0.138 e^{-(r-r_{\odot})/4.5 - \mid z\mid /0.325} \nonumber \\ 
&+& \left(0.79 e^{-(z/0.212)^{2}} + 0.21 e^{-(z/0.636)^{2}} \right) 
 \times 0.943 e^{-(r^{2} - r_{\odot}^{2})/6.8^{2}} \; r>3.7~\kpc \nonumber \\
 &+& \left(0.79 e^{-(z/0.212)^{2}} + 0.21 e^{-(z/0.636)^{2}} \right) \times 3.349 e^{-(r-3.7)^{2}/2.1^{2}} \; r<3.7~\kpc. 
\label{eq:FerriereDistr}
\end{eqnarray}
The relative normalization of the two populations is based on the averaged occurrence frequencies
of SNe Type I and Type II in other galaxies \cite{1989ApJ...345..752E, 1997A&A...322..431C}, while the spatial profiles are based on the assumption that Type I have a distribution similar to that of old disk stars \cite{1987ARA&A..25..603F}, while Type II (which are the most frequent) are tightly correlated to the arms. Especially in the inner 3.7 kpc the profile of the Type II is correlated to our Galaxy's pulsar distribution and thus sensitive to selection effects.

Also the distribution given in Eq.~\ref{eq:GalpropSource} has been used extensively in GALPROP \cite{Strong:1998pw,Strong:2007nh} as a conventional distribution for SNRs.
\be
f_{s}(r,z) = \left(r/r_{\odot}\right)^{2.35} e^{-5.56\left(r-r_{\odot}\right)/r_{\odot}} 
e^{-\mid z \mid/0.2},
\label{eq:GalpropSource}
\ee
This distribution parameterizations (``Source B") comes from actual observations of galactic SNe \cite{1996A&AS..120C.437C} with its detailed values being selected to agree with the diffuse fluxes of analyzed \textit{EGRET} $\gamma$-ray data \cite{1996A&A...308L..21S,Strong:1998pw}. As a result, this parametrization has to be taken with some care, in particular with respect to its prediction on the inner parts of the Galaxy.

Since both parametrizations of \cite{Ferriere:2001rg} and \cite{Strong:1998pw} are least predictive towards the inner regions of the Galaxy, we study also a third parametrization described by ``Source C": 
\be
f_{p}(r,z) = \left(0.078 + 2.57 e^{-\left(r/r_{\odot}\right)^{4}} \right) e^{-\mid z \mid/0.2}.
\label{eq:UlternativeSource}
\ee

\begin{figure}[tbp]
\begin{center}
\includegraphics[width=0.45\textwidth]{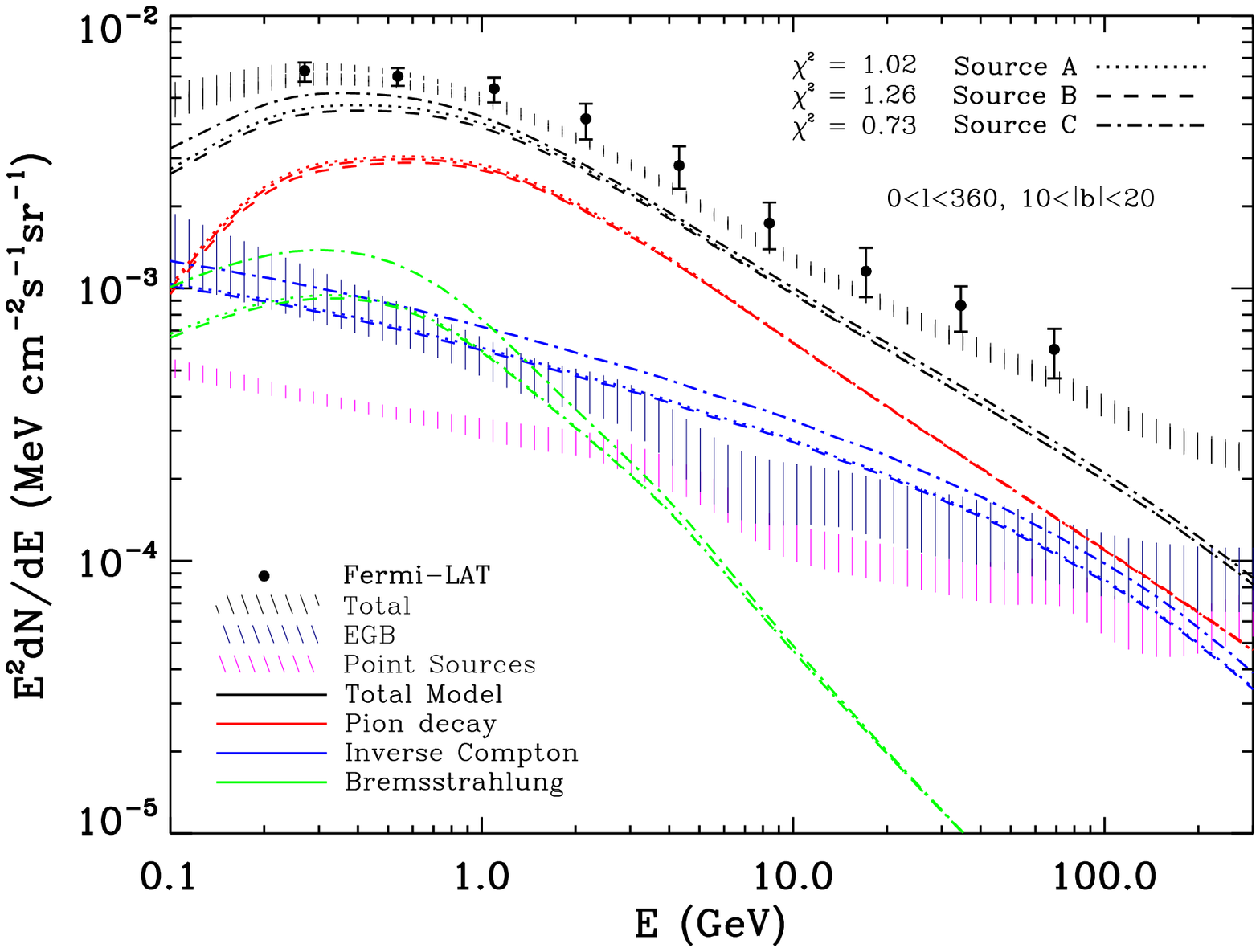}
\includegraphics[width=0.45\textwidth]{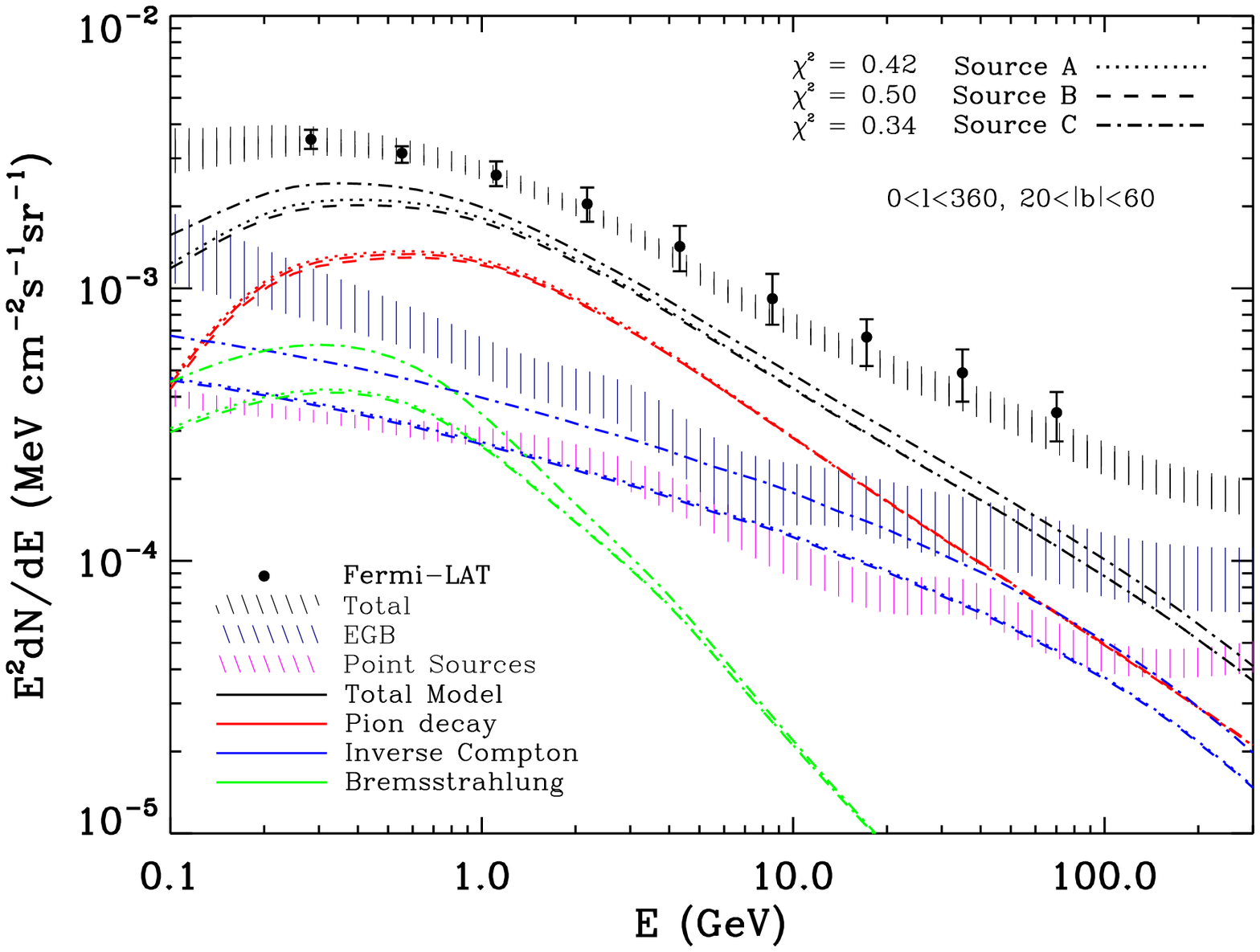}
\includegraphics[width=0.45\textwidth]{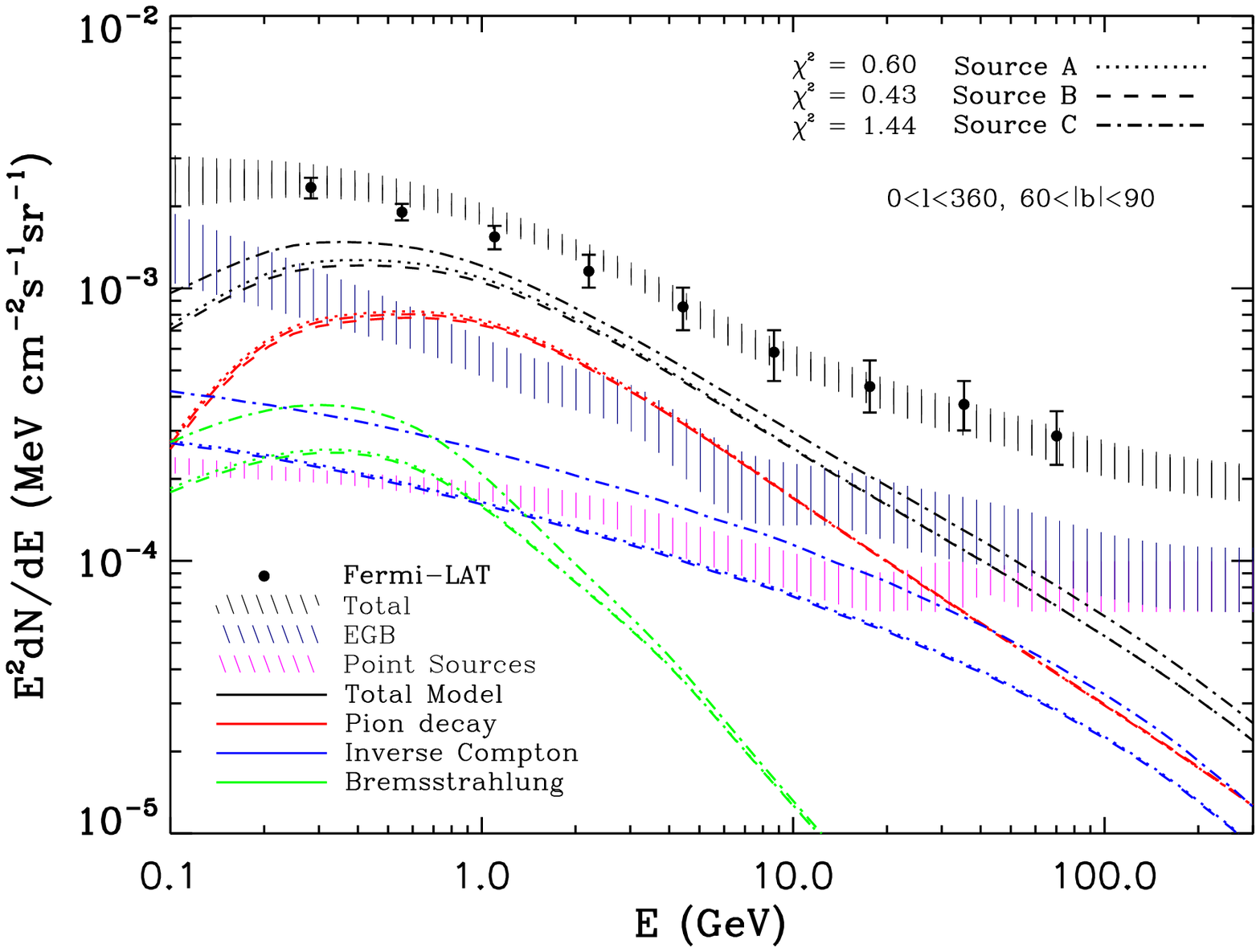}
\end{center}
\caption{Gamma-ray fluxes assuming different models for the SNR distribution. 
Plots refer to the different sky regions of our study. For each SNR model we 
refit the diffusion parameters, as is shown in Table~\ref{tab:Param}.
\emph{Dotted lines}: Source A of Eq.~\ref{eq:FerriereDistr} \cite{Ferriere:2001rg},
\emph{dashed lines}: Source B of Eq.~\ref{eq:GalpropSource} \cite{Strong:1998pw},
\emph{dashed-dotted lines}: Source C of Eq.~\ref{eq:UlternativeSource}.}\label{fig:SNRVary}
\end{figure}

As is also shown in Fig.~\ref{fig:SNRdistr}, this parametrization gives an almost constant radial distribution in the inner 3 kpc, 
while its averaged distribution at $r > 5$ kpc 
is similar to that of ``Source A" and ``Source B", with its vertical behavior being the same 
as that of ``Source B". Since it is expected that Type II SNe are correlated to the spiral arms, and since the inner few kpc are populated by old stars, we use ``Source C" as a probe of the maximal effect that the uncertainties in the SNRs distribution could have on the diffuse $\gamma$-ray data analysis, rather than as an optimal model for SNRs. Yet as it can be seen from Table~\ref{tab:Param} and Fig.~\ref{fig:SNRVary}, where we plot the total $\gamma$-ray fluxes for the three SNR models, ``Source C" indeed provides a better fit to the $\gamma$-ray data at low latitudes, that probe the inner parts of the Galaxy. That could be an indication that either ``Source A" and ``Source B", being connected to observations, indeed under-predict the ``recent" SNR density towards the inner part of the Galaxy, or simply, as suggested in section~\ref{sec:astro_results_RefModel}, that unresolved point sources are present. Also changing the gas assumption only in the inner few kpc could have an effect. To probe such uncertainties a study including lower latitude regions toward the GC would be very well suited.

\subsection{pp-Collision $\gamma$-ray spectra}   

\begin{figure}[tbp]
\begin{center}
\includegraphics[width=0.45\textwidth]{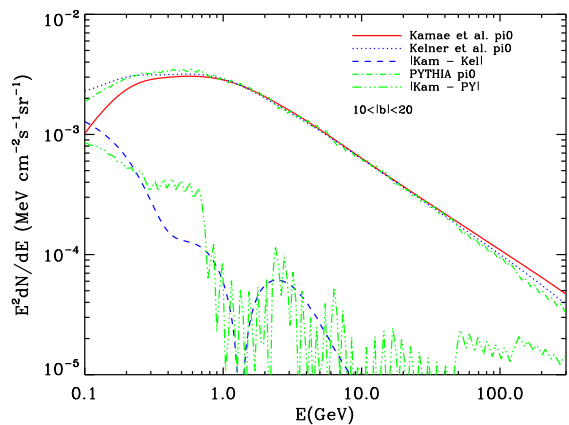}
\end{center}
\caption{The diffuse $\pi^0$ component of the flux using the Kamae {\it et al.} \cite{Kamae:2006bf}
parametrization for $pp$ collisions vs that from \cite{Kelner:2006tc} and \textsc{Pythia} runs.
\emph{Red solid}: Kamae {\it et al.} parametrization. \emph{Blue dotted}: using \cite{Kelner:2006tc} 
and \emph{blue dashed}: its difference from the \cite{Kamae:2006bf}. \emph{Green dashed-dotted}:
using \textsc{Pythia}, and \emph{green dashed-dotted-dotted-dotted} its difference from the \cite{Kamae:2006bf}. 
We normalize the $\pi^{0}$ fluxes at 10 GeV.}
\label{fig:ppColpi0}
\end{figure}
In Fig.~\ref{fig:ppColpi0} we show the $\pi^{0}$ component of the diffuse $\gamma$-ray flux at $10^\circ<\mid b \mid <20^\circ$ using three different parametrizations for the $\gamma$-ray spectra produced by $pp$ collisions. We clarify that we call these spectra ``$\pi^{0}$ spectra" since their contribution to the $\gamma$-ray
spectra is the dominant one, in all parameterizations. However also the $\gamma$-rays from the decay channels 
of other produced mesons such as $K^{\pm}$, $K^{0}$, $\eta$, $D^{\pm}$, $D^{0}$ are taken into account.

As a reference parametrization we use that of Kamae {\it et al.}~\cite{Kamae:2006bf, KamaePrivate} that was derived for the cross-sections of diffractive, non-diffractive and excitation of resonance processes, based on simulation and experimental data on $pp$ collisions.
We compare that parametrization to that from Kelner {\it et al.}~\cite{Kelner:2006tc} that was based on running SIBYLL \cite{Fletcher:1994bd} simulations of $pp$ collisions and also to the $\gamma$-ray spectrum from our own \textsc{Pythia} (version 6.4) simulations \cite{Sjostrand:2006za}.
For the Kamae {\it et al.} parametrization we use the updated tables information \cite{KamaePrivate} relevant to those of tables 2 and 3 of~\cite{Kamae:2006bf} that are used in eq.~5-14 of~\cite{Kamae:2006bf}. For the Kelner {\it et al.} parametrization~\cite{Kelner:2006tc} we used the information given in eq.~58-61 of~\cite{Kelner:2006tc}, while in our \textsc{Pythia} simulations we run $pp$ collisions with center of mass energy from 2.33 GeV up to 7 TeV, with subsequent decay of all mesons and including final state radiation. We keep the information for the 3D momenta of the final stable particles, which we re-boost to the proper observer frame (where a CR $p$ hits a practically stable ISM $p$).  
We see that the $\gamma$-ray spectra, normalized at 10 GeV, agree well from 100 GeV down to energies of 1 GeV where it is expected that the simulations from~\cite{Kelner:2006tc} and \textsc{Pythia} would be no longer reliable.
Thus it is safe to say that uncertainties in the $\gamma$-ray spectra produced by $pp$ collisions that could be due to missing processes in the parametrization of \cite{Kamae:2006bf}, are too small to have a strong impact on the constraints imposed on the ISM properties that we have described using the combined CR and $\gamma$-ray spectra.  

\section{Conclusions and Discussion}
\label{sec:Conclusions}

The results of the analysis show that combining the local CR measurements,
which are rather powerful to constrain the local averaged properties of propagation 
and source/gas distribution, with the diffuse gamma-ray fluxes at intermediate 
and high latitudes in the energy range currently covered by Fermi, can be useful
in giving some constraints also on the global galactic parameters.
We have tested a large subset of the benchmark models selected to depict rather 
diverse settings for modeling propagation and the ISM. While many of these models 
provide a good fit to both the CR and the $\gamma$-ray data, still there are a few 
scenarios that our analysis can disfavor.

In particular, we have studied rather extreme limits on CR diffusion galactic profiles,
ranging from essentially constant diffusion coefficient everywhere in the Galaxy, down to
exponentially suppressed vertical and radial profiles with 1 kpc scale hight and 5 kpc radial scale. The combined fit of CRs and
$\gamma$-rays suggests a slight preference for thicker diffusion zones, while there
is a weak dependence on the variation of the diffusion coefficient in the radial direction, which is however better probed and constrained by the antiproton spectrum.
While this result is not conclusive, it suggests a trend that better statistics and smaller
systematics on the $\gamma$-ray spectra (soon to come) will lead in testing specific models
for the position dependence of the diffusion coefficient $D(r,z)$ of CRs in the Galaxy.
The high statistics measurements of the local flux of radioactive isotopes by AMS-02 
\cite{Kounine:2010js}, placed on the International Space Station, will
add further information on the vertical thickness of the diffusion region, possibly allowing to break the 
degeneracy between thicker regions of emissivity populated by CRs diffusing out of the galactic disk
and exotic sources with an intrinsically thicker scale height, such as from DM.

Moreover, the current diffuse $\gamma$-ray spectra can discriminate 
(and even constrain) profiles for the ISM gasses. Before \textit{Fermi}-LAT $\gamma$-ray 
data, in order to place constraints on the ISM properties, the secondary to primary CR
spectra were used, with the best measured data sets coming from B/C and $\bar{p}/p$. Yet,
with only the CR data as a handle, a variation of the large scale gasses 
distributions could almost always be compensated by changing the diffusion properties 
(mainly the normalization of the diffusion coefficient). By exploiting the $\gamma$-ray diffuse 
fluxes above $\mid b \mid > 10^\circ$ and combining them with the CR data, we have shown that 
we can actually break the degeneracy between diffusion and ISM gas 
distribution. In fact, thanks to the expected improvement both in statistics and systematics 
errors of the $\gamma$-ray data from \textit{Fermi}-LAT, and even more with the CR spectral 
data up to Fe from 0.1 GeV/n to at least 100 GeV/n from AMS-02, we can be optimistic in 
further constraining the properties of the ISM gas distributions, within the next 
few years. \footnote{Smaller scale features are much better probed by synchrotron data, 
as for example has recently been done in \cite{Planck:2011aj}.}

On the other hand, we also find that CR and $\gamma$-ray data do not constrain strongly the diffusion spectral index $\delta$ 
within the range we considered.

Furthermore, we have discussed the implications from the recently found rigidity break 
in the protons and He CR spectra \cite{Adriani:2011cu} (confirmed also by 
\cite{Yoon:2011zz}). We have addressed the possibility of discriminating whether the break
is in the injection spectrum (connected to either acceleration effects in the sources, or to the presence
of an extra population of primary sources injecting CRs with harder spectra) or in the energy dependence of the diffusion coefficient.
We have found that the galactic diffuse $\gamma$-rays cannot be used to this aim, neither with the current nor with the near future projected accuracy of the spectra, 
leaving this task to other observables, such as antiprotons as suggested by~\cite{Evoli:2011id}. 

As a final remark, we have shown that our analysis is robust with respect to uncertainties in the parameterization of
the $\gamma$-ray spectra produced in $pp$ collisions.

Having achieved with these results a better understanding of the contributions of the astrophysical components to the diffuse $\gamma$-rays,
 a natural application will be to place limits on a possible exotic contribution to the high latitude $\gamma$-ray flux. 
Our forthcoming analysis devoted to the search for a DM signal will be of particular relevance in this respect.
%------------------------------------------------------------------------------
\section*{Acknowledgments}
We warmly thank T. Kamae and H. Lee for sharing with us their 
up-to-date parameterization of spectra resulting from $pp$ interactions. 
We are also thankful to M. Boezio, G. Dobler, D. Gaggero, S. Leach, P. D. Serpico and 
N. Weiner for valuable discussions we have shared.
LM acknowledges support from the State of Hamburg, through the Collaborative Research program ``Connecting Particles with the Cosmos'' within the framework of the LandesExzellenzInitiative (LEXI).

\begin{appendix}

\section{Synchrotron losses of CR electrons}
\label{sec:SynchLoss}

Cosmic Ray electrons as they propagate in the Galaxy, lose energy via
synchrotron and bremsstrahlung radiation, and by up-scattering low energy 
photons to higher energies (as X-rays or $\gamma$-rays) or by ionization
losses inside gasses. Adiabatic losses can also be important inside expanding
SNRs but do not matter when CRs are considered free to diffusively propagate 
in the interstellar medium (and not inside some expanding volume of matter).
Above energies of several GeVs, $e^{\pm}$ energy losses via inverse Compton 
scattering and synchrotron radiation become most important, while at lower 
energies and denser ISM environments bremsstrahlung and ionization can 
dominate.

\begin{figure*}
\begin{center}
\includegraphics[scale=0.90]{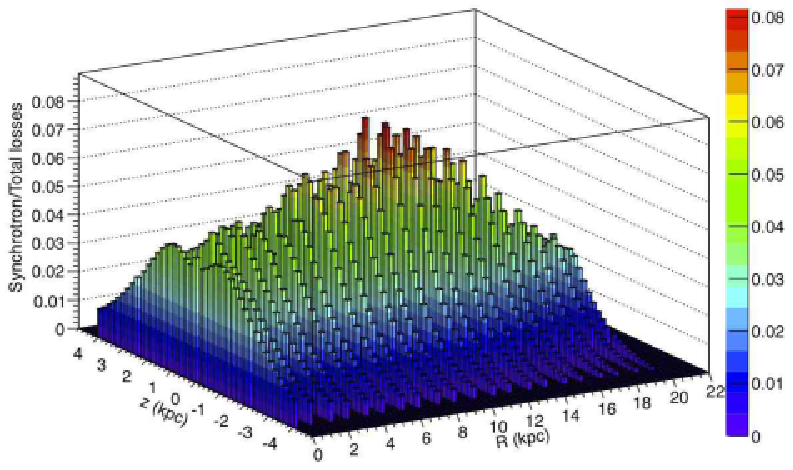}
\hspace{-0.2cm}
\includegraphics[scale=0.72]{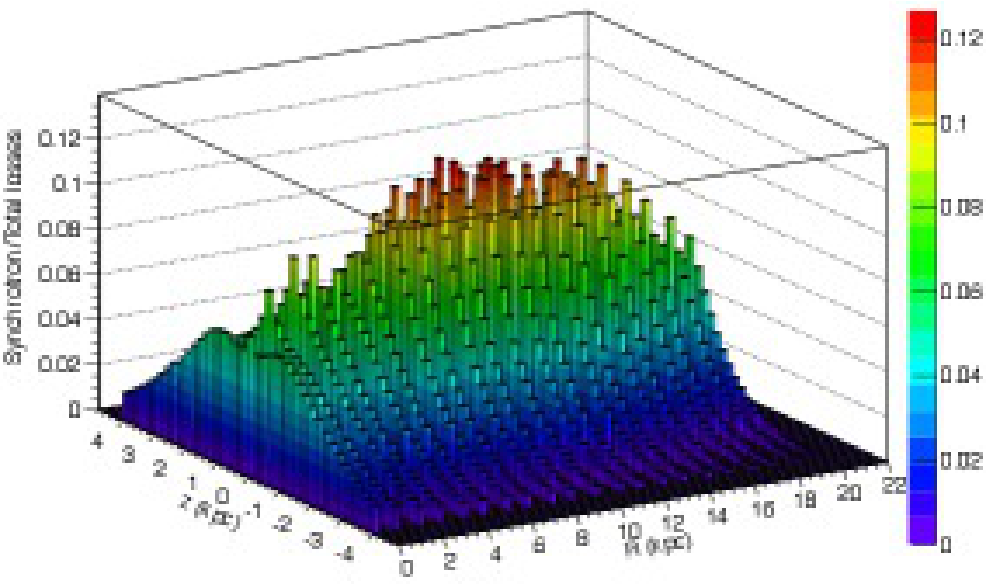}\\
\includegraphics[scale=0.90]{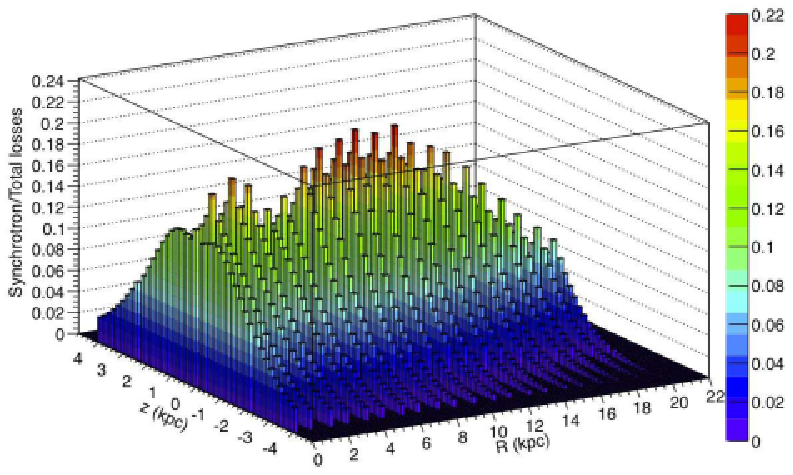}
\end{center}
\caption{Ratio of the energy loss rate of 1 GeV(\textit{top left}), 10 GeV 
(\textit{top right}) and 100 GeV (\textit{bottom}) electrons due to 
synchrotron radiation to the total energy loss rate.}
\label{fig:synch_to_tot_losses}
\end{figure*}

As we have described in section~\ref{sec:Diffusion_Bfields}, we expect a 
correlation between the large scale profile of the galactic B-fields and 
that of CR diffusion. Yet, 
in this work, for simplicity we have chosen to keep a specific parametrization 
of the B-field as given in eq.~\ref{eq:Bfield} (following WMAP 
\cite{MivilleDeschenes:2008hn}), while changing the 
diffusion coefficient's profile described by eq.~\ref{eq:DiffCoef}. 
Such a simplification is 
possible since the B-field assumptions in our code can influence only the
synchrotron emissivity calculations, which are not presented here and are left
for future work, and the synchrotron energy losses of CR electrons.
The latter are subdominant at all energies and positions of the Galaxy with 
respect to the energy losses due to up-scattering CMB, infrared and optical 
photons, or bremsstrahlung radiation, by at least a factor of 3 for typical 
values of galactic magnetic fields. For the magnetic field that we used in 
this paper, we show in Fig.~\ref{fig:synch_to_tot_losses} the ratio of 
synchrotron energy losses rate to the total energy losses rate for electrons 
of $E =$ 1, 10 and 100 GeV.

As can be seen, the synchrotron energy losses are in no part of the Galaxy, 
dominant, accounting at most up to 22$\%$ of the total electron energy losses
at 100 GeV, locally. The synchrotron radiation losses are more important 
at very high energies, where the ICS losses due to optical and infrared 
photons up-scattering become less efficient as the Klein-Nishina 
cross-section decreases away from the Thomson cross-section value.
Also, far away from the galactic disk, synchrotron radiation losses
(which scale with the square of the magnetic field strength) 
drop because of the exponential decrease of the B-field (eq.~\ref{eq:Bfield}),
while the energy losses due to CMB up-scattering remain the same.

Therefore, we can treat the effects of the magnetic fields on diffusion and 
on energy losses to a good approximation separately.

\section{Impact of a 2D vs a 3D ISM gas distribution on the diffuse $\gamma$-ray spectra}
\label{sec:2Dvs3D}

The \textit{Fermi}-LAT $\sim 1^{\circ}$ angular resolution (for energies above
few GeV), allows to trace in some detail the morphology of the $\gamma$-ray 
emissivity associated to the gas in the Galaxy. Thus one can use a 
3-dimensional ISM gas distribution to compare to the $\gamma$-ray data.
In this paper we have instead used 2-dimensional spatially smoothed gas
distributions, which does not account for any of the small-scale features in 
the $\gamma$-ray maps. 

\begin{figure}[tbp]
\begin{center}
\includegraphics[scale=.38]{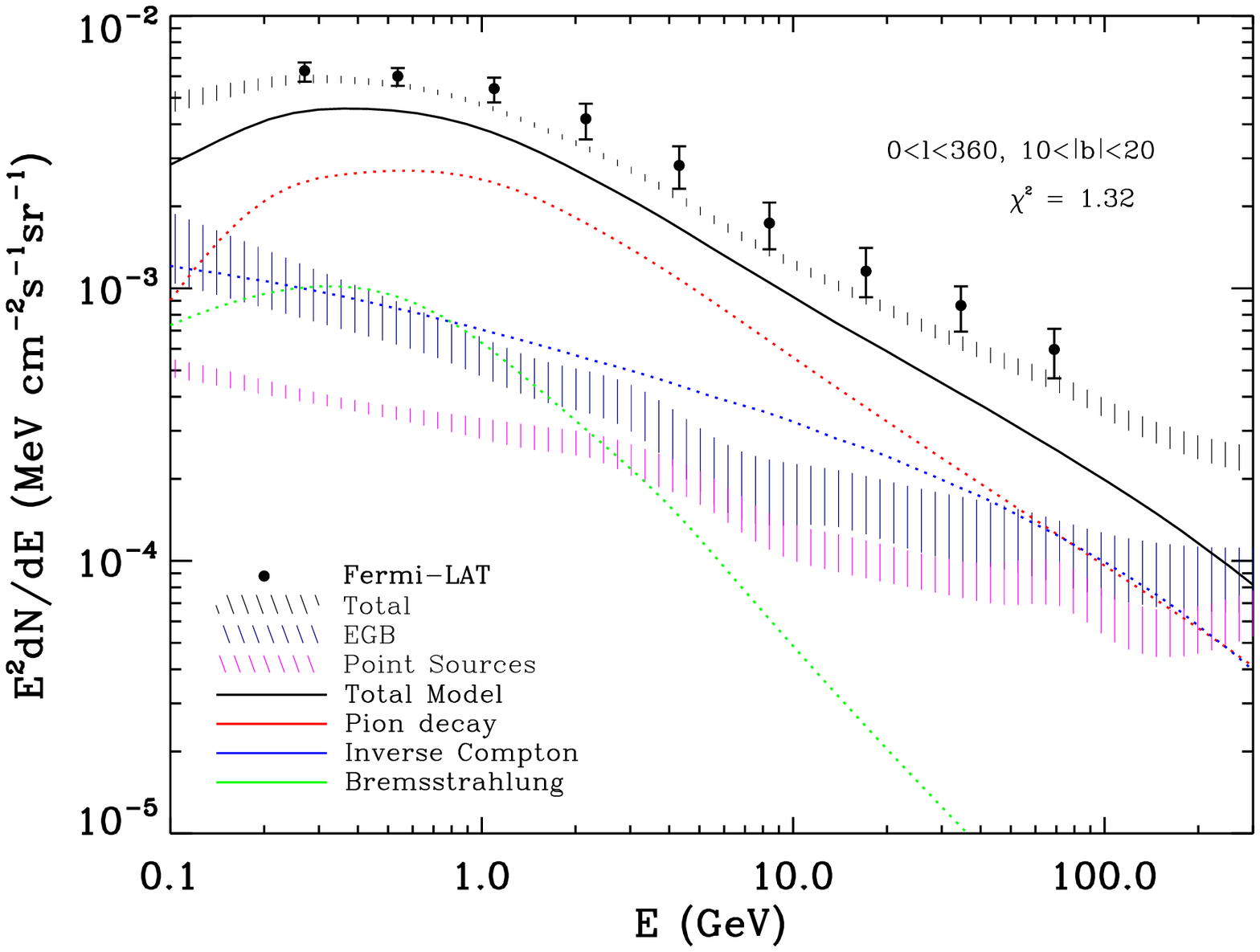}
\hspace{-0.2cm}
\includegraphics[scale=.38]{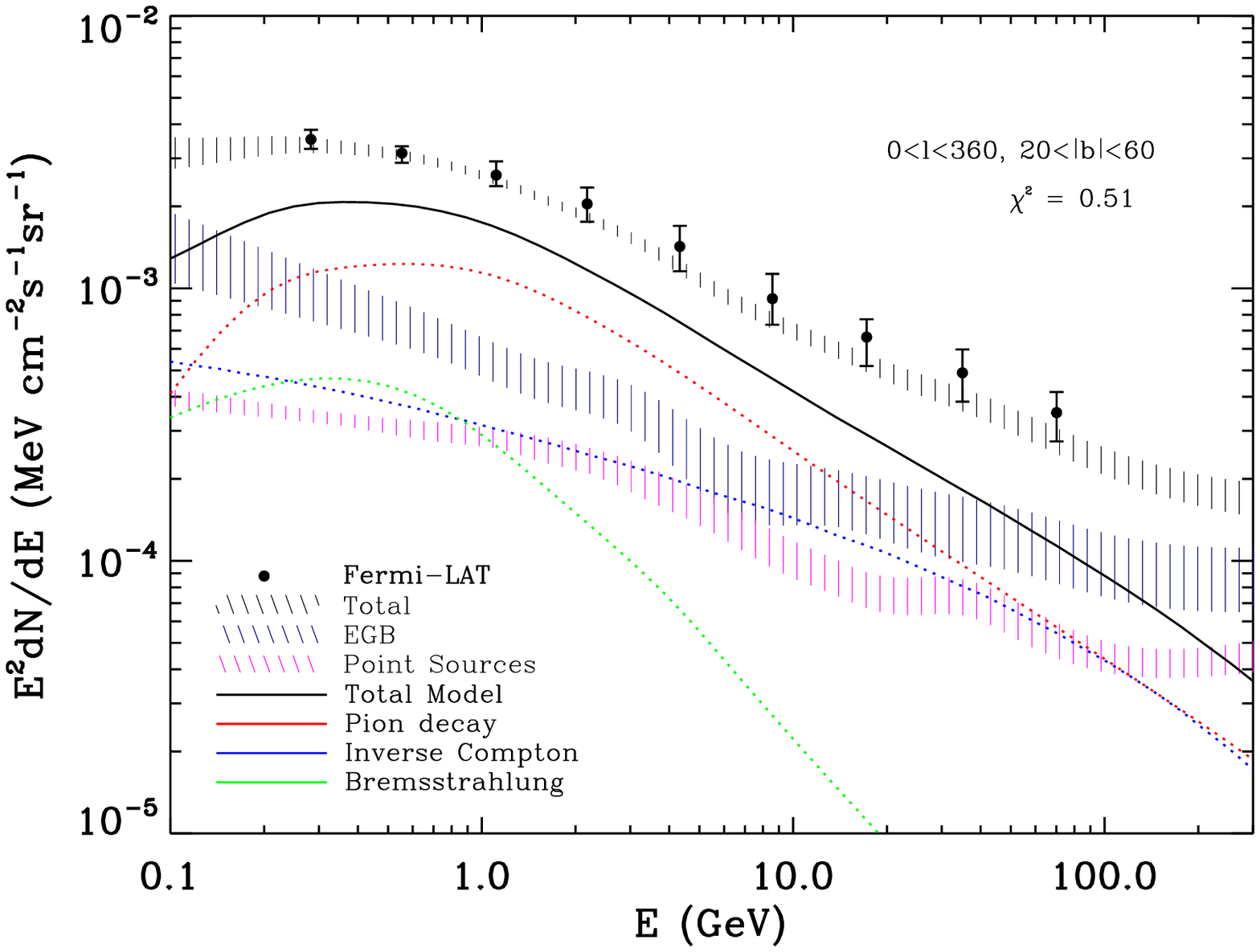}\\
\includegraphics[scale=.38]{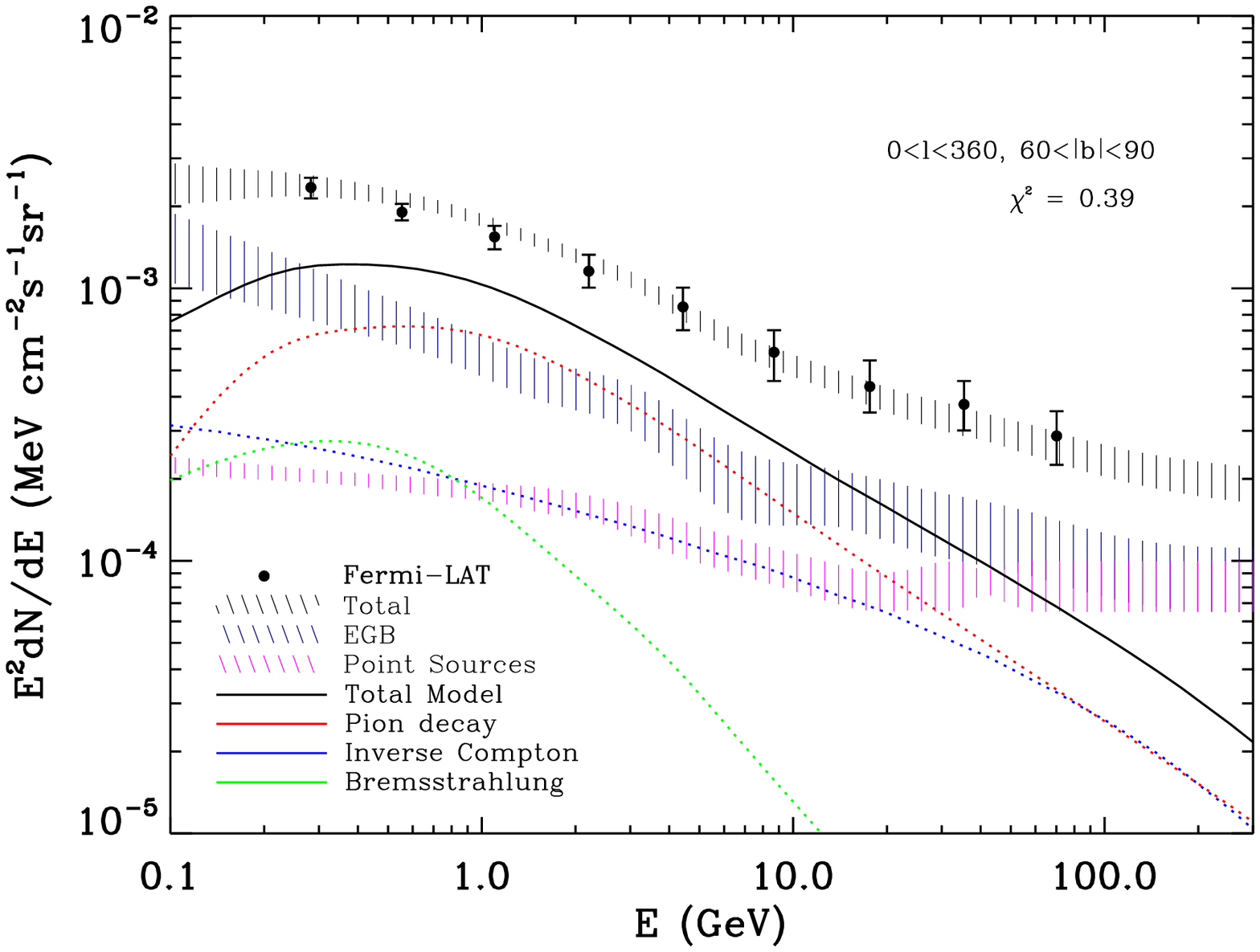}
\end{center}
\caption{Gamma-ray spectra for the 3 regions of interest using the 3D 
formalism in DRAGON.}
\label{fig:gamma_3D}
\end{figure}

Yet, our goal has \textit{not} been to study or interpret these structures,
but rather the larger scale properties of the Galaxy, that are incorporated
in the $\gamma$-ray spectra measured in the very wide angular windows that 
we use.
To illustrate the minimal impact on our analysis of having averaged out the 
small-scale features, we consider our reference propagation model and 
calculate in the three regions: 
$0^{\circ}<l<360^{\circ}$, $10^{\circ}<\mid b\mid <20^{\circ}$
/ $20^{\circ}<\mid b\mid <60^{\circ}$ / $\mid b\mid >60^{\circ}$, 
the diffuse $\gamma$-ray spectra using a 3D model for the HI and H2 gas,
equivalent to the 2D reference model; results are shown in 
Fig.~\ref{fig:gamma_3D} and should be compared to those in 
Fig.~\ref{fig:RefAstroModelGammas} obtained under the same configuration
for cosmic-ray propagation but with the 2D gas model.

\begin{figure}
\begin{center}
\includegraphics[scale=.53]{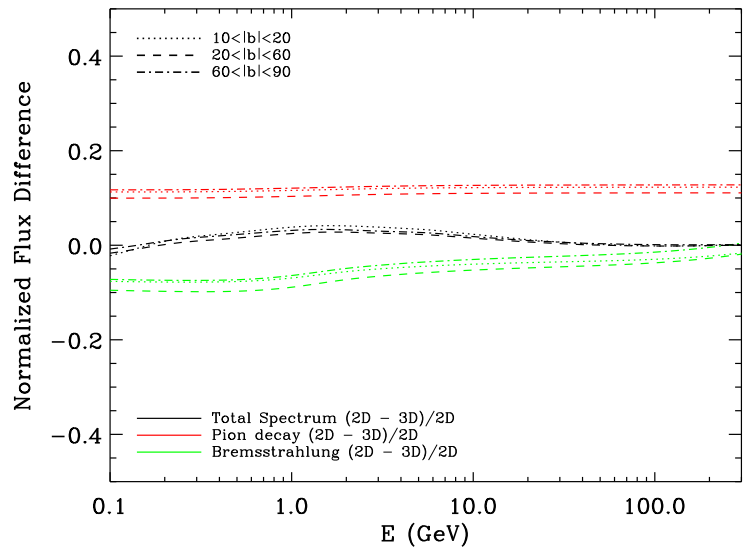}
\end{center}
\caption{Comparison between the $\gamma$-ray spectra between the 3D and 
the 2D formalism with DRAGON. Normalized differences are given for the three 
regions of interest and for the tow galactic diffuse components that are 
affected by the gas distribution ($\pi^0$ and bremsstrahlung). We also show 
the normalized difference in the total $\gamma$-ray spectra.}
\label{fig:gamma_3D_VS_2D}
\end{figure}

The differences in the $\pi^{0}$ and the bremsstrahlung (relevant for ISM gas) 
are at the 10$\%$ level, in the three parts of the sky that we study. This 
is most clearly seen in Fig.~\ref{fig:gamma_3D_VS_2D}, where we show the 
normalized differences in using the 2D and the 3D gas distributions. 

These results are well within the accuracy needed for our analysis, in which
we have studied models of HI and H2 gas distributions that can differ by up to 
$30\%$ in their prediction of the $\pi^0$ and bremsstrahlung diffuse 
fluxes (Fig.~\ref{fig:GasComponents} bottom right).

The total galactic diffuse 
model predictions between the 2D and the 3D cases are actually less than 
5$\%$, as there are compensations between the different components.
Furthermore, given that in deriving our physical conclusions we use the 
$\chi^{2}$ analysis carried between the total $\gamma$-ray \textit{Fermi} 
fluxes and the total $\gamma$-ray predicted fluxes, which include the EGBR and 
point sources, the impact of using a 2D ISM gas model is minimal.
Finally using our 3D ISM gas model, we have checked that the contribution 
from the "Fermi Bubbles" \cite{Su:2010qj} (that we do not include explicitly)
in the windows of $0^{\circ} < l < 360^{\circ}$ is of roughly the same magnitude 
as the numerical uncertainties in calculating the $\gamma$-ray fluxes.
\end{appendix}

%------------------------------------------------------------------------------
\bibliography{GammaRays}
\bibliographystyle{apsrev}
%------------------------------------------------------------------------------
\end{document}